\documentclass[11pt]{article}

\usepackage[margin=1in]{geometry}
\usepackage{setspace}
\usepackage{amsmath,amssymb,amsthm}
\usepackage{mathtools}
\usepackage{graphicx}
\usepackage{booktabs}
\usepackage{threeparttable}
\usepackage{enumitem}
\usepackage{float}
\usepackage{needspace}
\usepackage{placeins}

\usepackage{natbib}
\usepackage[colorlinks=true,allcolors=blue]{hyperref}

\newif\ifblind
\blindfalse    

\newcommand{\E}{\ensuremath{\mathbb{E}}}
\newcommand{\Pp}{\ensuremath{\mathbb{P}}}
\newcommand{\VaR}{\ensuremath{\mathrm{VaR}}}
\newcommand{\ES}{\ensuremath{\mathrm{ES}}}
\newcommand{\Cov}{\ensuremath{\operatorname{Cov}}}
\graphicspath{{figures/}}

\theoremstyle{plain}
\newtheorem{theorem}{Theorem}[section]

\newtheorem{proposition}[theorem]{Proposition}
\newtheorem{corollary}[theorem]{Corollary}

\theoremstyle{definition}
\newtheorem{definition}[theorem]{Definition}
\newtheorem{assumption}[theorem]{Assumption}
\newtheorem{remark}[theorem]{Remark}

\ifblind
  \title{Backtesting High-Dimensional Expected Shortfall}
  \author{Anonymous}
  \date{}
\else
  \title{Backtesting High-Dimensional Expected Shortfall%
  \thanks{Escanciano acknowledges research support from grants PID2021-127794NB-I00, funded by MCIN/AEI/10.13039/501100011033, PID2025-170712NB-I00, funded by  ICIU/AEI/10.13039/501100011033, and Comunidad de Madrid, grants EPUC3M11 (VPRICIT) and H2019/HUM-589. We thank participants at the 2026 Zaragoza and NBER–NSF Time Series Conferences for useful comments.}}

  \author{
    Zaichao Du\thanks{School of Economics, Fudan University, Shanghai, China. Email: zaichaodu@fudan.edu.cn.}
    \and
    Juan Carlos Escanciano\thanks{Department of Economics, Universidad Carlos III de Madrid, Getafe (Madrid), Spain. Email: jescanci@econ.uc3m.es.}
    \and
    Pei Pei\thanks{Central University of Finance and Economics, Beijing, China. Email: ppeimetrics@cufe.edu.cn.}
  }
  \date{November 10 2026}
\fi

\begin{document}

\begingroup
\onehalfspacing
\maketitle

\begin{abstract}
Forecast evaluation increasingly uses high-dimensional
predictors, but estimation and regularization of large predictive models can distort conventional
backtests. This issue is particularly important for forecasting downside risk, where the signal-to-noise ratio is low. We develop out-of-sample Expected Shortfall (ES) backtests that remove the large first-order
bias effects of estimating high-dimensional Value-at-Risk, ES, and the evaluation instrument. In a linear ES specification, the proposed methods are easily implemented as Double Lasso with a tail-size-adjusted score as a generated dependent variable. More generally, we characterize orthogonal ES backtesting moments, and establish oracle
equivalence and valid inference under time-series dependence and
repeated estimation with overlapping and expanding training windows. Simulations
show that conventional backtests can severely overreject ES optimal forecasts, whereas orthogonalization improves size control.
An application to downside-growth forecasting for G7 economies finds
evidence against a US growth-only ES specification, even though
forecast-oriented regularization selects no additional financial
predictors. The application also shows that accounting for estimation
can change rejection decisions in either direction. These findings
highlight the distinct roles of predictive variable selection and
backtesting in the evaluation of high-dimensional
tail-risk models.
\end{abstract}
\noindent\textit{Keywords:} Expected Shortfall; Forecast evaluation;
High-dimensional prediction; Machine learning; Neyman-orthogonal scores.
\par
\noindent\textit{JEL:} C14; C52; G17; G21.
\par

\clearpage
\endgroup

\section{Introduction}\label{sec:intro}

Financial institutions and policymakers use downside-risk forecasts to
guide capital allocation, stress testing, and financial surveillance.
Expected Shortfall (ES), which measures the average outcome beyond a
Value-at-Risk (VaR) threshold, plays a central role in market-risk
regulation \citep{BCBS2017,BCBS2019}. Increasingly, these forecasts draw
on large financial and macroeconomic information sets through
regularization and machine learning
\citep{IMF2024GFSR,adrian2025mlgar,FSB2024AI}. Applications include
random forests for extreme financial stress, recurrent neural networks
for VaR and ES, and regularized forecast combinations
\citep{AldasoroHoerdahlSchrimpfZhu2025PFMSML,QiuLazarNakata2024,TaylorWang2026}.
This creates a problem for model review: estimation and regularization
errors can cause an optimal forecast ES model to fail a conventional
backtest, prompting an unwarranted conclusion about its out-of-sample evaluation.

This paper develops ES backtests that account for
high-dimensional estimation. The main difficulty is that ES forecasts depend
on an estimated tail threshold and an estimated conditional tail mean.
A useful construction transforms the outcome using estimated VaR so
that its conditional expectation equals ES at the true quantile
\citep{Barendse2020TailIQE,Barendse2023ESLasso}. This transformation leads to a tail-size-adjusted score that removes first-order sensitivity to VaR estimation, but a backtest based
on the resulting ES residual remains sensitive to ES estimation.
With many predictors, regularization bias can be large and dominate the backtest's
sampling uncertainty. Holding evaluation observations out of the
training sample does not, by itself, remove this effect.

Our first contribution characterizes orthogonal backtesting moments
within a class formed from VaR and ES residuals. The characterization
leads to a simple correction: multiply the transformed ES residual by
a residualized test function. Residualization removes the component
that responds to local changes in the ES model. For linear ES
specifications, it requires a regression of the test function on the
full set of ES predictors, using regularization when needed. The
resulting moment is, to first order, insensitive to the estimation of VaR,
ES, and the projection, including when the test function depends on
estimated ES. The proposed method can be easily implemented as a three-step Double Lasso regression with a tail-size-adjusted score as a generated dependent variable, generalizing the two-step setting of \citet{chernozhukov2018double,chernozhukov2022locally} to a time series ES setting.
We allow flexible, including nonparametric and machine-learning,
estimation of VaR, which also helps isolate ES misspecification from
misspecification of the quantile model. With a consistently estimated
VaR, the backtest can target omitted predictors or nonlinear ES
dependence, giving rejections an interpretable direction for model
revision.

Our second contribution establishes oracle equivalence for backtests
with three estimated components---VaR, ES, and the residualized test
function---under dependent observations and overlapping and expanding estimation
windows. Nonsmooth losses and generated responses link the estimation
stages: changing VaR changes the responses used to estimate ES, which
can also change the evaluation instrument. We show when their combined
effect is asymptotically negligible, giving the feasible statistic the
same limiting distribution as if the population components were known.
Under the null of optimal forecasts, VaR and ES are correctly specified given the forecasting
information $X_t$, while the larger information set $\mathcal F_t$ may
contain additional predictive information. VaR hits and ES innovations
therefore need not be martingale differences. The oracle scores may be
serially correlated, and inference uses a long-run covariance estimator.
We verify sufficient conditions for sparse linear Lasso models under
polynomially decaying dependence and suitable finite-moment restrictions,
covering fixed, rolling, and recursive estimation. The technical contribution is a unified argument controlling how
estimation errors across the three linked stages accumulate in
the backtest when models are repeatedly updated. This establishes
valid inference under the stated dependence and moment conditions
while preserving the original forecasting procedures.

Our work connects two literatures. ES backtesting procedures use
realized tail losses and cumulative violations
\citep{AcerbiSzekely2014,DuEscanciano2017}, or build on the joint
elicitability of VaR and ES
\citep{FisslerZiegel2016,NoldeZiegel2017,BayerDimitriadis2022ESR}.
\citet{HogaDemetrescu2023} develop sequential monitoring procedures,
and \citet{WangWangZiegel2026} introduce model-free e-backtests valid
under optional stopping. Their ES e-backtest assesses whether reported
forecasts underestimate risk; our backtests assess the optimality of the
ES forecasts while accounting for high-dimensional estimation.
Estimation effects in low-dimensional settings are central to the
forecast-evaluation literature
\citep{west1996asymptotic} and are studied specifically
for VaR and ES backtesting by \citet{BarendseKoleVanDijk2023}.
We address the additional challenges posed by high dimensionality,
regularization, and the estimation of multiple nuisance components,
including the evaluation instrument.

The ES estimation literature develops robust inference with many
regressors in an independent and identically distributed (iid) setting;
see \citet{HeTanZhou2023,ZhangHeTanZhou2025,HondaPeng2026}. \citet{WuBradicTanZhou2025} develop robust high-dimensional
ES estimation and debiased inference for linear
combinations of ES coefficients under independent
sampling. Our inferential target is an out-of-sample dynamic
forecast evaluation problem using estimated forecasts
and instruments. We establish conditions for its validity
under serial dependence and repeated estimation with
overlapping and expanding training windows.
For time series, \citet{Barendse2023ESLasso} studies penalized ES
estimation under dependence and heavy tails.
We build on this approach to develop backtests that additionally
account for the estimation of residualized instruments and the
repeated estimation of forecasts using overlapping rolling or
recursive training windows. Forecast evaluation with smooth losses
that are already orthogonal and a fixed forecasting scheme is studied
by \citet{escanciano2025extending}; our baseline ES moments are
nonsmooth, three-step, and require an explicit orthogonality correction, while
our theory accommodates time-varying estimation under dependence.

Monte Carlo experiments connect the theory to finite-sample performance.
With dependent predictors and heavy-tailed innovations, baseline tests
that omit residualization can exhibit substantial size distortions,
whereas the orthogonal tests generally deliver rejection frequencies
closer to the nominal level. Omission experiments show that the proposed tests detect excluded
predictors, including designs with correct VaR and misspecified ES.
Test functions aligned with the omitted predictors often provide
greater power, illustrating the value of choosing economically
informative directions.

The empirical application asks whether macro-financial predictors
contain information about downside growth beyond recent growth history
in the G7 economies. Growth Expected Shortfall complements Growth-at-Risk
by measuring average outcomes in the lower tail \citep{adrian2019growth}.
Using 64 predictors and a common high-dimensional quantile first step,
we compare growth-only and expanded ES forecasting models. Under a
maintained high-dimensional quantile specification, the orthogonal backtest rejects the
US growth-only restriction in selected financial directions, even
after multiple-testing adjustment, although forecast-oriented
regularization selects no additional US ES predictors. The Canadian
results show that accounting for estimation can change rejection
decisions in either direction, depending on the test function.
Together, these findings illustrate why predictive variable selection
cannot substitute for backtesting, and why the latter must account for how the forecasts are estimated.

The broader contribution is to separate two sources of difficulty
in tail-risk model review: economically relevant departures from an
ES specification and estimation effects that can distort the evidence
against it. Our backtests remove the leading estimation effects
while allowing the choice of test functions to target departures
of substantive interest. The correction leaves the forecasts and
their estimation windows unchanged, making it an additional evaluation
step rather than a replacement for the forecasting procedure.
For risk managers and policymakers, the objective is not fewer
rejections, but interpretable evidence about which restrictions on
downside risk the data contradict.

The remainder of the paper is organized as follows.
Section~\ref{sec:tail} develops the orthogonal backtests and their
implementation. Section~\ref{sec:theory} establishes their asymptotic
properties and sufficient conditions for Lasso estimation.
Section~\ref{sec:mc} reports the Monte Carlo results, and
Section~\ref{sec:empirical} presents the empirical application.
Proofs and additional results appear in an Online Appendix.
\section{Backtesting High-Dimensional ES}\label{sec:tail}
\subsection{Set-up, targets, and two-step structure}\label{sec:tail:moments}

We observe a sample segment $\{(Y_{t+1},X_t):t=0,\ldots,T-1\}$ from a strictly stationary process, where $Y_{t+1}$ is the scalar
next-period outcome (e.g., a return, profit and loss, or GDP growth) and
$X_t\in\mathbb{R}^{p_T}$ is a possibly high-dimensional predictor vector
available at time $t$. Throughout, the predictor vector includes an intercept:
$X_t=(1,W_t')'$, where $W_t\in\mathbb R^{p_T-1}$ collects
the nonconstant predictors. Thus $p_T$ denotes the total
number of regressors, including the intercept. The vector $W_t$ may include the current outcome
$Y_t$ and its lags, together with other predictors; no disjointness between
$Y_t$ and $X_t$ is assumed. We allow $p_T$ to grow with the sample size.
For simplicity, we focus on one-step-ahead forecasts. The framework
extends to any fixed horizon $H>1$ under the corresponding
horizon-specific assumptions, replacing $Y_{t+1}$ with $Y_{t+H}$
and restricting estimation to training pairs $(X_s,Y_{s+H})$
with $s+H\le t$. Long-run covariance estimation accounts for
serial dependence, including that induced by overlapping
forecast horizons. 
Throughout, for a vector $v=(v_1,\ldots,v_d)'$, we write
$\|v\|=\|v\|_2=(\sum_{j=1}^d v_j^2)^{1/2}$
for the Euclidean norm,
$\|v\|_1=\sum_{j=1}^d |v_j|$ for the $\ell_1$ norm,
and $\|v\|_\infty=\max_{1\le j\le d}|v_j|$
for the supremum norm.
We also write $\|v\|_0$ for the number of nonzero
coordinates of $v$.
For a matrix $A=(A_{jk})$, $\|A\|_F$ and
$\|A\|_{\max}$ denote the Frobenius norm and the
largest absolute entry, respectively.
For a scalar random variable $Z$,
$\|Z\|_{L^q}=(\E|Z|^q)^{1/q}$, $q\ge1$.
Population prediction and empirical norms are defined
where they are first used. Henceforth, $\mathbf 1\{A\}$ is the indicator function of $A$, equal to one when $A$ is true and zero otherwise.

\paragraph{Conditional tail risk objects.}
Fix a tail probability level $\alpha\in(0,1)$ and focus on the left tail.
Let $F(\cdot\mid X_t)$ denote the conditional cumulative distribution function (CDF) of $Y_{t+1}$ given $X_t$.
Define the conditional $\alpha$-VaR and conditional $\alpha$-ES by\footnote{We report VaR and ES on the original scale of $Y_{t+1}$,
without reversing their signs to express them as loss measures.
Lower values therefore indicate greater downside risk.}
\[
\begin{aligned}
\VaR_{\alpha}(x)
  &:=\inf\{y\in\mathbb{R}:F(y\mid X_t=x)\ge \alpha\},\\
\ES_{\alpha}(x)
  &:=\E\!\left[Y_{t+1}\mid Y_{t+1}\le \VaR_{\alpha}(x),\,X_t=x\right].
\end{aligned}
\]

\paragraph{Two-step predictive models.}
We allow for general (possibly nonlinear) predictive specifications
\[
\VaR_{\alpha}(X_t)=q_{\alpha}(X_t,\gamma_0),
\qquad
\ES_{\alpha}(X_t)=e_{\alpha}(X_t,\beta_0),
\]
where $\gamma_0\in\Gamma\subseteq\mathbb{R}^{d_\gamma}$ and $\beta_0\in\mathcal{B}\subseteq\mathbb{R}^{d_\beta}$ may be high-dimensional.
A leading high-dimensional example for ES is the linear model $e_{\alpha}(X_t,\beta_0)=X_t'\beta_0$ (with $\beta_0$ sparse or approximately sparse); see \cite{Barendse2023ESLasso} and references therein. Thus, $d_\gamma$ and $d_\beta$ are also allowed to depend on and grow with $T$.

\paragraph{Conditional moment restrictions.}
Although ES is not elicitable by itself, it is jointly elicitable with VaR, which motivates
score-based estimation and backtesting for the pair; see, e.g., \cite{FisslerZiegel2016,patton2019dynamic}. These targets admit convenient conditional moment characterizations that we will use to build evaluation (backtesting) scores.

First, under Assumption~\ref{ass:CDF} below, an optimal VaR forecast satisfies the conditional
quantile condition
\begin{equation}\label{eq:VaRmoment_cond}
\E\!\left[\mathbf 1\{Y_{t+1}\le q_{\alpha}(X_t,\gamma_0)\}-\alpha \,\Bigm|\, X_t\right]=0 .
\end{equation}
Second, an optimal ES forecast is characterized by the conditional moment
\begin{equation}\label{eq:ESmoment_cond}
\E\!\left[\frac{Y_{t+1}}{\alpha}\,\mathbf 1\{Y_{t+1}\le q_{\alpha}(X_t,\gamma_0)\}-e_{\alpha}(X_t,\beta_0)\,\Bigm|\,X_t\right]=0 .
\end{equation}

\medskip

Define the filtration $\mathcal F_t := \sigma(Y_s,X_s:s\le t)$, and let $F_t$ denote the conditional CDF of $Y_{t+1}$ given $\mathcal F_t$.
The forecasting state $X_t$ need not summarize the whole filtration.
Accordingly, throughout the main theory ``optimal forecast'' refers
to the conditional tail-risk objects given $X_t$ in
\eqref{eq:VaRmoment_cond}--\eqref{eq:ESmoment_cond}.  We allow these same
restrictions to fail after conditioning on $\mathcal F_t$.  Thus omitted
predictive information in the larger filtration may generate serially
correlated VaR hits and ES innovations even though the forecasting models
are optimal relative to $X_t$.  We call this \emph{dynamic forecast
misspecification relative to $\mathcal F_t$}.  The backtest is designed to
evaluate the $X_t$-conditional forecasting model; dynamic dependence that
does not violate that model is accommodated through the long-run variance.

\begin{assumption}[Regularity of the conditional distribution]\label{ass:CDF}
Fix $\alpha\in(0,1)$.  The conditional CDF $F(\cdot\mid X_t)$ satisfies
the following properties almost surely.
\begin{enumerate}[label=(\roman*),leftmargin=*]
\item $F(\cdot\mid X_t)$ is absolutely continuous in a neighborhood of
its conditional $\alpha$-quantile, with conditional density
$f(\cdot\mid X_t)$.
\item There exists a constant $C>0$ such that $0<C\le f(y\mid X_t)\le C^{-1}<\infty$ for all $y$ in a neighborhood of $\VaR_{\alpha}(X_t)$. The density $f(\cdot\mid X_t)$ is continuous at
$\VaR_\alpha(X_t)$.
\item There exists $\delta>0$ such that
$\E[|Y_{t+1}|^{2+\delta}]<\infty$.
\end{enumerate}
\end{assumption}

Henceforth, for a random vector $X_t$, let $L_2(X_t)$ denote the space of square-integrable measurable functions of $X_t$, i.e., $L_2(X_t):=\{g(X_t):\E[g(X_t)^2]<\infty\}.$ For a fixed positive integer $d_w$, let $L_2^{d_w}(X_t)$ denote the
corresponding space of $d_w$-vectors.  Projections and moment restrictions
below are understood component by component when $d_w>1$.
\paragraph{Directional derivatives and derivative range.}
To remove first-order estimation effects, we construct backtesting
moments that are locally insensitive to small parameter changes
allowed by the model. We describe these changes through
\emph{admissible directions} and their effects on the forecasts
through directional derivatives.
Let $\mathcal H_\gamma \subseteq \mathbb R^{d_\gamma}$ and
$\mathcal H_\beta \subseteq \mathbb R^{d_\beta}$ denote sets of
admissible directions for the VaR and ES parameters, respectively. For any $h_\gamma\in\mathcal H_\gamma$, define the directional derivative
\begin{equation}\label{der_q}
\partial_\gamma q_\alpha(X_t,\gamma_0)[h_\gamma]
=
\left.\frac{d}{d\tau}
q_\alpha(X_t,\gamma_0+\tau h_\gamma)
\right|_{\tau=0}.
\end{equation}
Likewise, for any $h_\beta\in\mathcal H_\beta$,
\begin{equation}\label{der_e}
\partial_\beta e_\alpha(X_t,\beta_0)[h_\beta]
=
\left.\frac{d}{d\tau}
e_\alpha(X_t,\beta_0+\tau h_\beta)
\right|_{\tau=0}.
\end{equation}

\begin{assumption}[Predictive models]\label{ass:models}
\begin{enumerate}[label=(\roman*),leftmargin=*]
\item $\VaR_\alpha(X_t)=q_\alpha(X_t,\gamma_0)$ and $q_\alpha(\cdot,\gamma)$ is Gateaux differentiable at $\gamma_0$ tangentially to $\mathcal H_\gamma$, i.e., (\ref{der_q}) is well-defined.
\item $\ES_\alpha(X_t)=e_\alpha(X_t,\beta_0)$ and $e_\alpha(\cdot,\beta)$ is Gateaux differentiable at $\beta_0$ tangentially to $\mathcal H_\beta$, i.e., (\ref{der_e}) is well-defined.
\item The derivatives in \eqref{der_q}--\eqref{der_e} belong to $L_2(X_t)$.
\item For every admissible direction, the conditional moment maps used
below have integrable local envelopes that justify differentiation under
the outer expectation.
\end{enumerate}
\end{assumption}

\subsection{A locally robust ES backtesting score}\label{sec:tail:LRmoment}

We now formalize orthogonal (locally robust) ES backtesting moments.

\begin{definition}[Orthogonal ES backtesting moment]\label{def:orthmoment}
Let $S_{t+1}(\gamma,\beta)$ be a measurable function of $(Y_{t+1},X_t)$ such that
$S_{t+1}(\gamma,\beta)$ has finite variance for $(\gamma,\beta)$ in a neighborhood of $(\gamma_0,\beta_0)$.
The function $S_{t+1}$ is an \emph{orthogonal backtesting moment for ES} if it satisfies:
\begin{enumerate}[label=(\roman*),leftmargin=*]
\item \textbf{Moment validity:}\quad $\E\!\left[S_{t+1}(\gamma_0,\beta_0)\right]=0$.
\item \textbf{Local robustness at \ $(\gamma_0,\beta_0)$:}\quad for every pair of admissible perturbation
directions $h_\gamma\in\mathcal H_\gamma$ and $h_\beta\in\mathcal H_\beta$,
\[
\left.
\frac{d}{d\tau}
\E\!\left[
S_{t+1}(\gamma_0+\tau h_\gamma,\beta_0+\tau h_\beta)
\right]
\right|_{\tau=0}=0 .
\]
\end{enumerate}
\end{definition}

Define the moments
\[
\psi_{VaR,t+1}(\gamma):=\mathbf 1\{Y_{t+1}\le q_\alpha(X_t,\gamma)\}-\alpha,
\qquad
\gamma\in\Gamma,
\]
and
\[
\psi_{ES,t+1}(\theta):=\frac{Y_{t+1}}{\alpha}\mathbf 1\{Y_{t+1}\le q_\alpha(X_t,\gamma)\}-e_\alpha(X_t,\beta),
\qquad
\theta=(\gamma,\beta)\in\Gamma\times\mathcal B.
\]

\begin{theorem}[Characterization within the VaR--ES residual class]\label{thm:orth}
Suppose Assumptions \ref{ass:CDF}--\ref{ass:models} hold. Fix
$\lambda_1,\lambda_2\in L_2^{d_w}(X_t)$ such that the following score has
a finite second moment:
\[
S_{t+1}(\gamma,\beta)
=\lambda_1(X_t)\psi_{VaR,t+1}(\gamma)
+\lambda_2(X_t)\psi_{ES,t+1}(\gamma,\beta).
\]
The score is valid at $(\gamma_0,\beta_0)$.  It is orthogonal with
respect to $\gamma$ at $\gamma_0$ if and only if
\begin{equation}\label{orth0}
\E\!\left[
\left\{\lambda_1(X_t)+\frac{q_{0t}}{\alpha}\lambda_2(X_t)\right\}
f(q_{0t}\mid X_t)\,
\partial_\gamma q_\alpha(X_t,\gamma_0)[h_\gamma]
\right]=0
\quad\text{for every }h_\gamma\in\mathcal H_\gamma,
\end{equation}
where $q_{0t}=q_\alpha(X_t,\gamma_0)$.  It is orthogonal with respect to
$\beta$  at $\beta_0$ if and only if
\begin{equation}\label{orth2}
\E\!\left[
\lambda_2(X_t)\,\partial_\beta e_\alpha(X_t,\beta_0)[h_\beta]
\right]=0
\quad\text{for every }h_\beta\in\mathcal H_\beta.
\end{equation}
In particular, the pointwise restriction
\begin{equation}\label{orth1}
\lambda_1(X_t)+\frac{q_{\alpha}(X_t,\gamma_0)}{\alpha}\lambda_2(X_t)=0
\end{equation}
is sufficient for \eqref{orth0}.  
\end{theorem}

\paragraph{Pseudo-outcome representation and baseline score.}
We now use Theorem \ref{thm:orth} to obtain a tractable score representation. Using \eqref{orth1} in the generic
representation of $S_{t+1}$ and taking a common factor yields the score 
\[
S_{t+1}
=
\bigl\{Y^*_{t+1}(\gamma_0)-e_\alpha(X_t,\beta_0)\bigr\}\,\lambda_2(X_t),
\]

with the pseudo-outcome or tail-size-adjusted score 
\begin{equation}\label{eq:pseudoOutcome}
Y^*_{t+1}(\gamma)
=
q_\alpha(X_t,\gamma)
+
\frac{1}{\alpha}\,
\mathbf 1\{Y_{t+1}\le q_\alpha(X_t,\gamma)\}
\bigl(
Y_{t+1}-q_\alpha(X_t,\gamma)
\bigr),
\end{equation}
and $\lambda_2$ subject to \eqref{orth2} for orthogonality with respect to (w.r.t.)\ the ES step.

A baseline (non-residualized) ES backtesting score corresponds to taking a
user-chosen test function $w(\cdot)\in L_2^{d_w}(X_t)$:
\begin{equation}\label{eq:ESscore_naive}
S_{ES,t+1}(\theta)
=
\bigl\{Y^*_{t+1}(\gamma)-e_\alpha(X_t,\beta)\bigr\}\,w(X_t),
\qquad
\theta=(\gamma,\beta).
\end{equation}
Define the ES derivative range
\[
\mathcal R_e
=
\overline{\mathrm{span}}
\left\{
\partial_\beta e_\alpha(X_t,\beta_0)[h_\beta]:
h_\beta\in\mathcal H_\beta
\right\}
\subseteq L_2(X_t).
\]

\paragraph{Residualization and the orthogonal score $S^\perp$.}
To satisfy \eqref{orth2}, start from a user-chosen test function
$w(X_t)\in L_2^{d_w}(X_t)$ and remove its component in
$\mathcal R_e$:
\[
\lambda_2(X_t)=w(X_t)-\Pi_e w(X_t),
\]
where $\Pi_e w(X_t)$ denotes the componentwise $L_2(X_t)$
projection of $w(X_t)$ onto $\mathcal R_e$. We choose $w$ so
that $w-\Pi_e w\neq0$ in $L_2^{d_w}(X_t)$: the test function
must contain variation that cannot be reproduced by local
changes in the ES model.
When $e_\alpha(\cdot,\beta)$ is finite-dimensional, differentiable in $\beta$, and $\mathrm{span}(\mathcal H_\beta)=\mathbb R^{d_\beta}$, let
\[
\dot e_\alpha(X_t,\beta)
:=
\left(
\frac{\partial}{\partial \beta_1}e_\alpha(X_t,\beta),
\ldots,
\frac{\partial}{\partial \beta_{d_\beta}}e_\alpha(X_t,\beta)
\right)'
\]
and note that $\mathcal R_e$ is the closed linear span of the components of $\dot e_\alpha(X_t,\beta_0)$.
Hence $\Pi_e w(X_t)=\rho_0'\dot e_\alpha(X_t,\beta_0)$ for any population projection coefficient
\[
\rho_0 \in \arg\min_{\rho\in\mathbb R^{d_\beta\times d_w}}
\E\!\left[\bigl\|w(X_t)-\rho'\dot e_\alpha(X_t,\beta_0)\bigr\|^2\right].
\]
The coefficient need not be unique, but the fitted projection and the
residualized instrument are unique in $L_2$.  In the sparse linear theory
below, $\rho_0$ denotes a fixed sparse representative of this set.
This yields the orthogonal ES backtesting score
\begin{equation}\label{eq:ESscore_orth}
S^\perp_{ES,t+1}(\theta,\rho)
=
\bigl\{Y^*_{t+1}(\gamma)-e_\alpha(X_t,\beta)\bigr\}
\bigl\{w(X_t)-\rho'\dot e_\alpha(X_t,\beta)\bigr\}.
\end{equation}
Evaluated at $(\theta_0,\rho_0)$, this score is locally robust in the sense of Definition \ref{def:orthmoment}. We allow the initial test function $w$ to depend on $\beta$; in that case,
we write it as $w(X_t,\beta)$ and $w_0(X_t)=w(X_t,\beta_0)$. This is possible because the orthogonal ES backtesting score is also orthogonal to the residualized instrument, as shown in the following proposition.
\begin{proposition}[Orthogonality of the residualized score]
\label{prop:residualized_orthogonality}
Suppose Assumptions~\ref{ass:CDF} and \ref{ass:models} hold, and
$w(X_t,\beta)$ and $\dot e_\alpha(X_t,\beta)$ are Gateaux
differentiable at $\beta_0$ with integrable derivatives. If
$\rho_0'\dot e_\alpha(X_t,\beta_0)$ is the projection of $w_0(X_t)$
onto $\mathcal R_e$, then the score in \eqref{eq:ESscore_orth} is
orthogonal to $\gamma$, $\beta$, and $\rho$ at the truth. The same
conclusion holds for $w(X_t,\beta)=e_\alpha(X_t,\beta)^2$ whenever the
relevant products are integrable.
\end{proposition}
\begin{remark}[Direct orthogonality w.r.t.\ the VaR step]
\cite{Barendse2020TailIQE} shows that under \eqref{eq:VaRmoment_cond},
\[
\left.
\frac{d}{d\tau}
\E\!\left[
Y^{*}_{t+1}(\gamma_0+\tau h_\gamma)\bigm|\,
X_t
\right]
\right|_{\tau=0}=0,
\]
which corresponds to the direct-channel orthogonality encoded by \eqref{orth1} and the pseudo-outcome
representation \eqref{eq:pseudoOutcome}.
\end{remark}

\begin{remark}[Role of orthogonality]\label{rem:role_orth}
If \eqref{orth2} fails, the backtesting moment becomes sensitive to the estimation of the ES step, and can also inherit sensitivity to the first step through the dependence of $\beta_0$ on $\gamma_0$.
When $\gamma_0$ and/or $\beta_0$ are high-dimensional, this channel may introduce substantial regularization and
model-selection bias, potentially invalidating ES backtests. Residualization via \eqref{eq:ESscore_orth} removes
this first-order sensitivity by orthogonalizing the score with respect to the ES derivative space.
\end{remark}

\subsection{Out-of-sample implementation}\label{sec:tail:oos}

We now describe the out-of-sample (OOS) implementation of the proposed orthogonal ES backtest. Let $T$ denote the total sample size, $R$ the initial estimation window length, and $P$ the OOS size, so that
\[
T=R+P.
\]
The OOS evaluation period is indexed by the forecast origins
\[
\mathcal T_{\text{test}}:=\{R,\ldots,T-1\},
\qquad |\mathcal T_{\text{test}}|=P.
\]

At each forecast origin $t\in\mathcal T_{\text{test}}$, the forecaster uses information available up to time $t$ to estimate the predictive models and then evaluates the one-step-ahead forecast using the realized outcome $Y_{t+1}$ and predictors $X_t$.

\medskip

\noindent
\textbf{Training samples and forecast timing.}
For each $s=0,\ldots,T-1$, define the predictor--outcome observation $D_s:=(X_s,Y_{s+1}).$ At forecast origin $t$, the observations $D_s$ with $s\leq t-1$ are
available, whereas $D_t$ is not complete because $Y_{t+1}$ has not yet
been observed. Let
\[
\mathcal I_t\subseteq\{0,\ldots,t-1\}
\]
denote the indices of the observations used for estimation at origin $t$.
Thus, the corresponding training sample is
\[
\{D_s:s\in\mathcal I_t\}
=
\{(X_s,Y_{s+1}):s\in\mathcal I_t\}.
\]
We consider three standard estimation schemes:
\[
\begin{aligned}
\mathcal I_t&=\{0,\ldots,R-1\}
&&\text{(fixed)},\\
\mathcal I_t&=\{t-R,\ldots,t-1\}
&&\text{(rolling)},\\
\mathcal I_t&=\{0,\ldots,t-1\}
&&\text{(recursive or expanding)},
\end{aligned}
\qquad t=R,\ldots,T-1.
\]
Under the fixed scheme, the models are estimated once using the initial
$R$ observations. The rolling scheme always uses the most recent $R$
observations, whereas the recursive scheme adds each newly available
observation to the estimation sample.

At origin $t$, the models are first estimated on the selected training
sample. The fitted coefficients are then combined with the current
predictor $X_t$ to produce the VaR and ES forecasts for $Y_{t+1}$.
After $Y_{t+1}$ is observed, the forecasts are evaluated and the
backtesting score is computed. Thus, the evaluation observation
$D_t=(X_t,Y_{t+1})$ is never used to estimate its own forecast.

For the formal timing conditions used below, let
\[
\mathcal G_t:=\sigma(D_s:s\in\mathcal I_t)\subseteq\mathcal F_t.
\]
The fitted coefficients are $\mathcal G_t$-measurable. The forecasts
and feasible test function, which also use $X_t$, are
$\mathcal F_t$-measurable, whereas the feasible score becomes observable
only after $Y_{t+1}$ is realized.
\medskip

\noindent
\textbf{Nuisance estimation.}
Define the estimator at forecast origin $t$ by
\[
\widehat\theta_t
:=
\widehat\theta(\{(X_s,Y_{s+1}):s\in\mathcal I_t\}),
\qquad
\widehat\theta_t=(\widehat\gamma_t,\widehat\beta_t).
\]

Given $\mathcal I_t$, estimate:
\begin{enumerate}[label=(\roman*),leftmargin=*]
\item $\widehat\gamma_t$ from the VaR model $q_\alpha(X_s,\gamma)$ (see, e.g., (\ref{eq:linear_VaR_lasso}));
\item $\widehat\beta_t$ from the ES model $e_\alpha(X_s,\beta)$ using the pseudo-outcome
\[
Y_{s+1}^*(\widehat\gamma_t)
=
q_\alpha(X_s,\widehat\gamma_t)
+
\frac{1}{\alpha}
\mathbf 1\{Y_{s+1}\le q_\alpha(X_s,\widehat\gamma_t)\}
\bigl(Y_{s+1}-q_\alpha(X_s,\widehat\gamma_t)\bigr),
\qquad s\in\mathcal I_t;
\]
See, e.g., (\ref{eq:linear_ES_lasso}).
\item $\widehat\rho_t$ by projecting the training response \(\widehat w_{t,s}\) onto the derivative space of the ES model using
\(s\in\mathcal I_t\); e.g., the projection of $\widehat w_{t,s}=e_\alpha(X_s,\widehat\beta_t)^2$ on $\dot e_\alpha(X_s,\widehat\beta_t)
=
\partial_\beta e_\alpha(X_s,\widehat\beta_t)$ to obtain $\widehat\rho_t$ solving
\[
\widehat\rho_t
\in
\arg\min_{\rho}
\frac{1}{|\mathcal I_t|}
\sum_{s\in\mathcal I_t}
\Bigl\|
\widehat w_{t,s}-\rho'\dot e_\alpha(X_s,\widehat\beta_t)
\Bigr\|^2,
\]
or a penalized version thereof.

\end{enumerate}

\medskip

\noindent
\textbf{Example: linear high-dimensional models.}
Consider the high-dimensional linear models
\[
q_\alpha(X_t,\gamma)=X_t'\gamma,
\qquad
e_\alpha(X_t,\beta)=X_t'\beta.
\]
Write
\[
X_t=(1,W_t')'=(Z_t',V_t')',
\]
where $W_t$ collects the nonconstant predictors,
$Z_t\in\mathbb R^{d_Z}$ contains the intercept and any
additional regressors whose coefficients are left unpenalized,
and $V_t\in\mathbb R^{p_T-d_Z}$ contains the regressors whose
coefficients are penalized. The partition is prespecified
and used throughout the estimation windows, and $d_Z$ is fixed.
When only the intercept is unpenalized, $Z_t=1$ and $V_t=W_t$.
We retain the intercept explicitly and impose no centering
assumption on $X_t$. 

Partition the coefficient vectors conformably as
$\gamma=(\gamma_Z',\gamma_V')'$ and
$\beta=(\beta_Z',\beta_V')'$. Within each estimation stage,
the unpenalized and penalized coefficients are estimated
jointly.

\smallskip
\noindent
\emph{VaR estimation.}
Let $\rho_\alpha(u)=u(\alpha-\mathbf 1\{u<0\})$ denote the
check loss. The partially penalized quantile estimator is
\begin{equation}
\label{eq:linear_VaR_lasso}
\widehat\gamma_t
\in
\arg\min_{\gamma\in\mathbb R^{p_T}}
\left\{
\frac{1}{|\mathcal I_t|}
\sum_{s\in\mathcal I_t}
\rho_\alpha(Y_{s+1}-X_s'\gamma)
+
\lambda_{\gamma,t}\|\gamma_V\|_1
\right\}.
\end{equation}

A smoothed alternative replaces the check loss by its
convolution with a kernel \citep{TanWangZhou2022}. Let
$\mathcal K$ be a bounded, symmetric probability density
with finite second moment, and define
\begin{equation}
\label{eq:smoothed_check_loss}
\rho_{\alpha,\nu}(u)
:=
\int_{\mathbb R}
\rho_\alpha(u-\nu v)\mathcal K(v)\,dv,
\qquad \nu>0.
\end{equation}
For example, $\mathcal K$ may be the standard normal density.
At forecast origin $t$, the smoothed quantile estimator is
\begin{equation}
\label{eq:linear_VaR_smoothed}
\widehat\gamma_t^{\mathrm{sm}}
\in
\arg\min_{\gamma\in\mathbb R^{p_T}}
\left\{
\frac{1}{|\mathcal I_t|}
\sum_{s\in\mathcal I_t}
\rho_{\alpha,\nu_t}(Y_{s+1}-X_s'\gamma)
+
\lambda_{\gamma,t}\|\gamma_V\|_1
\right\},
\end{equation}
where $\nu_t>0$ is the smoothing bandwidth. Both programs
leave $\gamma_Z$ unpenalized. In the remaining stages,
$\widehat\gamma_t$ denotes the estimate obtained from the
chosen VaR program. Smoothing is confined to the VaR
estimation loss: the pseudo-outcome in
\eqref{eq:pseudoOutcome} retains its original indicator.

\smallskip
\noindent
\emph{ES estimation.}
Using the training pseudo-outcomes
$Y_{s+1}^*(\widehat\gamma_t)$, estimate
\begin{equation}
\label{eq:linear_ES_lasso}
\widehat\beta_t
\in
\arg\min_{\beta\in\mathbb R^{p_T}}
\left\{
\frac{1}{|\mathcal I_t|}
\sum_{s\in\mathcal I_t}
\bigl(Y_{s+1}^*(\widehat\gamma_t)-X_s'\beta\bigr)^2
+
\lambda_{\beta,t}\|\beta_V\|_1
\right\}.
\end{equation}

\smallskip
\noindent
\emph{Instrument estimation.}
For a $d_w$-dimensional test function, let
$\rho\in\mathbb R^{p_T\times d_w}$, and denote by
$\rho_{V,r}$ the coefficients on $V_t$ in column
$r=1,\ldots,d_w$. Estimate the projection coefficients by
\begin{equation}
\label{eq:linear_projection_lasso}
\widehat\rho_t
\in
\arg\min_{\rho\in\mathbb R^{p_T\times d_w}}
\left\{
\frac{1}{|\mathcal I_t|}
\sum_{s\in\mathcal I_t}
\bigl\|\widehat w_{t,s}-\rho'X_s\bigr\|^2
+
\lambda_{\rho,t}
\sum_{r=1}^{d_w}\|\rho_{V,r}\|_1
\right\}.
\end{equation}
Here $\widehat w_{t,s}$ is the training test function
defined above.
The program separates into $d_w$ partially penalized
least-squares regressions. Residualization uses the full
ES regressor vector $X_s$, including the intercept and
the other unpenalized regressors, rather than only the
regressors selected by Lasso. In the discussion above, $(\lambda_{\gamma,t},\lambda_{\beta,t},\lambda_{\rho,t})$ are penalty levels which are subject to the rate and stability requirements in Section~\ref{sec:theory}. For smoothed VaR
estimation, these requirements must also account for
smoothing bias and the choice of bandwidth $\nu_t$.

In the ES and projection stages, exact profiling of the
unpenalized coefficients gives an equivalent implementation.
When $Z_t=1$, this amounts to fitting the penalized slopes
using the demeaned response and regressors, and recovering
the intercept as the response mean minus the regressor
means multiplied by the fitted slopes. For a larger
unpenalized block, the corresponding operation partials
$Z_t$ out of both the response and the penalized regressors
before fitting the slopes. 
\medskip

\noindent
\textbf{Scores and backtest statistic.}
For each $t\in\mathcal T_{\text{test}}$, define
\[
S_{t+1} ^{\perp}:= S^\perp_{ES,t+1}(\theta_0,\rho_0),
\qquad
\widehat S_{t+1} ^{\perp}:= S^\perp_{ES,t+1}(\widehat\theta_t,\widehat\rho_t),
\]
where $\theta_0=(\gamma_0,\beta_0)$.

The OOS score average is
\[
\bar S_P^{\perp}
=
\frac{1}{P}\sum_{t\in\mathcal T_{\text{test}}}\widehat S_{t+1}^{\perp}.
\]

The ES backtest statistic is
\[
J_P
=
P\,\bar S_P^{\perp'}\widehat\Omega^{-1}\bar S_P^{\perp},
\]
where $\widehat\Omega$ is a HAC estimator of the long run variance
\[
\Omega_T
=
\sum_{k=-\infty}^{\infty}\mathrm{Cov}(S_{t+1}^{\perp},S_{t+1-k}^{\perp}).
\]
given in equation (\ref{eq:HAC_definition}). The null hypothesis of optimal VaR and ES forecasts relative to $X_t$ is rejected at level $\tau$ whenever
\[
J_P>\chi^2_{1-\tau,d_w},
\]
where $\chi^2_{1-\tau,d_w}$ is the $(1-\tau)$ quantile of a chi-square
distribution with $d_w$ degrees of freedom.
\section{Asymptotic Theory}\label{sec:theory}

\subsection{Oracle equivalence}
\label{sec:theory:general}

Write $q_{0t}=q_\alpha(X_t,\gamma_0)$,
$e_{0t}=e_\alpha(X_t,\beta_0)$, and
$w_{0t}=w(X_t,\beta_0)$, with $w_{0t}=w(X_t)$
for a fixed test function (we drop the dependence on $\alpha$ for simplicity of notation). Define
\[
\begin{aligned}
U_{t+1}(q)
&:=q+\alpha^{-1}\mathbf 1\{Y_{t+1}\le q\}(Y_{t+1}-q),
\qquad q\in\mathbb R,\\
\varepsilon_{t+1}
&:=U_{t+1}(q_{0t})-e_{0t},
\qquad
k_{0t}:=w_{0t}-\rho_0'\dot e_\alpha(X_t,\beta_0).
\end{aligned}
\]
With this notation in place, the oracle score is $S_{t+1}^{\perp}=k_{0t}\varepsilon_{t+1}$,
with fixed dimension $d_w$. Its feasible counterpart replaces
the unknown components by estimates from the fixed, recursive,
or rolling training window in Section~\ref{sec:tail:oos}. Throughout this section, the prefix OA identifies assumptions collected
in the Online Appendix to save space in the main text.

\begin{assumption}[Optimal forecasts relative to the forecasting information]
\label{ass:VaRcorrect}
At each forecast origin, $\E[\mathbf 1\{Y_{t+1}\le q_{0t}\}-\alpha\mid X_t]=0,
$ and $\E[\varepsilon_{t+1}\mid X_t]=0
$ a.s.
\end{assumption}

Assumption~\ref{ass:VaRcorrect} makes precise the information set against
which the forecasts are evaluated.  The VaR and ES forecasts are optimal relative to the forecasting state $X_t$, but $X_t$ need not exhaust all
predictive information in $\mathcal F_t$. 

We write $A\succ0$ to mean that $A$ is symmetric positive
definite: $A=A'$ and $v'Av>0$ for every nonzero vector $v$.
The oracle score $S_{t+1} ^{\perp}$ may depend on $T$ through the growing
predictor dimension; we suppress this dependence except
in its long-run covariance $\Omega_T$.

\begin{assumption}[Regularity of the forecasting pipeline]
\label{ass:general_pipeline}
The fitted VaR and ES forecasts and the residualized instrument
are $\mathcal F_t$-measurable. For each $T$, the oracle score
process is strictly stationary and $ \beta$-mixing, with mixing
coefficients bounded uniformly in $T$ by $\beta(j)$.
For some $q>4$,
\[
\sum_{j=1}^{\infty}\beta(j)^{1-2/q}<\infty,
\qquad
\sup_T \E\|S_{t+1}^{\perp}\|^q<\infty.
\]
Its long-run covariance exists and satisfies $\Omega_T \longrightarrow \Omega\succ0$, as $T\rightarrow \infty$.
The remainder conditions
\eqref{eq:OA_HL1}--\eqref{eq:OA_HL4}
in the Online Appendix also hold.
\end{assumption}

The measurability, moment, and mixing restrictions are standard
regularity conditions for time-series inference. Positive
definiteness of $\Omega$ rules out degenerate score directions;
in particular, the residualized instrument $k_{0t}$ cannot be
identically zero. The appendix conditions control the weighted
estimation errors and their interaction with serially dependent
innovations, which prediction rates alone need not control.
Section~\ref{sec:theory:lasso} provides primitive rate and
stability conditions for Lasso estimators.

For completeness, let $K$ be a bounded kernel supported on $[-1,1]$,
continuous at zero, with $K(0)=1$.  Define
\[
\widehat\Gamma_\ell=P^{-1}\sum_{t=R+\ell}^{T-1}
\widehat S_{t+1}^{\perp}\widehat S_{t+1-\ell}^{\perp\prime},
\]
and
\begin{equation}\label{eq:HAC_definition}
\widehat\Omega=
\widehat\Gamma_0+
\sum_{\ell=1}^{L_P}K(\ell/L_P)
(\widehat\Gamma_\ell+\widehat\Gamma_\ell').
\end{equation}
For the HAC bandwidth restrictions, define
\[
A_P:=1+\sum_{j=1}^{P}\beta(j)^{1-4/q}.
\]
Let $\Rightarrow$ and $\xrightarrow{p}$ denote convergence in distribution and probability, respectively.
\begin{theorem}[Oracle equivalence and backtest validity]
\label{thm:main_asymptotics}
Suppose $P\to\infty$. Under Assumptions~\ref{ass:VaRcorrect}
and \ref{ass:general_pipeline}, for fixed, recursive,
and rolling estimation,
\[
P^{-1/2}\sum_{t\in\mathcal T_{\mathrm{test}}}
\{\widehat S_{t+1}^{\perp}-S_{t+1}^{\perp}\}=o_p(1),
\qquad
\sqrt P\,\bar S_P^{\perp}\Rightarrow N(0,\Omega).
\]
If, in addition,
\[
r_{S,T}^2
:=
\frac{1}{P}
\sum_{t\in\mathcal T_{\mathrm{test}}}
\left\|
\widehat S_{t+1}^{\perp}-S_{t+1}^{\perp}
\right\|^2
=o_p(1),
\]
and
\[
L_P\to\infty,
\qquad
\frac{L_P^3}{P}\to0,
\qquad
\frac{L_P^2A_P}{P}\to0,
\qquad
L_Pr_{S,T}=o_p(1),
\]
then $\widehat\Omega\xrightarrow{p}\Omega$ and
\[
J_P\Rightarrow\chi_{d_w}^2.
\]
\end{theorem}

Under full-information optimality the oracle score is a martingale
difference and the off-diagonal autocovariances vanish.  We do not impose
that stronger restriction.  Under Assumption~\ref{ass:VaRcorrect}, the
oracle score is mean zero because $k_{0t}$ is a function of $X_t$, but it
may be serially correlated.  Hence $\Omega$ is a genuine long-run
variance and HAC estimation is part of the substantive inference.

\subsection{Uniform Lasso estimation and backtest validity}
\label{sec:theory:lasso}

We verify the preceding conditions for the penalized estimators in \eqref{eq:linear_VaR_lasso},
\eqref{eq:linear_ES_lasso}, and
\eqref{eq:linear_projection_lasso}. Throughout this subsection,
VaR estimation uses the unsmoothed check loss (which is harder to prove than the smoothed one). Retain the
partition $X_t=(Z_t',V_t')'$, where $Z_t$ contains the intercept
and the other unpenalized regressors, with fixed dimension $d_Z$.
The population forecasts and residualized instrument are
\[
q_{0t}=X_t'\gamma_0,\qquad
e_{0t}=X_t'\beta_0,\qquad
k_{0t}=w_{0t}-\rho_0'X_t,
\]
and the projection normal equations give $\E[k_{0t}X_t']=0$.

We consider either a fixed test function, with
$w_{0t}=w(X_t)$ and $\widehat w_{t,s}=w(X_s)$, or the scalar
generated test function
\begin{equation}\label{eq:generated_test_function}
w_{0t}=e_{0t}^2,\qquad
\widehat w_{t,s}=(X_s'\widehat\beta_t)^2,
\quad s\in\mathcal I_t\cup\{t\},
\end{equation}
where $\widehat w_{t,t}=\widehat w_t$. Other examples could be entertained.

\paragraph{Why uniformity over training windows matters.}
An out-of-sample backtest averages scores over $P$ evaluation
dates. With rolling or recursive estimation, the estimates
are recomputed as the training sample changes. A convergence
rate for one training sample does not, by itself, control
the sequence of estimates entering the backtest when the
number of evaluation dates also grows.

Our analysis begins with empirical fluctuation bounds that
hold simultaneously over all consecutive training windows
of length at least $R/2$. Under the design conditions below,
these bounds yield uniform rates for the VaR, ES, and
projection estimators. This provides a common estimation
argument for fixed, rolling, and recursive forecasting
schemes. We first define the quantities entering these
rates and then explain their implications for inference.

\paragraph{Sparsity and prediction norms.}
Let $\mathcal J_Z=\{1,\ldots,d_Z\}$ index the unpenalized
coordinates. For $b\in\mathbb R^{p_T}$, define its effective
support by
\[
S(b):=\mathcal J_Z\cup\{j>d_Z:b_j\ne0\}.
\]
Write $S_\gamma=S(\gamma_0)$, $S_\beta=S(\beta_0)$, and
$S_{\rho,r}=S(\rho_{0,r})$, where $\rho_{0,r}$ is column $r$
of $\rho_0$. Define
\[
s_\gamma=|S_\gamma|,\qquad
s_\beta=|S_\beta|,\qquad
s_\rho=\max_{1\le r\le d_w}|S_{\rho,r}|,\qquad
s_\circ=s_\gamma+s_\beta+s_\rho.
\]
These sparsity counts may grow with $T$ and include every
unpenalized coordinate, even when its true coefficient is
zero. Since $d_w$ is fixed, the maximum column sparsity
suffices for the projection estimator.
For coefficient errors, use the population prediction norm
\[
\|b\|_\Sigma:=(b'\Sigma b)^{1/2},
\qquad
\Sigma:=\E[X_tX_t'].
\]
Here $\Sigma$ is the regressor second-moment matrix, including
the intercept.

\paragraph{Separate penalties and their common rate.}
The penalties $\lambda_{\gamma,t}$, $\lambda_{\beta,t}$, and
$\lambda_{\rho,t}$ may differ across estimation stages and
forecast origins. The sufficient conditions require them
to have comparable orders, expressed through a common
reference rate.

Let $q>4$ and $\mu>2$ denote the moment order and polynomial
mixing exponent used below. Choose $\mu'\in(0,\mu)$ and, for
sufficiently large $n$, set
\[
a_n=\lceil n^{1/(1+\mu')}\rceil,\qquad
d_n=\lfloor n/(2a_n)\rfloor.
\]
Here $a_n$ is a block length and $d_n$ is the effective number
of blocks. For an array with $m$ coordinates, define
\begin{equation}\label{eq:Lambda_main}
\begin{aligned}
\Lambda_n(m,q)
&:=
\frac{(ma_n)^{2/q}}{d_n^{(q-1)/q}}
\vee
\sqrt{\frac{\log(ma_n)}{d_n}},\\
\overline\Lambda_n(m,q)
&:=
\Lambda_n(m,q)+\{md_n\beta(a_n)\}^{1-2/q}.
\end{aligned}
\end{equation}
Here $x\vee y=\max\{x,y\}$. The rate
$\Lambda_n(m,q)$ controls the maximum deviation
of centered sample averages across $m$ coordinates
and training windows. The block length $a_n$ and the
effective number of blocks $d_n$ account for serial
dependence. The square-root term corresponds to
Gaussian-type concentration, while the polynomial
term accounts for heavy tails under finite $q$-th
moments. The additional term in
$\overline\Lambda_n(m,q)$ bounds the effect of
coupling dependent blocks with independent copies.
It is used to control moments of averages over
the short blocks deleted in the oracle-equivalence
proof.

Set $\lambda_T:=\Lambda_R(p_T,q)$ and require
\begin{equation}\label{eq:lasso_penalty_comparability}
c_\lambda\lambda_T
\le \lambda_{j,t}\le C_\lambda\lambda_T,
\qquad j\in\{\gamma,\beta,\rho\},
\end{equation}
at every forecast origin, for fixed positive constants
$c_\lambda$ and $C_\lambda$. This does not require the
three penalties to be numerically equal. Under $R,P\asymp T$,
all training-window lengths are of order $R$, so $\lambda_T$
also has the same order as
$\Lambda_{|\mathcal I_t|}(p_T,q)$ and $\Lambda_T(p_T,q)$.

\paragraph{From uniform estimation to inference.}
Under the conditions stated below, the estimation rates hold
simultaneously over forecast origins. For example,
\[
\max_{t\in\mathcal T_{\mathrm{test}}}
\|\widehat\gamma_t-\gamma_0\|_\Sigma
=O_p(\sqrt{s_\gamma}\lambda_T).
\]
The maximum makes the uniformity explicit: the bound controls
the largest prediction error across the sequence of fitted
VaR models, rather than the error at one particular
forecast origin. Corresponding bounds hold for the ES and
projection estimators.

With rolling or recursive estimation, models are repeatedly fitted
using recent observations that may remain dependent on subsequent
forecast outcomes. Small estimation errors alone therefore do not
ensure that their accumulated effect on the backtest is negligible.
Our proof addresses this issue by comparing the actual estimates
with auxiliary estimates that exclude a short stretch of the most
recent training observations. The resulting time gap allows us to
control dependence, while estimator stability ensures that this
comparison has a negligible effect on the backtest under the stated
conditions. The auxiliary estimates are used only in the proof;
the implemented forecasts retain their original training windows.

Together with orthogonality and the growth conditions below,
these results imply, for each of the fixed, rolling, and
recursive estimation schemes, the asymptotic oracle equivalence
\[
\frac{1}{\sqrt P}
\sum_{t\in\mathcal T_{\mathrm{test}}}
\bigl(\widehat S_{t+1}^{\perp}-S_{t+1}^{\perp}\bigr)
=o_p(1).
\]
This equivalence means that the feasible
and oracle score averages differ by $o_p(P^{-1/2})$,
negligibly relative to the oracle backtest's sampling
fluctuations. Thus repeated estimation introduces no
additional first-order term in the limiting distribution.
The theorem below also controls the feasible--oracle score
distance needed for long-run covariance estimation.

\begin{assumption}[Conditions for partially penalized Lasso]
\label{ass:primitive_lasso}
The distributional, population-design, penalty, and
growth conditions in
Assumption~\ref{ass:OA_primitive_arrays} hold. For rolling and recursive estimation, the proof block
length and delay satisfy
\eqref{eq:OA_quantile_score_growth} and
\eqref{eq:OA_delay_rates}.
For fixed estimation, these restrictions are replaced
by \eqref{eq:OA_fixed_rate}.
\end{assumption}

The conditions concern the data distribution and
population parameters. 

\begin{theorem}[Uniform Lasso rates and oracle equivalence]
\label{thm:primitive_lasso}
Suppose Assumptions~\ref{ass:CDF}, \ref{ass:VaRcorrect},
and \ref{ass:primitive_lasso} hold. Uniformly over
$t\in\mathcal T_{\mathrm{test}}$, the estimation errors
satisfy the following rates, with the maximum also taken
over $1\le r\le d_w$ for the projection estimator:
\[
\begin{array}{c|cc}
 & \ell_1\text{ norm} & \Sigma\text{ norm}\\ \hline
\widehat\gamma_t-\gamma_0
 & O_p(s_\gamma\lambda_T)
 & O_p(\sqrt{s_\gamma}\lambda_T)\\
\widehat\beta_t-\beta_0
 & O_p((s_\beta+s_\gamma)\lambda_T)
 & O_p(\sqrt{s_\beta+s_\gamma}\lambda_T)\\
\widehat\rho_{t,r}-\rho_{0,r}
 & O_p(s_\circ\lambda_T)
 & O_p(\sqrt{s_\circ}\lambda_T)
\end{array}
\]
For a fixed test function, $s_\circ$ can be replaced
by $s_\rho$ in the projection rates.

If $\sqrt P\,s_\circ\lambda_T^2\to0$ and
\eqref{eq:OA_cubic_rate} holds when the test function
is generated, the oracle-equivalence and Gaussian
conclusions of Theorem~\ref{thm:main_asymptotics} hold.
Moreover,
\begin{equation}\label{eq:primitive_score_L2}
\left[
\frac1P\sum_{t\in\mathcal T_{\mathrm{test}}}
\|\widehat S_{t+1}^{\perp}-S_{t+1}^{\perp}\|^2
\right]^{1/2}
=O_p(b_{\circ,T}K_T),
\end{equation}
where $b_{\circ,T}=Cs_\circ\lambda_T$ and $K_T$
is defined in \eqref{eq:OA_inference_envelope}.

If, additionally,
\[
L_P\to\infty,\quad
L_P^3/P\to0,\quad
L_P^2A_P/P\to0,\quad
L_Pb_{\circ,T}K_T\to0,
\]
the HAC estimator is consistent and
$J_P\Rightarrow\chi^2_{d_w}$.
\end{theorem}

The quadratic-rate condition controls products of prediction
errors, while squared generated instruments require an additional
cubic-rate condition. The score-distance bound ensures feasible
HAC inference under the stated bandwidth restrictions.

\begin{corollary}[An admissible growing-dimension regime]
\label{cor:primitive_nonvacuous}
Suppose Assumption~\ref{ass:CDF} and conditions
(i), (ii), and (iv) of
Assumption~\ref{ass:OA_primitive_arrays} hold,
together with the remaining conditions in (iii).
Let $R,P\asymp T$, $s_\circ=O(1)$, $q>4$,
$\mu>2$, and
\[
\sup_{T,j}\E|X_{jt}|^{4q}<\infty.
\]
Suppose
\begin{equation}\label{eq:OA_dimension_regime}
p_T=O(T^\kappa),
\qquad
0<\kappa<
\min\left\{
\frac{5q-16}{60},
\frac{q-2}{15},
\frac{(q-2)(\mu-2)}{6(2q-3)}
\right\}.
\end{equation}
Then the conclusions of
Theorem~\ref{thm:primitive_lasso} hold with
$\mu'=2$, deterministic penalties satisfying
\eqref{eq:lasso_penalty_comparability}
with sufficiently large fixed constants,
and $L_P=\lfloor\log P\rfloor$.
The choice $L_P=\lfloor P^{1/8}\rfloor$
is also valid if $\mu(1-4/q)>1/4$.
\end{corollary}

The corollary illustrates the tradeoff between moment conditions,
temporal dependence, and dimension growth. For example, $q=8$
and $\mu=8$ permit every $\kappa<2/5$, while $q=20$ and $\mu=20$
permit every $\kappa<6/5$. Both examples admit the $P^{1/8}$
bandwidth. The coordinate-moment condition requires regressor
moments of order $4q$, namely $32$ and $80$, respectively.
These strong conditions provide a sufficient route to controlling
fourth-order regressor products uniformly over coordinates and
training windows, as required by our verification of the linked
estimation stages. The examples establish admissible regimes,
including dimensions growing faster than the sample size;
they do not characterize the weakest moment restrictions or
the largest admissible dimension.
\subsection{From estimation bounds to valid backtesting}
\label{sec:theory:strategy}

\citet{BelloniChernozhukov2011} establish quantile-Lasso
estimation rates for iid data. \citet{Barendse2023ESLasso} develops
ES-Lasso bounds with generated outcomes under dependence
and heavy tails.
Our contribution goes beyond estimation rates: we control
the accumulated effects of estimating VaR, ES, and the
evaluation instrument in a backtest with overlapping and expanding
training windows. Related estimation results for high-dimensional panel
quantile models with temporal dependence and latent
structure are provided by \citet{Bellonietal2023}.

Starting from the concentration inequality in Lemma~5 of
\citet{Barendse2023ESLasso}, we obtain bounds uniform over
training windows and propagate estimation errors through
the three stages. Uniform rates alone do not control the
interaction between changing estimates and current
innovations. We address this through proof-only deletion
of recent observations, estimator stability, and beta
mixing. This controls the first-order terms in the
backtesting expansion without requiring innovations to
be martingale differences relative to $\mathcal F_t$.
Together with orthogonality and the stated rate conditions,
the resulting bounds establish aggregate oracle equivalence
and valid HAC studentization. The implemented forecasting
procedure uses the original training windows throughout.
\section{Monte Carlo Simulations}\label{sec:mc}

We investigate the finite-sample size and power of the proposed
backtests. Our main objective is to assess whether orthogonalization
improves inference when ES forecasts are estimated by Lasso.
We compare the baseline (NLR) and orthogonal (LR) backtests,
holding the forecasts and initial test functions fixed.
We first examine rejection frequencies under correct specification
and then consider alternatives involving nonlinear conditional
tail risk and omitted predictors. Results for OLS and adaptive
Lasso ES estimation are reported in
Online Appendix~\ref{sec:OA_mc}.

\subsection{Size}

We first examine whether the backtests attain their nominal
5\% rejection probability when the conditional VaR and ES
models are correctly specified.

\paragraph{Data-generating process.}
The outcome follows
\begin{equation}\label{eq:dgp}
Y_{t+1}=\delta'X_t+\sigma u_{t+1},
\qquad X_t=(1,W_t')',
\end{equation}
where $\delta=(0,\delta_W')'$ is the conditional-mean
coefficient vector. The slope vector $\delta_W$ has
$s=\lfloor\sqrt{p_T-1}\rfloor$ nonzero coefficients,
with randomly selected locations and values drawn
independently from $\{-1,1\}$.
The innovations $u_{t+1}$ are independent of the predictors
and follow a Student-$t_5$ distribution standardized to
unit variance. We set $\sigma=1$ and $\alpha=0.10$.

Let $q_\alpha^u$ and $e_\alpha^u$ denote the
$\alpha$-quantile and ES of $u_{t+1}$, respectively.
The true conditional VaR and ES coefficient vectors are
\[
\gamma_0=\delta+\sigma q_\alpha^u\iota,
\qquad
\beta_0=\delta+\sigma e_\alpha^u\iota,
\qquad
\iota=(1,0,\ldots,0)'.
\]
Thus both forecasting models are correctly specified
and share the same slopes, but have different intercepts.

The predictors follow the sparse VAR process
\begin{equation}\label{eq:var}
W_t=AW_{t-1}+0.5\,\xi_t,
\end{equation}
where each row of $A$ contains $s$ randomly located
nonzero entries drawn from $\mathrm{Uniform}(0,1)$.
The matrix is rescaled to have spectral radius $0.5$.
The components of $\xi_t$ are independent Weibull
variables with shape $0.5$ and scale $1$, introducing
heavy-tailed predictors and serial dependence.
We discard the first 200 observations as burn-in.

\paragraph{Simulation design.}
We consider
\[
p_T-1\in\{25,50,100,150\},
\qquad
T\in\{250,500,1000,5000\},
\]
giving 16 configurations. The evaluation sample contains
$P=\lfloor0.3T\rfloor$ observations. Forecasts are estimated
once using the initial training sample and held fixed
during evaluation. Each configuration uses 200 Monte Carlo
replications.

VaR is estimated by $\ell_1$-penalized smoothed quantile
regression, as in \eqref{eq:linear_VaR_smoothed}, and ES
by Lasso regression of the pseudo-outcome in
\eqref{eq:pseudoOutcome}. Intercepts are unpenalized.
Penalties are selected using blocked cross-validation
and the additional rules described in
Online Appendix~\ref{sec:OA_mc_tuning}.
Instrument projections use adaptive Lasso with ridge
initial estimates.

The test function is $w(X_t)=(\delta'X_t)^2$,
which is nonlinear in the predictors. We compare
the baseline (NLR) and orthogonal (LR) scores in
\eqref{eq:ESscore_naive} and \eqref{eq:ESscore_orth},
respectively, using the HAC-based backtests of
Section~\ref{sec:tail:oos}.
Both procedures use identical forecasts and the same
initial test function; LR additionally residualizes
the instrument to remove first-order estimation effects.

\paragraph{Results.}
Table~\ref{table:size} reports rejection frequencies
under correct specification. The LR backtest rejects
between 3.0\% and 8.5\% of the time, close to the
nominal 5\% level. In contrast, NLR rejects between
11.5\% and 39.5\%, with substantial overrejection
even in the largest samples.

Because the forecasts and initial test function are
identical, these differences isolate the effect of
orthogonalization. The results illustrate its importance
for reliable backtesting with high-dimensional estimated
forecasts, particularly under heavy tails and serial
dependence.

\begin{table}[H]
\centering
\caption{Empirical size of ES backtests with Lasso estimation}
\label{table:size}
\small
\begin{tabular}{@{}r*{8}{r}@{}}
\toprule
& \multicolumn{2}{c}{$p_T-1=25$}
& \multicolumn{2}{c}{$p_T-1=50$}
& \multicolumn{2}{c}{$p_T-1=100$}
& \multicolumn{2}{c}{$p_T-1=150$} \\
\cmidrule(lr){2-3}
\cmidrule(lr){4-5}
\cmidrule(lr){6-7}
\cmidrule(l){8-9}
$T$ & LR & NLR & LR & NLR & LR & NLR & LR & NLR \\
\midrule
250  & 0.055 & 0.115 & 0.065 & 0.120
     & 0.030 & 0.195 & 0.075 & 0.395 \\
500  & 0.050 & 0.145 & 0.050 & 0.140
     & 0.075 & 0.215 & 0.040 & 0.280 \\
1000 & 0.060 & 0.130 & 0.035 & 0.190
     & 0.085 & 0.220 & 0.065 & 0.305 \\
5000 & 0.070 & 0.155 & 0.065 & 0.200
     & 0.060 & 0.285 & 0.045 & 0.335 \\
\bottomrule
\end{tabular}

\medskip
\begin{minipage}{0.96\textwidth}
\footnotesize
\textit{Notes.}
Entries are rejection frequencies under $H_0$ at the
nominal 5\% level, based on 200 Monte Carlo replications.
VaR is estimated by $\ell_1$-penalized smoothed quantile
regression and ES by ordinary Lasso.
LR and NLR use the orthogonal and baseline scores,
respectively.
The evaluation sample has size $P=\lfloor0.3T\rfloor$;
$p_T-1$ is the number of nonconstant predictors.
\end{minipage}
\end{table}

\subsection{Power}

We examine three forms of misspecification: nonlinear
conditional tail risk ($H_1$), predictors omitted from
both VaR and ES ($H_2$), and predictors omitted from
ES alone ($H_3$). We report rejection frequencies
for the orthogonal LR backtest with Lasso ES estimation.

\paragraph{Nonlinear conditional scale ($H_1$).}
The first alternative introduces conditional heteroskedasticity
while retaining all predictors in both forecasting models:
\begin{equation}\label{eq:dgp1}
Y_{t+1}=\delta'X_t+\sigma_{t+1}u_{t+1},
\qquad
\sigma_{t+1}^2=1+\eta(2\delta'X_t-1)^2.
\end{equation}
All other features of the null DGP are unchanged.
The parameter $\eta\ge0$ controls the strength of
conditional heteroskedasticity, with $\eta=0$
corresponding to the null.

The true conditional VaR and ES are
\[
\begin{aligned}
\VaR_\alpha(X_t)
&=\delta'X_t+\sigma_{t+1}q_\alpha^u,\\
\ES_\alpha(X_t)
&=\delta'X_t+\sigma_{t+1}e_\alpha^u.
\end{aligned}
\]
Thus, when $\eta>0$, both depend nonlinearly on
the predictors, whereas the fitted models remain
linear. We use squared fitted ES as the test
function, as in \eqref{eq:generated_test_function}.

Table~\ref{tab:mc_h1} reports the results.
Rejection frequencies are generally modest for
small samples but increase with the magnitude
of the alternative and sample size. At $T=5000$
and $\eta=0.03$, rejection frequencies reach
25.0\% and 39.5\% for 50 and 150 predictors,
respectively.

\begin{table}[!htbp]
\centering
\caption{Rejection frequencies under nonlinear conditional scale}
\label{tab:mc_h1}
\small
\begin{tabular}{@{}r*{6}{r}@{}}
\toprule
& \multicolumn{3}{c}{$p_T-1=50$}
& \multicolumn{3}{c}{$p_T-1=150$} \\
\cmidrule(lr){2-4}\cmidrule(l){5-7}
& \multicolumn{3}{c}{$\eta$}
& \multicolumn{3}{c}{$\eta$} \\
$T$ & 0.001 & 0.003 & 0.030
    & 0.001 & 0.003 & 0.030 \\
\midrule
250  & 0.060 & 0.050 & 0.085 & 0.055 & 0.050 & 0.085 \\
500  & 0.065 & 0.095 & 0.130 & 0.045 & 0.065 & 0.170 \\
1000 & 0.070 & 0.120 & 0.175 & 0.095 & 0.125 & 0.180 \\
5000 & 0.125 & 0.225 & 0.250 & 0.225 & 0.315 & 0.395 \\
\bottomrule
\end{tabular}
\par\medskip
\begin{minipage}{0.90\textwidth}
\footnotesize
\textit{Notes.} LR rejection frequencies at the nominal
5\% level under $H_1$, with ordinary Lasso ES estimation
and squared fitted ES as the test function.
The parameter $\eta$ controls conditional heteroskedasticity
in \eqref{eq:dgp1}; $p_T-1$ excludes the intercept.
\end{minipage}
\end{table}

\paragraph{Omitted predictors ($H_2$ and $H_3$).}
The remaining experiments examine whether the backtest
detects predictors that are relevant for tail risk but
excluded from the fitted ES model.

Starting from the null DGP, select three predictors
$W_{j_1,t},W_{j_2,t},W_{j_3,t}$ with nonzero coefficients
in $\delta_W$, using a fixed index ordering.
Multiply their coefficients by $\eta\ge0$, leaving
all other coefficients unchanged. Denoting the
resulting vector by $\delta^{(\eta)}$, generate
\begin{equation}\label{eq:dgp2}
Y_{t+1}=X_t'\delta^{(\eta)}+\sigma u_{t+1}.
\end{equation}
The true conditional VaR and ES remain linear
in the full predictor vector:
\[
\begin{aligned}
\VaR_\alpha(X_t)
&=X_t'\delta^{(\eta)}+\sigma q_\alpha^u,\\
\ES_\alpha(X_t)
&=X_t'\delta^{(\eta)}+\sigma e_\alpha^u.
\end{aligned}
\]

We deliberately exclude either one or three of these
predictors from the forecasting models. Let
$X_{-k,t}$ denote $X_t$ without
$W_{j_1,t},\ldots,W_{j_k,t}$, for $k\in\{1,3\}$.
The two alternatives differ only in the information
used for forecasting:
\[
\begin{array}{c|cc}
& \text{VaR regressors} & \text{ES regressors}\\ \hline
H_2 & X_{-k,t} & X_{-k,t}\\
H_3 & X_t & X_{-k,t}
\end{array}
\]

Under $H_2$, both forecasts omit relevant predictors.
Under $H_3$, VaR uses the full information set and is
correctly specified, while ES omits relevant predictors.
Thus $H_3$ isolates ES forecast misspecification without
VaR misspecification.

To illustrate, when $k=1$ the true conditional ES is
\[
\ES_\alpha(X_t)
=X_{-1,t}'\delta_{-1}^{(\eta)}
+\eta\delta_{j_1}W_{j_1,t}
+\sigma e_\alpha^u,
\]
where $\delta_{-1}^{(\eta)}$ excludes the coefficient
of $W_{j_1,t}$. The fitted ES model omits the term
$\eta\delta_{j_1}W_{j_1,t}$.
At $\eta=0$, the omitted predictor has no effect
and the null holds; increasing $\eta$ strengthens
the misspecification. The distinction between
$H_2$ and $H_3$ is whether the VaR model also
omits this predictor.

To detect the omission, we use
$w(X_t)=W_{j_1,t}$ as the initial test function.
The orthogonal score residualizes this instrument
against the included ES regressors $X_{-k,t}$,
retaining the information in the omitted predictor
not explained by those regressors.

\begin{figure}[!htbp]
\centering
\includegraphics[width=\textwidth,height=0.42\textheight,
keepaspectratio]{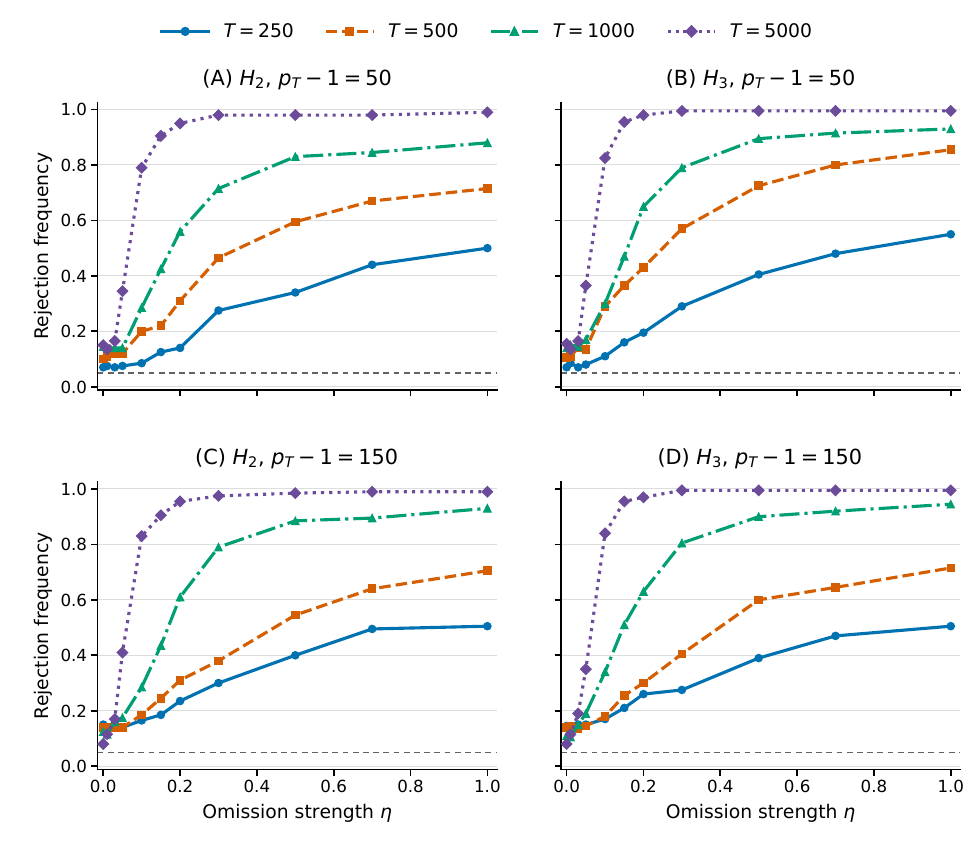}
\caption{Power against omitted predictors.
Columns compare $H_2$ (VaR and ES omissions) and
$H_3$ (ES-only omissions); rows correspond to
50 and 150 predictors. Curves distinguish sample sizes,
$\eta$ measures the omitted effect, and the dashed
line denotes the nominal 5\% level.}
\label{fig:mc_omission_power}
\end{figure}

Figure~\ref{fig:mc_omission_power} reports power against
one omitted predictor. Rejection frequencies generally
increase with the omitted effect and sample size.
Under $H_3$, the test detects ES misspecification
even when VaR is correctly specified.

The figure reports unadjusted rejection frequencies.
At $\eta=0$, these range from 7\% to 15.5\%,
indicating some finite-sample size distortion.
Consequently, rejection frequencies above 5\%
cannot be interpreted entirely as power gains.

Online Appendix~\ref{sec:OA_mc} reports additional
dimensions, three-predictor omissions, alternative
ES estimators, and test functions. Squared fitted ES has limited power against these omission
alternatives. This illustrates a general feature of the proposed
backtests: power depends on the association between the
residualized test function and the departure from the ES model.
A nonlinear transformation of fitted ES need not capture
information in an excluded predictor, whereas using that
predictor directly targets its omission. More generally, a
fixed collection of test functions can have little or no local
power against departures weakly correlated or uncorrelated
with the residualized instruments. Orthogonality controls
first-order estimation effects; power additionally requires
test functions informative about the alternatives of interest.

\FloatBarrier
\section{Empirical Application: Growth Expected Shortfall}
\label{sec:empirical}

We investigate whether macro-financial indicators contain
information about downside GDP growth beyond recent growth
history, and whether high-dimensional forecasting procedures
adequately exploit that information. Growth-at-Risk (GaR)
measures a lower quantile of future growth, while Growth
Expected Shortfall (GES) measures average growth in that
lower tail \citep{adrian2019growth}. Both are relevant for
monitoring vulnerabilities to severe economic downturns.

Our application to the G7 economies distinguishes two
questions: whether financial predictors improve forecasting
accuracy, and whether excluding them leaves detectable
misspecification. These questions need not have the same
answer. In particular, regularized forecasting methods may
exclude predictors that contain relevant information about
conditional tail risk. Neyman-orthogonal backtesting reduces
the first-order effects of estimation and variable-selection
errors on inference, allowing such restrictions to be
evaluated more reliably.

\subsection{Data and forecasting design}
\label{sec:empirical:data}

The outcome $Y_{t+1}$ is year-on-year growth in seasonally
adjusted quarterly real GDP, measured in percentage points.
We forecast one quarter ahead for the United States (USA),
Canada (CAN), Germany (DEU), the United Kingdom (UK),
Italy (ITA), Japan (JPN), and France (FRA).

Our information set combines recent GDP growth with
macro-financial indicators:
\[
Z_t=(1,Y_t,Y_{t-1},Y_{t-2},Y_{t-3})',
\qquad X_t=(Z_t',V_t')',
\]
where $V_t$ contains four lags of 15 additional series
covering financial conditions, credit, asset markets,
investment, inflation, and uncertainty. The full model
therefore contains 64 nonconstant predictors and an
intercept ($p_T=65$). This information set follows the
macro-financial forecasting literature
\citep{adrian2019growth,BrownleesSouza2021}.

We use expanding estimation windows. The initial training
sample contains $R=154$ observations, and the evaluation
period is 2002Q3--2019Q4 ($P=70$). For Canada, $R=152$
and the evaluation period is 2003Q1--2019Q4 ($P=68$).
At each forecast origin, estimation uses only preceding
predictor--outcome observations. This is a pseudo-out-of-sample
exercise based on historical data rather than real-time
vintages. Online Appendix~\ref{sec:OA_empirical} provides
additional data and implementation details, including
fixed-estimation results.

\subsection{High-dimensional GaR and GES forecasts}
\label{sec:empirical:estimation}

We set $\alpha=0.10$, focusing on the lower 10\% tail
of the conditional growth distribution. At each forecast
origin, we estimate GaR using penalized quantile regression
with the full predictor vector $X_t$. The intercept and
four growth lags are unpenalized, while the 60 additional
macro-financial coefficients are subject to an $\ell_1$
penalty.

Using this common GaR estimate, we construct the ES
pseudo-outcome in \eqref{eq:pseudoOutcome} and compare
two GES forecasting models:
\begin{equation}\label{eq:empirical_nested}
\begin{aligned}
\mathrm{M0}:&\quad
e_\alpha^{(0)}(X_t,\beta)=Z_t'\beta,\\
\mathrm{M1}:&\quad
e_\alpha^{(1)}(X_t,\beta)=X_t'\beta.
\end{aligned}
\end{equation}
M0 uses only recent growth history to forecast GES,
while M1 additionally allows the 60 macro-financial
predictors to enter. M0 is estimated by ordinary
least squares on the training pseudo-outcomes. M1
uses Lasso, leaving the intercept and growth lags
unpenalized. Both models use the same high-dimensional
GaR first step. Thus, M0 excludes financial predictors
from GES but not from GaR.

\begin{figure}[!htbp]
\centering
\includegraphics[
  width=0.95\textwidth,
  height=0.66\textheight,
  keepaspectratio
]{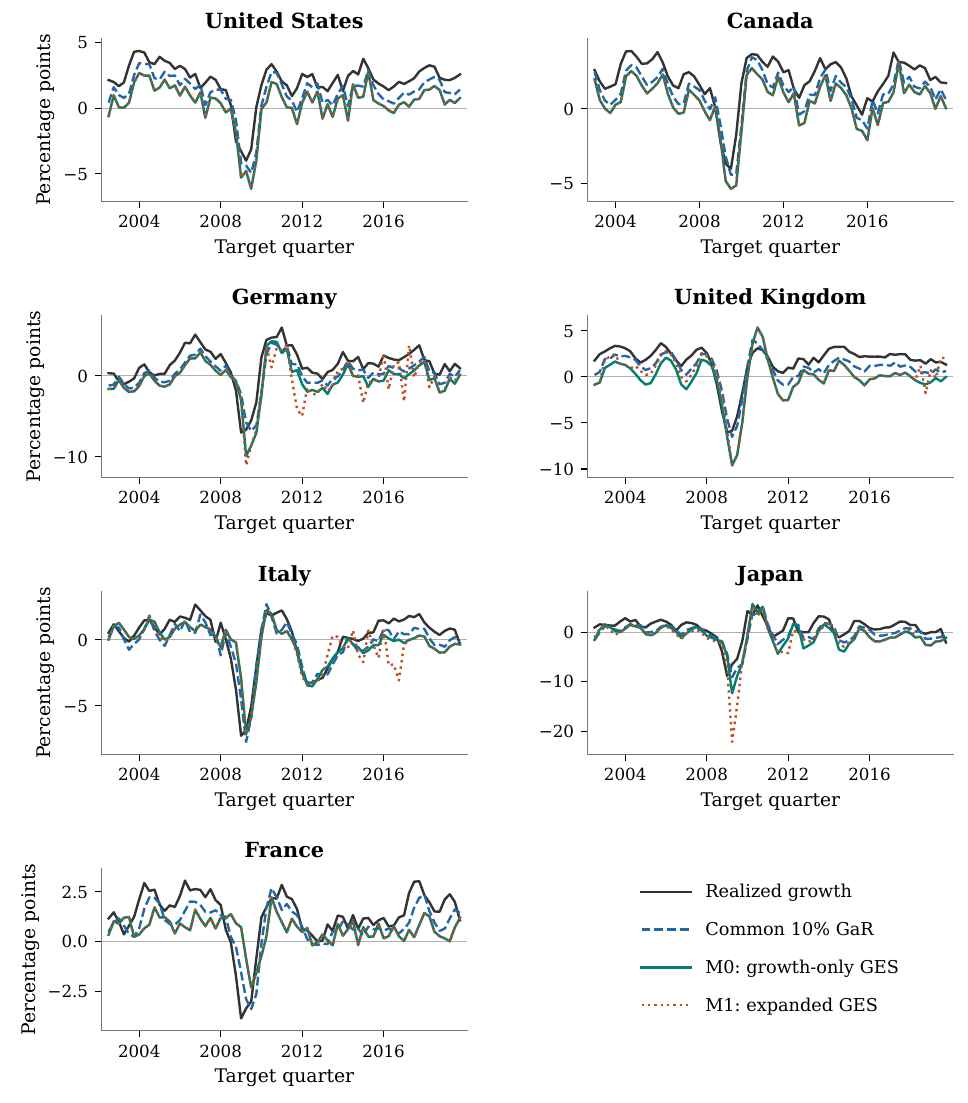}
\caption{Out-of-sample G7 downside-growth forecasts.
The common 10\% GaR, growth-only GES (M0), and expanded
GES (M1) use expanding estimation. The outcome is
year-on-year GDP growth, forecast one quarter ahead.
All evaluation dates are retained. Country scales differ.}
\label{GESALL}
\end{figure}

Predictors are standardized within each training window,
and penalties are selected by temporal cross-validation,
recomputing scaling and GaR estimates within the relevant
training folds. No minimum number of financial predictors
is imposed. Figure~\ref{GESALL} displays the resulting
GaR and GES forecasts alongside realized growth.

\FloatBarrier
\subsection{Backtesting financial information}
\label{sec:empirical:backtests}

Our first question is whether excluding financial
predictors produces suboptimal GES forecasts.
Under the maintained conditional GaR specification,
M0 imposes
\[
\E\!\left[
Y^*_{t+1}(\gamma_0)-Z_t'\beta_{0,0}
\mid X_t
\right]=0
\]
for some $\beta_{0,0}$. The corresponding restriction
for M1 replaces $Z_t'\beta_{0,0}$ with
$X_t'\beta_{1,0}$. Rejection provides evidence against
the conditional ES restriction in the directions tested.

\paragraph{Backtest construction.}
For M0, we test whether three financial indicators
contain information omitted from GES forecasts:
\[
w^{(0)}(X_t)
=(\mathrm{nfci}_t,\mathrm{inv}_t,\mathrm{ts}_t)',
\]
where the variables denote the latest included
financial-conditions index, investment growth, and
term spread, respectively. We jointly test these
three directions after projecting them on the
growth regressors $Z_t$ using training-sample OLS.
Projecting on the full $X_t$ would eliminate the
financial directions we wish to test.

For M1, which already incorporates linear financial
effects, we examine two nonlinear directions:
\[
w^{(1)}(X_t)
=(\mathrm{nfci}_t^2,\mathrm{nfci}_tY_t)',
\]
projected on $X_t$ by regularized regression.
We also consider the scalar test function
$\widehat w_t=(\widehat e_{t,\alpha}^{(1)})^2$.

We compare conventional baseline (NLR) and locally
robust (LR) backtests. Both use identical forecasts
and initial test functions. LR additionally
residualizes the test functions against the
derivatives of the fitted GES model, using $Z_t$
for M0 and $X_t$ for M1. Its score is
\[
\widehat S_{t+1}^{\perp,(j)}
=
\bigl\{Y^*_{t+1}(\widehat\gamma_t)
-\widehat e_{t,\alpha}^{(j)}\bigr\}
\bigl\{\widehat w_t^{(j)}
-\widehat\rho_{j,t}'
\dot e_\alpha^{(j)}(X_t,\widehat\beta_{j,t})\bigr\}.
\]
The baseline score uses the unprojected test function.

This distinction is particularly relevant with Lasso
forecasting. Regularization may shrink coefficients,
exclude relevant predictors, or select irrelevant ones.
Under the maintained assumptions, Neyman orthogonality
eliminates the first-order effect of such nuisance
estimation errors on the backtest. Valid inference
therefore does not require perfect variable selection.
At the same time, the test retains sensitivity to
omitted information that violates the conditional
ES restriction. Thus, robustness to estimation errors
does not imply robustness to genuine misspecification.

We use the HAC-based statistic $J_P$ from
Section~\ref{sec:tail:oos}, with a centered Bartlett
covariance estimator and four lags, retaining all
cross-covariances. The primary joint tests have
$d_w=3$ for M0 and $d_w=2$ for M1; the squared-GES
diagnostic has $d_w=1$. Table~\ref{bktLRNLR}
reports the corresponding $p$-values and Holm
adjustments for multiple testing across countries.

\begin{table}[!htbp]
\centering
\begin{threeparttable}
\caption{Backtesting financial information in GES forecasts}
\label{bktLRNLR}
\small
\setlength{\tabcolsep}{8pt}
\begin{tabular}{lrrrr}
\toprule
\multicolumn{5}{l}{Panel A: M0, financial predictors excluded from GES}\\
Country & Baseline & LR & LR: Holm 7 & LR: Holm 14\\
\midrule
USA & 0.0001 & 0.0009 & 0.0060 & 0.0120 \\
CAN & 0.6278 & 0.4393 & 0.7080 & 1.0000 \\
DEU & 0.1046 & 0.2432 & 0.7080 & 1.0000 \\
UK  & 0.2769 & 0.0874 & 0.4371 & 1.0000 \\
ITA & 0.0373 & 0.0152 & 0.0913 & 0.1979 \\
JPN & 0.2200 & 0.1650 & 0.6601 & 1.0000 \\
FRA & 0.6638 & 0.2360 & 0.7080 & 1.0000 \\
\midrule
\multicolumn{5}{l}{Panel B: M1, expanded GES specification}\\
& \multicolumn{2}{c}{Nonlinear joint test}
& \multicolumn{2}{c}{Squared GES}\\
Country & Baseline & LR & Baseline & LR\\
\midrule
USA & 0.2883 & 0.7775 & 0.1829 & 0.2154 \\
CAN & 0.0002 & 0.8125 & 0.7776 & 0.0061 \\
DEU & 0.1940 & 0.4047 & 0.5673 & 0.2473 \\
UK  & 0.6565 & 0.5078 & 0.3806 & 0.3927 \\
ITA & 0.2151 & 0.0954 & 0.1749 & 0.2893 \\
JPN & 0.5570 & 0.3098 & 0.3600 & 0.3478 \\
FRA & 0.4113 & 0.4326 & 0.0780 & 0.1824 \\
\bottomrule
\end{tabular}
\begin{tablenotes}[flushleft]\footnotesize
\item Entries are $p$-values from expanding-window
backtests using centered Bartlett HAC with four lags.
Panel A jointly tests financial conditions, investment
growth, and the term spread, residualized on $Z_t$
for LR ($d_w=3$). Panel B tests two nonlinear
financial directions, residualized on $X_t$
($d_w=2$), and separately squared fitted GES
($d_w=1$). Baseline scores use unprojected functions.
Holm 7 adjusts the M0 LR tests across countries;
Holm 14 adjusts both primary LR joint tests across
countries. The squared-GES diagnostic is a separate
family; its Canadian LR Holm-7 $p$-value is 0.0428.
Canada's nonlinear baseline Holm-14 $p$-value is
0.0031 in the corresponding baseline family.
\end{tablenotes}
\end{threeparttable}
\end{table}

\paragraph{Evidence of omitted financial information.}
Panel A provides strong evidence against the
growth-only GES specification in the United States.
The orthogonal joint test rejects the exclusion
of financial predictors ($p=0.0009$), and the
rejection survives Holm adjustment across all
14 primary country--model tests ($p=0.0120$).
Investment growth has the smallest individual
LR $p$-value ($0.0015$). The finding is also
robust to the two- and eight-lag HAC specifications
reported in the appendix.

Italy rejects the exclusion restriction at the
individual 5\% level ($p=0.0152$), although not
after multiple-testing adjustment. The other five
countries do not reject at 5\%. These findings
provide evidence that financial indicators contain
information about US downside-growth risk beyond
recent growth history, conditional on the maintained
GaR specification.

\paragraph{The importance of orthogonalization.}
Panel B illustrates how accounting for estimation
errors can materially change backtesting conclusions.
For Canada, the conventional nonlinear joint test
strongly rejects the expanded GES model
($p=0.0002$), whereas the orthogonal test does not
($p=0.8125$). The conventional rejection survives
multiple-testing adjustment. Because both procedures
use identical forecasts and initial test functions,
the difference arises from residualization and
the associated covariance estimation.

This contrast is consistent with the estimation-induced
overrejection documented in our simulations. In the
empirical application, however, the true specification
is unknown, so the comparison cannot establish that
the baseline rejection is false.

Orthogonalization does not invariably weaken evidence
against a model. Canada's squared-GES test rejects
under LR ($p=0.0061$) but not under the baseline
score ($p=0.7776$). The LR rejection survives Holm
adjustment across the seven squared-GES tests.
By contrast, the expanded nonlinear LR test rejects
none of the G7 countries at 5\%. These diagnostics
examine different departures from the conditional
ES restriction; nonrejection in one set of directions
does not establish correct specification.

\FloatBarrier
\subsection{Variable selection, forecast accuracy, and implications}
\label{sec:empirical:selection}

The backtests evaluate conditional ES restrictions,
but they do not directly establish that additional
predictors improve forecasting accuracy. We therefore
examine which variables the forecasting procedures
select, how well the resulting forecasts perform,
and whether targeted financial predictors can
improve them.

Table~\ref{tab:empirical_forecasts} addresses three
questions. First, how many additional macro-financial
coefficients does Lasso select in GaR and GES?
Second, are the resulting forecasts consistent with
basic tail-risk diagnostics? We report the proportion
of realized outcomes below estimated GaR, which
should be approximately 10\% under unconditional
quantile calibration, and the number of forecasts
for which GES exceeds GaR. Third, does including
financial predictors improve out-of-sample accuracy?
We compare M1 with M0 using a strictly consistent
joint VaR--ES loss from \citet{FisslerZiegel2016},
specified in Online Appendix~\ref{sec:OA_empirical}.
Because both models share the same GaR forecast,
their loss difference isolates the effect of
changing the GES forecast.

\begin{table}[!htbp]
\centering
\begin{threeparttable}
\caption{Variable selection, tail diagnostics, and GES forecast accuracy}
\label{tab:empirical_forecasts}
\small
\setlength{\tabcolsep}{5pt}
\begin{tabular}{lrrrrrr}
\toprule
& & \multicolumn{2}{c}{Selected $V_t$ coefficients}
& & & \\
\cmidrule(lr){3-4}
Country & $P$ & GaR & GES M1
& Hits (\%) & $e>q$: M0/M1 & $\Delta$ loss\\
\midrule
USA & 70 & 0.60 & 0.00 & 7.1 & 1/1 & 0.0000 \\
CAN & 68 & 2.10 & 0.00 & 8.8 & 0/0 & 0.0000 \\
DEU & 70 & 2.63 & 16.19 & 11.4 & 3/9 & 0.0849 \\
UK  & 70 & 2.60 & 4.24 & 12.9 & 3/11 & 0.0140 \\
ITA & 70 & 2.49 & 5.20 & 22.9 & 28/30 & 0.0385 \\
JPN & 70 & 1.19 & 5.33 & 17.1 & 8/7 & -0.0151 \\
FRA & 70 & 0.63 & 0.00 & 21.4 & 21/21 & 0.0000 \\
\bottomrule
\end{tabular}
\begin{tablenotes}[flushleft]\footnotesize
\item Expanding estimation, $\alpha=0.10$.
$P$ is the number of out-of-sample forecasts.
The GaR and GES columns report average numbers
of selected additional macro-financial coefficients
across forecast origins; the intercept and four
growth lags are unpenalized. Hits is the percentage
of outcomes below the common GaR forecast.
The $e>q$ column counts GES-above-GaR violations
under M0/M1. $\Delta$ loss is the average joint
VaR--ES loss for M1 minus M0, using
\eqref{eq:OA_empirical_loss}; negative values favor M1.
Loss differences are descriptive, not significance tests.
\end{tablenotes}
\end{threeparttable}
\end{table}

\paragraph{Variable selection versus specification.}
The most striking finding concerns the United States.
Lasso selects no additional financial predictors for
GES, so M1 produces the same forecasts as the
growth-only model M0. Nevertheless, the orthogonal
backtest strongly rejects M0's exclusion of financial
information.

This illustrates an important distinction between
variable selection and specification testing.
Lasso selects variables to optimize a penalized
forecasting objective, not to guarantee that conditional
ES restrictions hold. It may exclude predictors
whose contribution is small relative to the penalty
but whose omission leaves detectable information
in the forecast errors. Orthogonal backtesting is
particularly useful in this setting: it reduces
the first-order impact of estimation and selection
errors without requiring exact recovery of the
relevant predictors, while retaining sensitivity
to omitted conditional information.

Canada and France also select no additional financial
GES predictors, so M0 and M1 produce identical
forecasts. In Germany, the UK, Italy, and Japan,
M1 selects an average of 16.19, 4.24, 5.20, and
5.33 additional coefficients, respectively. Yet
greater model complexity does not systematically
improve forecasting performance. The expanded
model lowers average joint VaR--ES loss only in
Japan, while M0 performs better in Germany,
the UK, and Italy.

These results illustrate why variable selection,
forecast accuracy, and specification testing should
be distinguished. Even when M0 and M1 produce
identical forecasts, their backtests examine different
restrictions and use different test functions.
Nonrejection of M1's nonlinear restrictions therefore
does not overturn a rejection of M0's financial
exclusions.

\paragraph{Can the omitted predictors improve forecasts?}
To examine the forecasting relevance of the financial
information identified by the backtests, we consider
a targeted extension of M0. We retain the common
GaR first step and the four growth lags, but focus
on the three financial indicators used in the
exclusion tests: financial conditions, investment
growth, and the term spread.

Model M2 includes all three indicators without
penalization, while M3 selects among their eight
possible subsets using chronological validation.
Table~\ref{tab:empirical_targeted} in the Online
Appendix reports the results.

M3 achieves lower average joint VaR--ES loss and
lower squared pseudo-outcome loss than M0 in the
United States, Canada, Germany, the UK, and France.
For the United States, squared pseudo-outcome loss
falls from 4.4244 to 4.2781, a 3.31\% reduction,
while joint loss falls from $-1.2724$ to $-1.2876$.

These findings reinforce the practical relevance
of the US exclusion rejection. Although the
high-dimensional Lasso selects no additional
financial GES predictors, a targeted specification
using the tested indicators achieves lower forecasting
loss. This illustrates how orthogonal backtesting
can complement regularized variable selection
by identifying potentially useful omitted information.

The targeted exercise was motivated by previously
inspected backtesting results, however, and the
improvements are not uniform across countries or
diagnostics. These forecasting comparisons should
therefore be regarded as exploratory rather than
independent confirmation of predictive gains.

\paragraph{Forecast diagnostics and implications.}
Table~\ref{tab:empirical_forecasts} also reveals
limitations of the estimated forecasts. GaR
exceedance frequencies range from 7.1\% to 22.9\%,
with particularly high values in Italy and France.
Some fitted GES forecasts also exceed GaR, especially
in these countries. Although the population risk
measures satisfy $\ES_\alpha(X_t)\leq\VaR_\alpha(X_t)$,
our separate regressions do not impose this inequality.
Linear projection of the ES pseudo-outcome need not
preserve the ordering, and M0 additionally uses
a restricted information set. These crossings
therefore cannot be attributed to regularization
bias alone.

Our methodology addresses the effect of high-dimensional
estimation on backtesting inference, not bias or
coherence of the forecasts themselves. Its validity
also relies on the maintained GaR conditions.
The Online Appendix reports further forecast
diagnostics and comparisons with empirical-AR
and Student-$t$ AR--GARCH benchmarks.

Overall, the application demonstrates the value
of separating forecasting from specification assessment.
Regularized estimation can produce parsimonious forecasts
while leaving relevant financial information unused.
Orthogonal backtesting reduces sensitivity to nuisance
estimation and variable-selection errors, providing
evidence about conditional ES restrictions that
predictive selection alone does not supply.
For practitioners, this offers a complementary tool
for validating high-dimensional downside-risk forecasts
and identifying specifications that merit further review.
\FloatBarrier
\section{Conclusions}\label{sec:conclusions}

This paper develops out-of-sample ES backtests that account for
high-dimensional estimation of VaR, ES, and the evaluation instrument.
For linear ES models, the orthogonality correction requires an additional
regression of the test function on the ES regressors. Under the stated
moment, dependence, and nuisance-estimation conditions, the feasible
backtest is asymptotically equivalent to its oracle counterpart, including
with overlapping rolling or expanding training windows. Correctness is
required relative to the forecasting information $X_t$; HAC studentization
allows serial correlation arising from additional predictive information
in the larger filtration $\mathcal F_t$.

The simulations show that orthogonalization improves size control relative
to baseline ES backtests, while sensitivity to omitted predictors depends
on the test function. Remaining null overrejection in the omission designs
and the degenerate-instrument stress test clarify the limits of the
finite-sample evidence. The G7 application finds evidence against the US
growth-only ES restriction even when forecast-oriented regularization
selects no additional financial predictors. The exploratory small-predictor
extensions show that revisiting these exclusions can improve forecasts in
some countries, but gains are not uniform and coherence violations remain.
Accounting for estimation can also change rejection decisions in either
direction, as illustrated by Canada.

These findings distinguish predictive selection, comparative forecast
performance, and backtesting of conditional ES restrictions. Useful
extensions include more powerful choices of test functions, inference
across larger families of forecasting models, and multi-horizon or
multivariate tail-risk forecasts.


\clearpage
\bibliographystyle{plainnat}
\bibliography{references}

\clearpage
\appendix
\setcounter{table}{0}
\setcounter{figure}{0}
\renewcommand{\thetable}{OA.\arabic{table}}
\renewcommand{\thefigure}{OA.\arabic{figure}}
\renewcommand{\theHtable}{OA.\arabic{table}}
\renewcommand{\theHfigure}{OA.\arabic{figure}}
\renewcommand{\theHsection}{OA.\arabic{section}}
\renewcommand{\thesection}{OA.\arabic{section}}
\renewcommand{\thesubsection}{\thesection.\arabic{subsection}}
\numberwithin{equation}{section}

\section*{Online Appendix}
\addcontentsline{toc}{section}{Online Appendix}


\theoremstyle{plain}
\newtheorem{OAtheorem}{Theorem}[section]
\newtheorem{OAlemma}{Lemma}[section]
\newtheorem{OAproposition}{Proposition}[section]
\newtheorem{OAcorollary}{Corollary}[section]

\theoremstyle{definition}
\newtheorem{OAassumption}{Assumption}[section]

\theoremstyle{remark}
\newtheorem{OAremark}{Remark}[section]
\section{Proofs for orthogonal backtesting moments}
\label{sec:OA_proofs}

\subsection{Proof of Theorem~\ref{thm:orth}}

\begin{proof}
For the residuals defined in Section~\ref{sec:tail:LRmoment},
conditioning on $X_t$ gives
\[
\begin{aligned}
\E[\psi_{VaR,t+1}(\gamma)\mid X_t]
&=F(q_\alpha(X_t,\gamma)\mid X_t)-\alpha,\\
\E[\psi_{ES,t+1}(\gamma,\beta)\mid X_t]
&=\alpha^{-1}
\int_{(-\infty,q_\alpha(X_t,\gamma)]}
y\,dF(y\mid X_t)-e_\alpha(X_t,\beta).
\end{aligned}
\]
Both expressions vanish at $(\gamma_0,\beta_0)$.
The assumed integrability and iterated expectations therefore
give $\E[S_{t+1}(\gamma_0,\beta_0)]=0$.

For any admissible direction $h_\gamma$, differentiation at
$q_{0t}=q_\alpha(X_t,\gamma_0)$ yields
\begin{align}
\left.
\frac{d}{d\tau}
\E[\psi_{VaR,t+1}(\gamma_0+\tau h_\gamma)\mid X_t]
\right|_{\tau=0}
&=
f(q_{0t}\mid X_t)
\partial_\gamma q_\alpha(X_t,\gamma_0)[h_\gamma],
\label{eq:OA_var_derivative}\\
\left.
\frac{d}{d\tau}
\E[\psi_{ES,t+1}(\gamma_0+\tau h_\gamma,\beta_0)\mid X_t]
\right|_{\tau=0}
&=
\frac{q_{0t}}{\alpha}f(q_{0t}\mid X_t)
\partial_\gamma q_\alpha(X_t,\gamma_0)[h_\gamma].
\label{eq:OA_es_derivative}
\end{align}
Assumption~\ref{ass:models}(iv) justifies differentiation
under the outer expectation. Hence
\[
\begin{aligned}
&\left.
\frac{d}{d\tau}
\E[S_{t+1}(\gamma_0+\tau h_\gamma,\beta_0)]
\right|_{\tau=0}\\
&\quad=
\E\!\left[
\left\{\lambda_1(X_t)+\frac{q_{0t}}{\alpha}\lambda_2(X_t)\right\}
f(q_{0t}\mid X_t)
\partial_\gamma q_\alpha(X_t,\gamma_0)[h_\gamma]
\right].
\end{aligned}
\]
This derivative vanishes for every $h_\gamma$ if and only if
\eqref{orth0} holds; the pointwise restriction \eqref{orth1}
is sufficient.

Similarly, since only the ES residual depends on $\beta$,
\[
\left.
\frac{d}{d\tau}
\E[S_{t+1}(\gamma_0,\beta_0+\tau h_\beta)]
\right|_{\tau=0}
=
-\E\!\left[
\lambda_2(X_t)
\partial_\beta e_\alpha(X_t,\beta_0)[h_\beta]
\right].
\]
This vanishes for every admissible $h_\beta$ if and only if
\eqref{orth2} holds, completing the proof.
\end{proof}

\subsection{Proof of Proposition~\ref{prop:residualized_orthogonality}}

\begin{proof}
Define
\[
k_t(\beta,\rho)
:=w(X_t,\beta)-\rho'\dot e_\alpha(X_t,\beta),
\]
with $w$ independent of $\beta$ for a fixed test function.
At the truth, write
\[
w_{0t}=w(X_t,\beta_0),\qquad
k_{0t}=k_t(\beta_0,\rho_0),\qquad
\varepsilon_{t+1}=Y^*_{t+1}(\gamma_0)-e_{0t}.
\]
Since $\E[\varepsilon_{t+1}\mid X_t]=0$, moment validity
follows by iterated expectations.

For every admissible $h_\gamma$, the pseudo-outcome satisfies
\[
\left.
\frac{d}{d\tau}
\E[Y^*_{t+1}(\gamma_0+\tau h_\gamma)\mid X_t]
\right|_{\tau=0}
=
\left\{1-\frac{F(q_{0t}\mid X_t)}{\alpha}\right\}
\partial_\gamma q_\alpha(X_t,\gamma_0)[h_\gamma]
=0.
\]
Thus the score is orthogonal to $\gamma$.
For a direction $h_\beta$, the derivative of its expectation is
\[
-\E\!\left[
k_{0t}\partial_\beta e_\alpha(X_t,\beta_0)[h_\beta]
\right]
+
\E\!\left[
\varepsilon_{t+1}
\partial_\beta k_t(\beta_0,\rho_0)[h_\beta]
\right]
=0.
\]
The first term vanishes because $k_{0t}$ is orthogonal to
$\mathcal R_e$ and the ES derivative belongs to that space.
The second vanishes by conditioning on $X_t$.
Likewise, a matrix direction $h_\rho$ gives
\[
-\E\!\left[
\varepsilon_{t+1}h_\rho'\dot e_\alpha(X_t,\beta_0)
\right]=0.
\]
The stated envelope and integrability conditions justify
interchanging derivatives and expectations.
For $w(X_t,\beta)=e_\alpha(X_t,\beta)^2$, its derivative is
$2e_{0t}\partial_\beta e_\alpha(X_t,\beta_0)[h_\beta]$,
so the same argument applies under the stated product
integrability conditions.
\end{proof}
\section{Oracle equivalence and studentization}
\label{sec:OA_general}
\subsection{Score decomposition and remainder conditions}
\label{sec:OA_remainders}

Throughout this subsection, sums are over
$t\in\mathcal T_{\mathrm{test}}$, and limits are taken as
$T\to\infty$. Using the notation of
Section~\ref{sec:theory:general}, define the estimated
residualized instrument
\[
\widehat k_t
:=
\widehat w_t-\widehat\rho_t'
\dot e_\alpha(X_t,\widehat\beta_t),
\]
and the estimation errors
\[
\begin{aligned}
\delta_{qt}
&:=q_\alpha(X_t,\widehat\gamma_t)-q_{0t},\\
B_t
&:=e_\alpha(X_t,\widehat\beta_t)-e_{0t},\\
D_t
&:=\widehat k_t-k_{0t},\\
G_{t+1}
&:=U_{t+1}(q_{0t}+\delta_{qt})-U_{t+1}(q_{0t}).
\end{aligned}
\]
Here $\delta_{qt}$ and $B_t$ are the VaR and ES forecast
errors, $D_t$ is the residualized-instrument error, and
$G_{t+1}$ is the change in the pseudo-outcome induced by
VaR estimation.

Since
$\widehat S_{t+1}^{\perp}
=(k_{0t}+D_t)(\varepsilon_{t+1}+G_{t+1}-B_t)$,
the feasible--oracle score difference admits the exact
decomposition
\begin{equation}\label{eq:OA_master_decomposition}
\widehat S_{t+1}^{\perp}-S_{t+1}^{\perp}
=
k_{0t}G_{t+1}
-k_{0t}B_t
+\varepsilon_{t+1}D_t
+D_t(G_{t+1}-B_t).
\end{equation}
The first three terms isolate the effects of estimating
VaR, ES, and the residualized instrument; the last contains
their interactions.

The remainder conditions imposed in
Assumption~\ref{ass:general_pipeline} require these
components to be negligible after aggregation:
\begin{align}
\left\|
P^{-1/2}\sum_t k_{0t}G_{t+1}
\right\|
&=o_p(1),
\label{eq:OA_HL1}\\
\left\|
P^{-1/2}\sum_t D_t\varepsilon_{t+1}
\right\|
&=o_p(1),
\label{eq:OA_HL2}\\
\left\|
P^{-1/2}\sum_t k_{0t}B_t
\right\|
&=o_p(1),
\label{eq:OA_HL3}\\
\left\|
P^{-1/2}\sum_t D_t(G_{t+1}-B_t)
\right\|
&=o_p(1).
\label{eq:OA_HL4}
\end{align}
Together with \eqref{eq:OA_master_decomposition}, these
conditions imply aggregate oracle equivalence. They do
not impose martingale-difference restrictions on the
innovations. Their verification for the Lasso estimators
uses the uniform estimation and stability results
developed below.

For studentization, define the empirical score distance
\[
r_{S,T}
:=
\left\{
\frac1P\sum_t
\|\widehat S_{t+1}^{\perp}-S_{t+1}^{\perp}\|^2
\right\}^{1/2}.
\]
The additional conditions in
Theorem~\ref{thm:main_asymptotics} are
\begin{equation}\label{eq:OA_HAC_rate}
\begin{gathered}
r_{S,T}=o_p(1),
\qquad
L_P\to\infty,
\qquad
\frac{L_P^3}{P}\to0,\\
\frac{L_P^2A_P}{P}\to0,
\qquad
L_Pr_{S,T}=o_p(1).
\end{gathered}
\end{equation}
where $L_P$ is the HAC truncation lag. The aggregate
remainder conditions control the score average, whereas
$r_{S,T}$ controls the effect of replacing oracle scores
by feasible scores in the covariance estimator.

\subsection{Proof of Theorem~\ref{thm:main_asymptotics}}

\begin{proof}
The decomposition \eqref{eq:OA_master_decomposition} and
\eqref{eq:OA_HL1}--\eqref{eq:OA_HL4} immediately give
\[
P^{-1/2}\sum_{t\in\mathcal T_{\mathrm{test}}}
(\widehat S_{t+1}^{\perp}-S_{t+1}^{\perp})=o_p(1).
\]
Moreover, Assumption~\ref{ass:VaRcorrect} implies
$\E S_{t+1}^{\perp}=0$.

We first establish the oracle central limit theorem,
allowing the score distribution to depend on $T$.
Reindex the test observations as
$S_{T,i}:=S_{R+i}^{\perp}$, $i=1,\ldots,P$.
Fix $v\in\mathbb R^{d_w}\setminus\{0\}$ and put
$Z_{T,i}=v'S_{T,i}$ and $\vartheta=1-2/q$.
The covariance inequality for mixing sequences gives
\[
|\Cov(Z_{T,i},Z_{T,i-j})|
\le C_v\beta(j)^\vartheta.
\]
Consequently, the variance of a sum over any $m$ distinct
indices is bounded by $C_vm$, uniformly in $T$.
The summable covariance envelope also gives
\[
\frac1P\operatorname{Var}\left(\sum_{i=1}^P Z_{T,i}\right)
=v'\Omega_Tv+o(1)\longrightarrow v'\Omega v>0.
\]

For completeness, the required triangular-array CLT
follows by blocking. Let $\beta_*(j)$ be the supremum,
over $T$, of the oracle score mixing coefficients.
This envelope is nonincreasing, satisfies
$\beta_*(j)\le\beta(j)$, and therefore obeys
\[
\beta_*(j)=o(j^{-\nu}),
\qquad
\nu:=\vartheta^{-1}.
\]
Set
\[
a:=\frac1{1+\nu}=\frac{q-2}{2(q-1)}.
\]
Choose $u_P\to\infty$ sufficiently slowly that, with
\[
b_P=\lfloor P^a/u_P\rfloor,
\qquad
c_P=\lfloor P^a/u_P^2\rfloor,
\]
we have $c_P\to\infty$ and
$(P/b_P)\beta_*(c_P)\to0$.
Such a choice exists because
$a(1+\nu)=1$ and $j^\nu\beta_*(j)\to0$.

Partition the observations into big blocks of length
$b_P$, separated by gaps of length $c_P$, and discard
the final incomplete block. The number of discarded
indices is $o(P)$, so their normalized sum is $o_p(1)$
by the variance bound above. Absolute regularity couples
the retained block vectors to independent copies, with
total failure probability at most
$C(P/b_P)\beta_*(c_P)=o(1)$.

Let $B_{P,k}^*$ denote the corresponding independent
scalar block sums. Minkowski's inequality gives
$\E|B_{P,k}^*|^q\le C_vb_P^q$, and hence
\[
P^{-q/2}\sum_k\E|B_{P,k}^*|^q
\le C_vP^{1-q/2}b_P^{q-1}
\le C_vu_P^{-(q-1)}\longrightarrow0.
\]
Their total variance divided by $P$ converges to
$v'\Omega v$. Indeed, discarding the gaps changes the
normalized variance by $o(1)$, while removing
cross-block covariances changes it by at most
\[
C_v\sum_{j\ge c_P}\beta(j)^\vartheta=o(1).
\]
The independent-array Lyapunov theorem, coupling, and
the negligible discarded sum establish the scalar CLT.
Cram\'er--Wold and oracle equivalence now yield
\[
P^{-1/2}\sum_{i=1}^P S_{T,i}\Rightarrow N(0,\Omega),
\qquad
\sqrt P\,\bar S_P^\perp\Rightarrow N(0,\Omega).
\]

We next prove oracle HAC consistency. Define
\[
\widetilde\Gamma_\ell
:=\frac1P\sum_{i=\ell+1}^P S_{T,i}S_{T,i-\ell}',
\qquad
\Gamma_{T,\ell}:=\E[S_{T,i}S_{T,i-\ell}'],
\]
and let $\widetilde\Omega$ be the estimator
\eqref{eq:HAC_definition} with
$\widehat\Gamma_\ell$ replaced by
$\widetilde\Gamma_\ell$.

Each coordinate product in $\widetilde\Gamma_\ell$
has uniformly bounded $L^{q/2}$ norm.
For two such products whose time indices differ by
$r>\ell$, their covariance is bounded in absolute
value by
$C\beta(r-\ell)^{1-4/q}$.
For $r\le\ell$, Cauchy--Schwarz gives a uniform
constant bound. Summing these bounds shows that,
for every matrix entry $(j,k)$,
\[
\operatorname{Var}
\bigl(\widetilde\Gamma_\ell^{(j,k)}\bigr)
\le \frac{C(\ell+A_P)}P.
\]
Since the dimension and $\|K\|_\infty$ are fixed,
Minkowski's inequality therefore gives
\[
\begin{aligned}
\left\{\E\|
\widetilde\Omega-\E\widetilde\Omega
\|_F^2\right\}^{1/2}
&\le
\frac C{\sqrt P}\sum_{\ell=0}^{L_P}
\sqrt{\ell+A_P}\\
&\le C\left\{
\frac{L_P^{3/2}}{\sqrt P}
+L_P\sqrt{\frac{A_P}{P}}
\right\}
=o(1).
\end{aligned}
\]
Furthermore,
$\E\widetilde\Gamma_\ell=(1-\ell/P)\Gamma_{T,\ell}$.
The covariance inequality gives the summable envelope
\[
\|\Gamma_{T,\ell}\|_F
\le C\beta(\ell)^{1-2/q},
\qquad \ell\ge1.
\]
A finite-lag and tail decomposition, using boundedness
of $K$, continuity at zero, $K(0)=1$, and
$L_P\to\infty$, yields
\[
\|\E\widetilde\Omega-\Omega_T\|_F\to0.
\]
Thus $\widetilde\Omega\xrightarrow{p}\Omega$.

Finally, set
\[
M_{S,P}
:=\left(\frac1P\sum_{i=1}^P\|S_{T,i}\|^2\right)^{1/2}
=O_p(1).
\]
Expanding the feasible--oracle autocovariance difference
and applying Cauchy--Schwarz gives
\[
\max_{0\le\ell\le L_P}
\|\widehat\Gamma_\ell-\widetilde\Gamma_\ell\|_F
\le 2r_{S,T}M_{S,P}+r_{S,T}^2.
\]
Both index sets in every lagged sum are subsets of the
test sample, which justifies using the full-sample
norms on the right. Therefore,
\[
\begin{aligned}
\|\widehat\Omega-\widetilde\Omega\|_F
&\le
(1+2L_P\|K\|_\infty)
(2r_{S,T}M_{S,P}+r_{S,T}^2)\\
&=O_p\{L_Pr_{S,T}(1+r_{S,T})\}
=o_p(1).
\end{aligned}
\]
Hence $\widehat\Omega\xrightarrow{p}\Omega$.
Since $\Omega\succ0$, continuity of matrix inversion
and Slutsky's theorem imply
\[
J_P
=P\,\bar S_P^{\perp\prime}
\widehat\Omega^{-1}\bar S_P^\perp
\Rightarrow\chi^2_{d_w}.
\]
\end{proof}
\section{Uniform Lasso estimation and backtest validity}
\label{sec:OA_primitive_proofs}

\subsection{Notation and primitive conditions}

We retain the notation of Section~\ref{sec:theory:lasso},
including $X_t=(Z_t',V_t')'$, the effective supports
$S_\gamma,S_\beta,S_{\rho,r}$, and the sparsity counts
$s_\gamma,s_\beta,s_\rho,s_\circ$.
The forecasts and residualized instrument are
\[
q_{0t}=X_t'\gamma_0,\qquad
e_{0t}=X_t'\beta_0,\qquad
k_{0t}=w_{0t}-\rho_0'X_t.
\]
Define
\[
A_{t+1}:=
1-\alpha^{-1}\mathbf 1\{Y_{t+1}\le q_{0t}\},
\qquad
\Xi_t:=k_{0t}X_t'.
\]
Conditional correctness and the projection normal equations
give $\E[A_{t+1}\mid X_t]=0$ and $\E[\Xi_t]=0$.

For an interval $I$, write
\[
\widehat\Sigma_I
:=\frac1{|I|}\sum_{s\in I}X_sX_s',
\qquad
\|v\|_I^2:=v'\widehat\Sigma_Iv,
\qquad
\|v\|_\Sigma^2:=v'\Sigma v,
\quad \Sigma:=\E[X_tX_t'].
\]
For vectors or matrices, $\|\cdot\|_{\max}$ denotes the
largest absolute coordinate; absolute values of matrices
are understood entrywise.

\paragraph{Training samples and partial penalization.}
For rolling and recursive estimation, let
\[
\widetilde{\mathcal I}_t^{(h)}
:=
\mathcal I_t\setminus\{t-h,\ldots,t-1\},
\qquad 1\le h<R/2.
\]
These shortened samples are used only in the proof.
Let $\mathfrak I_t$ contain $\mathcal I_t$ and, for rolling
and recursive estimation, $\widetilde{\mathcal I}_t^{(h)}$.
All samples in $\mathfrak I_t$ have length at least $R/2$.
For $I\in\mathfrak I_t$, define
\[
\mathcal G_I:=\sigma\{(X_s,Y_{s+1}):s\in I\}.
\]
Thus $\mathcal G_{\mathcal I_t}=\mathcal G_t$, and
$\mathcal G_{\widetilde{\mathcal I}_t^{(h)}}
\subseteq\mathcal F_{t-h}$.

Every fit at origin $t$ uses the prespecified unpenalized
block $Z_t$ and a penalty on the $V_t$ coefficients only.
The shortened-sample fit uses the same stage-specific
penalty as the original fit. Penalties may differ across
stages and forecast origins.

For squared-loss regressions, multiplying both the loss
and its penalty by $1/2$ leaves the minimizer unchanged.
We use this normalization in the proofs and retain
$\lambda_{\beta,t}$ and $\lambda_{\rho,t}$ for the rescaled
penalties, adjusting the fixed comparability constants
accordingly.

For an effective support $S$, define the cone
\[
\mathcal C(S)
:=
\{v\in\mathbb R^{p_T}:
\|v_{S^c}\|_1\le3\|v_S\|_1\}.
\]
Every effective support includes all unpenalized coordinates.

\begin{OAassumption}[Primitive conditions for Lasso verification]
\label{ass:OA_primitive_arrays}
Constants below are independent of $T$, forecast origins,
and coordinates. The effective supports include all
unpenalized regressors, and $R,P\asymp T$.

\begin{enumerate}[label=(\roman*),leftmargin=*]

\item \emph{Sampling and conditional density.}
The triangular array
$\mathsf W_t=(X_t,Y_{t+1})$ is strictly stationary and
beta mixing, with coefficients bounded uniformly in $T$ by
\[
\beta(j)\le Cj^{-\mu},
\qquad \mu>2.
\]
The beta coefficient between $\mathcal F_{t-h}$ and
$\sigma(\mathsf W_t)$ is bounded by $C\beta(h)$,
allowing an immaterial fixed shift of the index.
The operational estimators are $\mathcal F_t$-measurable,
and the shortened-sample estimators are
$\mathcal F_{t-h}$-measurable.

Assumption~\ref{ass:VaRcorrect} holds.
For a fixed $r_f>0$, the conditional density of
$Y_{t+1}$ given $X_t$ is uniformly bounded above,
bounded away from zero, and Lipschitz in its first
argument on
\[
|y-q_{0t}|\le r_f.
\]

\item \emph{Population curvature.}
For every truth support
$S\in\{S_\gamma,S_\beta,S_{\rho,1},\ldots,S_{\rho,d_w}\}$,
\[
\|v\|_\Sigma\ge\underline\kappa\|v_S\|_2,
\qquad v\in\mathcal C(S),
\]
for some $\underline\kappa>0$.

There exist constants
$0<\underline\phi\le\overline\phi<\infty$ such that,
for every fixed $M\ge1$ and all sufficiently large $T$,
\[
\underline\phi\|v\|_2^2
\le v'\Sigma v
\le\overline\phi\|v\|_2^2
\qquad
\text{whenever }0<\|v\|_0\le Ms_\circ.
\]
The eigenvalue constants do not depend on $M$.
Also, $\max_j\Sigma_{jj}\le C$.

\item \emph{Unconditional directional moments.}
For some $q>4$, define
\[
\mathcal V_M
:=
\mathcal C(S_\gamma)
\cup\mathcal C(S_\beta)
\cup\bigcup_{r=1}^{d_w}\mathcal C(S_{\rho,r})
\cup\{v:\|v\|_0\le Ms_\circ\}.
\]
For every fixed $M$, assume
\begin{equation}
\label{eq:OA_directional_primitive}
\|X_t'v\|_{L^{4q}}
\le C_M\|v\|_\Sigma,
\qquad v\in\mathcal V_M.
\end{equation}
The true coefficient vectors satisfy
\[
\|\gamma_0\|_\Sigma+\|\beta_0\|_\Sigma
+\max_{1\le r\le d_w}\|\rho_{0,r}\|_\Sigma
\le C.
\]
For a fixed test function, assume
$\|\|w_0(X_t)\|\|_{L^{4q}}\le C$.
Alternatively, the generated test function is
$w_{0t}=e_{0t}^2$.

\item \emph{Innovation moments and nondegeneracy.}
The ES innovation satisfies
\[
\E\!\left[
|\varepsilon_{t+1}|^q\mid\mathcal F_t
\right]\le C
\quad\text{a.s.}
\]
The projection normal equations hold, and
$\Omega_T\to\Omega\succ0$.
The conditional moment restriction bounds the magnitude
of the innovation; it does not require
$\E[\varepsilon_{t+1}\mid\mathcal F_t]=0$.

\item \emph{Penalties and growth.}
The penalties are deterministic and satisfy
\eqref{eq:lasso_penalty_comparability}, with sufficiently
large fixed constants. A shortened-sample fit uses the
same penalty as the corresponding operational fit.

Let
\[
g_T:=\Lambda_T(p_T^2,q),
\qquad
g_{4,T}:=\Lambda_T(p_T^4,q).
\]
Require
\begin{equation}
\label{eq:OA_primitive_growth}
p_T^2d_T\beta(a_T)\to0,
\qquad
s_\circ g_T\to0,
\qquad
s_\circ^2g_{4,T}\to0,
\qquad
s_\circ\lambda_T\to0.
\end{equation}

\end{enumerate}
\end{OAassumption}

Condition~\eqref{eq:OA_directional_primitive} is an
unconditional restriction on the population distribution.
It does not condition on the training sample or refer to
estimated coefficient-error directions.
When $s_\circ=O(1)$, it follows from uniform coordinate
$4q$ moments and the population curvature conditions.
Indeed, on each class $\mathcal V_M$,
\[
\|v\|_1\le C_M\|v\|_\Sigma,
\]
and therefore Minkowski's inequality gives
\[
\|X_t'v\|_{L^{4q}}
\le
\sum_{j=1}^{p_T}|v_j|\,\|X_{jt}\|_{L^{4q}}
\le C_M\|v\|_\Sigma.
\]
Thus this sufficient condition permits unbounded
regressors. For growing sparsity,
\eqref{eq:OA_directional_primitive} is maintained as an
explicit population restriction.
\begin{OAlemma}[Consequences of the primitive design conditions]
\label{lem:OA_primitive_design}
Under Assumption~\ref{ass:OA_primitive_arrays},
the following conclusions hold uniformly over consecutive
intervals $I$ of length at least $cT$, for any fixed $c>0$.

First,
\[
\|\widehat\Sigma_I-\Sigma\|_{\max}=O_p(g_T).
\]
Consequently, the sample restricted eigenvalues on the
truth cones and the sample sparse eigenvalues on supports
of size at most $Ms_\circ$, for every fixed $M$, satisfy
the corresponding population bounds up to fixed factors,
with probability tending to one.

Second, for every fixed $M$, with probability tending to
one,
\begin{align}
\max_{j,k}P_I X_{jt}^2X_{kt}^2
&\le C,
\label{eq:OA_empirical_coordinate_fourth}\\
P_I(X_t'v)^4
&\le C_M\|v\|_\Sigma^4,
\qquad v\in\mathcal V_M,
\label{eq:OA_empirical_directional_fourth}\\
P_I e_{0t}^2(X_t'v)^2
&\le C_M\|v\|_\Sigma^2,
\qquad v\in\mathcal V_M.
\label{eq:OA_empirical_weighted_second}
\end{align}
Here $P_I f_t:=|I|^{-1}\sum_{t\in I}f_t$.
These conclusions include the test interval.

Third, on the quantile truth cone,
\begin{align}
\max_j\|X_{jt}X_t'v\|_{L^q}
&\le C\|v\|_\Sigma,
\label{eq:OA_local_moment_1}\\
\max_j\|X_{jt}(X_t'v)^2\|_{L^{q/2}}
&\le C\|v\|_\Sigma^2,
\label{eq:OA_local_moment_2}\\
\E|X_t'v|^3
&\le C\|v\|_\Sigma^3.
\label{eq:OA_local_moment_3}
\end{align}
The same conclusions hold on the other truth cones.
\end{OAlemma}

\begin{proof}
Applying \eqref{eq:OA_directional_primitive} to coordinate
vectors and the sparse true coefficient vectors gives
uniformly bounded $4q$ moments of $X_{jt}$, $q_{0t}$,
$e_{0t}$, and $\rho_{0,r}'X_t$.
Hence $k_{0t}$ has uniformly bounded $2q$ moments,
including when $w_{0t}=e_{0t}^2$.

H\"older's inequality implies that every coordinate of
\[
X_tX_t',
\qquad
e_{0t}^2X_tX_t',
\qquad
X_t^{\otimes4}
\]
has a uniformly bounded $q$th moment.
The maximal partial-sum argument in
Lemma~\ref{lem:OA_uniform_FN} therefore gives
\[
\sup_I\|\widehat\Sigma_I-\Sigma\|_{\max}
=O_p(g_T)
\]
and
\[
\sup_I
\left\|
P_I e_{0t}^2X_tX_t'
-\E[e_{0t}^2X_tX_t']
\right\|_{\max}
=O_p(g_T).
\]
Applied to the $p_T^4$-coordinate fourth-order tensor,
the same argument gives
\[
\sup_I
\left\|
P_I X_t^{\otimes4}-\E X_t^{\otimes4}
\right\|_{\max}
=O_p(g_{4,T}).
\]
Whole vector blocks are coupled once. Thus the coupling
failure probability is $O(d_T\beta(a_T))$, without a
factor $p_T^4$. The coordinate union bound applies only
to the independent-copy tails.

Population curvature implies
\[
\|v\|_1
\le C_M\sqrt{s_\circ}\|v\|_\Sigma,
\qquad v\in\mathcal V_M.
\]
It follows that
\[
\left|
v'(\widehat\Sigma_I-\Sigma)v
\right|
\le
O_p(g_T)\|v\|_1^2
\le
O_p(s_\circ g_T)\|v\|_\Sigma^2
\]
uniformly over the relevant classes.
Because $s_\circ g_T\to0$, this establishes the sample
restricted and sparse eigenvalue bounds.

Similarly,
\[
\left|
P_I(X_t'v)^4-\E(X_t'v)^4
\right|
\le
O_p(g_{4,T})\|v\|_1^4
\le
O_p(s_\circ^2g_{4,T})\|v\|_\Sigma^4.
\]
The population fourth moment is bounded by
\eqref{eq:OA_directional_primitive}, and
$s_\circ^2g_{4,T}\to0$.
This proves
\eqref{eq:OA_empirical_directional_fourth}.
Repeated tensor indices give
\eqref{eq:OA_empirical_coordinate_fourth}.

For the weighted second moment,
\[
\E[e_{0t}^2(X_t'v)^2]
\le
\|e_{0t}\|_{L^4}^2\|X_t'v\|_{L^4}^2
\le C_M\|v\|_\Sigma^2.
\]
The empirical perturbation is bounded by
\[
O_p(g_T)\|v\|_1^2
\le O_p(s_\circ g_T)\|v\|_\Sigma^2.
\]
This proves \eqref{eq:OA_empirical_weighted_second}.

Finally, H\"older's inequality and
\eqref{eq:OA_directional_primitive} yield
\eqref{eq:OA_local_moment_1}--\eqref{eq:OA_local_moment_3}.
All interval bounds follow from the maximal partial-sum
bound, because an interval sum is the difference of two
partial sums. No additional union bound over fitted
models is needed.
\end{proof}
\paragraph{Reference rates.}
The following deterministic quantities summarize the
coefficient and prediction rates established below:
\[
\begin{aligned}
b_{\gamma,T}&:=s_\gamma\lambda_T,
&
a_{\gamma,T}&:=\sqrt{s_\gamma}\lambda_T,\\
b_{\beta,T}&:=(s_\beta+s_\gamma)\lambda_T,
&
a_{\beta,T}&:=\sqrt{s_\beta+s_\gamma}\lambda_T.
\end{aligned}
\]
For projection estimation, set
\[
a_{w,T}:=
\begin{cases}
0, & \text{for a fixed test function},\\
a_{\beta,T}, & \text{for }w_{0t}=e_{0t}^2,
\end{cases}
\]
and define
\[
a_{\rho,T}^2:=s_\rho\lambda_T^2+a_{w,T}^2,
\qquad
b_{\rho,T}:=s_\rho\lambda_T+a_{w,T}^2/\lambda_T.
\]
These rates apply columnwise when $d_w>1$.
Also put
\[
m_{\beta,T}:=C(s_\beta+s_\gamma),\qquad
b_{\circ,T}:=Cs_\circ\lambda_T,\qquad
\zeta_{\gamma,T}:=s_\gamma\lambda_T(\lambda_T+g_T).
\]
Here $m_{\beta,T}$ is a reference support-size bound,
$b_{\circ,T}$ bounds the coefficient-error rates across
stages, and $\zeta_{\gamma,T}$ controls the effect of the
generated pseudo-outcome on regression scores.

\paragraph{Deletion and separation rates.}
For $h<R/2$, define
\begin{equation}\label{eq:OA_Dcirc}
\begin{aligned}
D_{\circ,T}(h)
:={}&
s_\circ\Bigg[
\frac hR\Big\{
\overline\Lambda_h(p_T^2,q)+\lambda_T
+\big(1+\overline\Lambda_h(p_T^2,q)\big)b_{\circ,T}
\Big\}
+\zeta_{\gamma,T}
\Bigg]\\
&+s_\gamma r_T^{\mathrm{sc}}
+\mathbf1_{\mathrm{gen}}\,
s_\circ\frac hR b_{\circ,T}^2p_T^{3/(2q)},
\end{aligned}
\end{equation}
where $r_T^{\mathrm{sc}}$ is defined in
\eqref{eq:OA_quantile_score_remainder}, and
$\mathbf1_{\mathrm{gen}}$ equals one for the generated
test function and zero for a fixed test function.
This is the common reference bound used below for the
change in the fitted coefficients after deleting recent
training observations. It accommodates the generated
pseudo-outcome and, when used, the generated test function.

Let $K_{\Xi,T}\ge1$ bound the $L^q$ norms of the
maximum absolute coordinates of
\[
X_t\varepsilon_{t+1},\qquad
e_{0t}X_t\varepsilon_{t+1},\qquad
k_{0t}X_t',\qquad
k_{0t}A_{t+1}X_t'.
\]
With $M_t=\|X_t\|_\infty$, define
\begin{equation}\label{eq:OA_inference_envelope}
K_T
:=
1+K_{\Xi,T}+K_{\mathrm{pred},T}
+\||e_{0t}|M_t\|_{L^{2q}},
\qquad
C_{3,T}:=\E M_t^3.
\end{equation}
The primitive moment conditions give
$K_T=O(p_T^{1/q})$ and
$C_{3,T}=O(p_T^{3/(2q)})$.
The bounds below retain the actual envelopes.

For rolling and recursive estimation, require
$h\to\infty$, $h=o(R)$, and some
$0<\eta<1-1/q$ such that
\begin{equation}\label{eq:OA_delay_rates}
\begin{aligned}
p_T^2d_h\beta(a_h)&\to0,\\
K_Tb_{\circ,T}\sqrt h&\to0,\\
\sqrt P\,K_Tb_{\circ,T}\beta(h)^\eta&\to0,\\
\sqrt P\,K_TD_{\circ,T}(h)&\to0.
\end{aligned}
\end{equation}
The quantile score restrictions
\eqref{eq:OA_quantile_score_growth} also hold.

For fixed estimation, replace these delay and quantile
score restrictions by
\begin{equation}\label{eq:OA_fixed_rate}
K_Tb_{\circ,T}\sqrt P\,
\Lambda_P(d_wp_T,q)\to0.
\end{equation}
Both sets of scheme-specific restrictions imply
$b_{\circ,T}K_T\to0$.

For inference with the generated test function, also
require
\begin{equation}\label{eq:OA_cubic_rate}
\sqrt P\,b_{\circ,T}^3C_{3,T}\to0.
\end{equation}
This controls the cubic prediction-error term arising
from the squared generated instrument.
It is unnecessary for a fixed test function.
\subsection{Uniform concentration over forecasting windows}

The first lemma controls all sufficiently long training windows
simultaneously, together with averages over the shorter blocks
deleted inside the proof. The second lemma extends the uniform
control to local loss and pseudo-outcome increments.

\begin{OAlemma}[Partial sums, training windows, and deleted blocks]
\label{lem:OA_uniform_FN}
Let $\mathsf H_t\in\mathbb R^m$ be a mean-zero, strictly
stationary, beta-mixing array with mixing coefficients bounded
by $\beta(j)$ and
$\max_j\|\mathsf H_{t,j}\|_{L^q}\le C$, uniformly in $T$,
for some $q>4$. The dimension $m$ may grow with $T$.
Suppose $R\asymp T$, and let $\mathcal W_T$ be the collection
of consecutive intervals in $\{0,\ldots,T-1\}$ with length
at least $R/2$. If $d_T\beta(a_T)\to0$, then
\begin{align}
\max_{0\le k\le T}
\left\|
\frac1T\sum_{s=0}^{k-1}\mathsf H_s
\right\|_\infty
&=O_p\{\Lambda_T(m,q)\},
\label{eq:OA_seq_FN}\\
\max_{I\in\mathcal W_T}
\left\|
\frac1{|I|}\sum_{s\in I}\mathsf H_s
\right\|_\infty
&=O_p\{\Lambda_T(m,q)\}.
\label{eq:OA_window_FN}
\end{align}
Moreover, both quantities are bounded by a sufficiently
large fixed multiple of $\Lambda_T(m,q)$ with probability
tending to one.

For $\mathcal D_t(h)=\{t-h,\ldots,t-1\}$, suppose
$h\to\infty$, $h<R/2$, and $md_h\beta(a_h)\to0$. Then
\begin{equation}\label{eq:OA_moving_block}
\frac1P\sum_{t\in\mathcal T_{\mathrm{test}}}
\left\|
\frac1h\sum_{s\in\mathcal D_t(h)}\mathsf H_s
\right\|_\infty
=O_p\{\overline\Lambda_h(m,q)\}.
\end{equation}
If $\omega_t\ge0$ and
$\sup_t\|\omega_t\|_{L^q}\le K_{\omega,T}$, then
\[
\frac1P\sum_{t\in\mathcal T_{\mathrm{test}}}
\omega_t
\left\|
\frac1h\sum_{s\in\mathcal D_t(h)}\mathsf H_s
\right\|_\infty
=O_p\{K_{\omega,T}\overline\Lambda_h(m,q)\}.
\]
No independence between $\omega_t$ and the deleted block
is required.
\end{OAlemma}

\begin{proof}
For a segment of length $n$, define
\[
M_n:=
\max_{0\le k\le n}
\left\|
\frac1n\sum_{s=0}^{k-1}\mathsf H_s
\right\|_\infty.
\]
Partition the segment into alternating blocks of length
$a_n$, with a remainder of length less than $2a_n$.
Couple each family of vector-valued blocks to independent
copies, preserving the distribution within each block.
The probability that either coupling fails is at most
$C d_n\beta(a_n)$; no factor $m$ is needed because the
whole vector block is coupled at once.

For the independent copies, fix a coordinate and a position
within a block. The observations across blocks are independent
and centered. Etemadi's maximal inequality and the
Fuk--Nagaev bound used in Lemma~5 of
\citet{Barendse2023ESLasso} give
\[
\Pp\!\left(
\max_{r\le d_n}
\left|\sum_{b=1}^r\xi_b\right|>x
\right)
\le C\left\{\frac{d_n}{x^q}
+\exp(-c x^2/d_n)\right\}.
\]
A union bound over the $m$ coordinates and $a_n$ positions,
followed by normalization using $n\asymp a_nd_n$, yields
the corresponding bound for complete partial-block sums.

Incomplete blocks are controlled by
\[
\E\left(\sum_{s\in B}|\mathsf H_{s,j}|\right)^q
\le C a_n^q
\]
and a union bound over coordinates and blocks. Combining
these bounds and adding the coupling-failure probability
gives, for $u>0$,
\begin{equation}\label{eq:OA_maximal_tail}
\Pp(M_n>u)
\le
Cma_n\left\{
\frac{1}{u^qd_n^{q-1}}
+\exp(-cu^2d_n)
\right\}
+C d_n\beta(a_n).
\end{equation}
The probability inequalities are applied to the independent
copies before adding the coupling error; independence is
not asserted conditional on successful coupling.

At $u=C_0\Lambda_n(m,q)$, the polynomial term is at most
$C/(C_0^qma_n)$ and the exponential term is at most
$C(ma_n)^{1-cC_0^2}$. Thus a sufficiently large fixed $C_0$
makes both terms vanish. This proves
\eqref{eq:OA_seq_FN} and its probability-tending-to-one
version.

Every interval sum is the difference of two partial sums.
Consequently,
\[
\max_{I\in\mathcal W_T}
\left\|
\frac1{|I|}\sum_{s\in I}\mathsf H_s
\right\|_\infty
\le \frac{4T}{R}M_T,
\]
which proves \eqref{eq:OA_window_FN} without a further
union bound over training windows.

For the deleted-block result, put $r=q/(q-1)$ and let
$\mathcal E_n$ denote successful coupling. Integrating
the non-coupling terms in \eqref{eq:OA_maximal_tail} gives
\[
\|M_n\mathbf 1_{\mathcal E_n}\|_{L^r}
\le C\Lambda_n(m,q).
\]
Also, Minkowski's inequality gives
$\|M_n\|_{L^q}\le C m^{1/q}$. Hence
\[
\begin{aligned}
\|M_n\mathbf 1_{\mathcal E_n^c}\|_{L^r}
&\le
C m^{1/q}\{d_n\beta(a_n)\}^{1-2/q}\\
&\le
C\{md_n\beta(a_n)\}^{1-2/q},
\end{aligned}
\]
where the last inequality uses $q>4$ and $m\ge1$.
It follows that
\[
\|M_n\|_{L^r}\le C\overline\Lambda_n(m,q).
\]

By stationarity, the same bound applies to the average
over each deleted block of length $h$. Averaging expectations
over forecast origins and applying Markov's inequality proves
\eqref{eq:OA_moving_block}. Finally, H\"older's inequality gives
\[
\E\!\left[
\omega_t
\left\|
\frac1h\sum_{s\in\mathcal D_t(h)}\mathsf H_s
\right\|_\infty
\right]
\le
C K_{\omega,T}\overline\Lambda_h(m,q),
\]
which proves the weighted conclusion.
\end{proof}

For a noncentered array, the lemma applies after subtracting
its population mean. The mean contribution must be retained;
in particular, the lemma does not assert a vanishing bound
for uncentered absolute arrays.

\begin{OAlemma}[Uniform contraction over training windows]
\label{lem:OA_contraction}
Suppose Assumption~\ref{ass:OA_primitive_arrays} holds.
The empirical design bounds used below are established
in Lemma~\ref{lem:OA_primitive_design}. Let $\ell(y,v)$ be one-Lipschitz in $v$, and let
$B_T>0$ be deterministic with $B_T=o(1)$.
Define
\[
\begin{aligned}
f_{s,\delta}
&:=
\ell(Y_{s+1},q_{0s}+X_s'\delta)
-\ell(Y_{s+1},q_{0s}),\\
g_{s,\delta}
&:=
X_s\{U_{s+1}(q_{0s}+X_s'\delta)-U_{s+1}(q_{0s})\}.
\end{aligned}
\]
Then
\begin{align}
\max_{I\in\mathcal W_T}
\sup_{\|\delta\|_1\le B_T}
\left|
\frac1{|I|}\sum_{s\in I}
\{f_{s,\delta}-\E f_{s,\delta}\}
\right|
&=O_p(B_T\lambda_T),
\label{eq:OA_loss_contraction}\\
\max_{I\in\mathcal W_T}
\sup_{\|\delta\|_1\le B_T}
\left\|
\frac1{|I|}\sum_{s\in I}
\{g_{s,\delta}-\E g_{s,\delta}\}
\right\|_\infty
&=O_p(B_Tg_T).
\label{eq:OA_U_contraction}
\end{align}
Expectations are taken with $\delta$ fixed.
The conclusions remain valid when the suprema are
restricted further by $\|\delta\|_\Sigma\le r_T$.
\end{OAlemma}

\begin{proof}
Both classes have the form
\[
c_{j,s}\varphi_s(X_s'\delta),
\qquad
\varphi_s(0)=0,
\]
with a common Lipschitz constant $L<\infty$.
For the first class, there is one output coordinate and
$c_{1,s}=1$. For the second, $c_{j,s}=X_{js}$,
$j=1,\ldots,p_T$, and
\[
|U_{s+1}(q)-U_{s+1}(q')|
\le(1+\alpha^{-1})|q-q'|.
\]

It suffices to control maximal partial sums, since every
long-window sum is the difference of two partial sums.
Use the vector-block coupling from the preceding proof,
whose failure probability is $o(1)$. For one independent
block family, write
\[
F_b(j,\delta)
:=
\sum_{s\in B_b}c_{j,s}\varphi_s(X_s'\delta).
\]
Conditional on the blocks, let $\epsilon_b$ be independent
Rademacher variables. L\'evy's maximal inequality reduces
the maximum over partial-block sums to the complete sum:
\[
\begin{aligned}
&\E_\epsilon
\max_{r\le d_T}\max_j
\sup_{\|\delta\|_1\le B_T}
\left|\sum_{b=1}^r\epsilon_bF_b(j,\delta)\right|\\
&\qquad\le
2\E_\epsilon
\max_j\sup_{\|\delta\|_1\le B_T}
\left|\sum_{b=1}^{d_T}\epsilon_bF_b(j,\delta)\right|.
\end{aligned}
\]

For a fixed $j$,
\[
|F_b(j,\delta)-F_b(j,\delta')|
\le
L\sqrt{a_T}
\left\{
\sum_{s\in B_b}
c_{j,s}^2\bigl(X_s'(\delta-\delta')\bigr)^2
\right\}^{1/2}.
\]
Apply Theorem~3 of \citet{Maurer2016} to the union of
the coordinate-indexed classes. Disjoint copies of the
within-block prediction vector handle different output
coordinates and both signs of the absolute value.
This is valid because $\varphi_s(0)=0$: differences across
two copies are bounded by the sum of their norms, and hence
by $\sqrt2$ times the norm of their concatenation.

The contraction inequality, $\ell_1$--$\ell_\infty$
duality, and the Rademacher maximal bound therefore give
\[
\begin{aligned}
&\E_\epsilon
\max_j\sup_{\|\delta\|_1\le B_T}
\left|\sum_b\epsilon_bF_b(j,\delta)\right|\\
&\qquad\le
C B_T\sqrt{a_T\log(2Np_T)}
\max_{j,k}
\left\{
\sum_b\sum_{s\in B_b}c_{j,s}^2X_{ks}^2
\right\}^{1/2},
\end{aligned}
\]
where $N=1$ for the loss class and $N=p_T$ for the
pseudo-outcome class. The parameter suprema can be taken
over countable dense subsets because the increments are
continuous in $\delta$.

The sample moment conditions imply
\[
Q_{1,T}^2:=
\max_k\frac1T\sum_{s=0}^{T-1}X_{ks}^2=O_p(1),
\qquad
Q_{2,T}^2:=
\max_{j,k}\frac1T\sum_{s=0}^{T-1}
X_{js}^2X_{ks}^2=O_p(1).
\]
Indeed, by stationarity and $R\asymp T$, the full segment
can be covered by a fixed number of length-$R$ intervals.
The same bounds hold for the coupled blocks and independent
ghost copies, since the coupling-failure probability vanishes.

To pass from the conditional Rademacher bound to the
centered empirical process, use independent-block
symmetrization in probability. Its variance requirement
follows from
\[
\sup_{j,\|\delta\|_1\le B_T}
\sum_b\E[F_b(j,\delta)^2]
\le C T a_T B_T^2,
\]
using the population second-moment bounds for $X_{ks}$
and $X_{js}X_{ks}$. Conditional Markov's inequality and
$Q_{1,T},Q_{2,T}=O_p(1)$ then give normalized bounds
\[
O_p\!\left(
B_T\sqrt{\frac{a_T\log(2p_T)}{T}}
\right),
\qquad
O_p\!\left(
B_T\sqrt{\frac{a_T\log(2p_T^2)}{T}}
\right)
\]
for the two classes, uniformly over complete partial-block
sums.

An incomplete block contributes at most
$O_p(B_T\sqrt{a_T/T})+O(B_Ta_T/T)$ after normalization,
uniformly over its endpoints, by Cauchy--Schwarz and the
same moment bounds. Adding the two block families and
the vanishing coupling-failure probability therefore
establishes the maximal partial-sum bounds.

Finally, $T\asymp a_Td_T$ and $R\asymp T$ imply
\[
\sqrt{\frac{a_T\log(2p_T)}{T}}
\lesssim\lambda_T,
\qquad
\sqrt{\frac{a_T\log(2p_T^2)}{T}}
\lesssim g_T.
\]
The interval-to-partial-sum bound from the preceding
lemma proves \eqref{eq:OA_loss_contraction} and
\eqref{eq:OA_U_contraction}. Restricting the parameter
sets cannot increase either supremum.
\end{proof}

\subsection{Uniform nuisance rates}

Throughout this subsection, $\max_{t,I}$ denotes the maximum
over $t\in\mathcal T_{\mathrm{test}}$ and $I\in\mathfrak I_t$.
Write $\widehat\gamma_{t,I}$, $\widehat\beta_{t,I}$, and
$\widehat\rho_{t,I}$ for the estimates computed on $I$ using
the stage-specific penalties at origin $t$. In particular,
$\widehat\gamma_{t,\mathcal I_t}=\widehat\gamma_t$, and
similarly for the other stages. All coefficients are
estimated jointly within each stage, with penalties applied
only to the $V$-coordinates. Under fixed estimation, the
training sample and tuning parameters are held constant
across forecast origins.

\begin{OAlemma}[Quantile Lasso and generated pseudo-outcome]
\label{lem:OA_quantile_lasso}
Under the maintained primitive conditions, the unsmoothed
partially penalized quantile estimator satisfies
\begin{equation}\label{eq:OA_qr_rates}
\begin{aligned}
\max_{t,I}\|\widehat\gamma_{t,I}-\gamma_0\|_1
&=O_p(b_{\gamma,T}),\\
\max_{t,I}\|\widehat\gamma_{t,I}-\gamma_0\|_\Sigma
&=O_p(a_{\gamma,T}).
\end{aligned}
\end{equation}
For any two quantile estimates $\gamma_1,\gamma_2$ from
these fits, uniformly over their indices,
\begin{equation}\label{eq:OA_generated_score}
\max_{I\in\mathcal W_T}
\left\|
\frac1{|I|}\sum_{s\in I}X_s
\{U_{s+1}(X_s'\gamma_1)-U_{s+1}(X_s'\gamma_2)\}
\right\|_\infty
=O_p(\zeta_{\gamma,T}).
\end{equation}
Moreover,
\begin{equation}\label{eq:OA_pseudo_L2}
\max_{t,I}\frac1{|I|}\sum_{s\in I}
\{U_{s+1}(X_s'\widehat\gamma_{t,I})
-U_{s+1}(q_{0s})\}^2
=O_p(a_{\gamma,T}^2).
\end{equation}
\end{OAlemma}

\begin{proof}
Fix $(t,I)$ and write
$\delta=\widehat\gamma_{t,I}-\gamma_0$,
$S=S_\gamma$, and $\lambda=\lambda_{\gamma,t}$.
Knight's identity is
\[
\rho_\alpha(u-v)-\rho_\alpha(u)
=
-v\{\alpha-\mathbf 1\{u\le0\}\}
+\int_0^v
\{\mathbf 1\{u\le z\}-\mathbf 1\{u\le0\}\}\,dz.
\]
Let $Q_I(v)$ be the empirical average of the integral term
with $u=Y_{s+1}-q_{0s}$ and scalar displacement $X_s'v$.
For deterministic $v$, its expectation is
\[
Q(v)
=
\E\int_0^{X_s'v}
\{F(q_{0s}+u\mid X_s)-\alpha\}\,du.
\]
On the uniform density neighborhood, the density lower
bound and Lipschitz condition imply a quadratic lower
bound with a cubic remainder. Outside that neighborhood,
the integrand has the sign of $u$, so the integral is
nonnegative. Enlarging the cubic constant therefore gives
\[
Q(v)\ge
\frac{f_{\min}}2\|v\|_\Sigma^2-C\E|X_s'v|^3.
\]
By \eqref{eq:OA_local_moment_3}, for $v\in\mathcal C(S)$
and $\|v\|_\Sigma\le c_0$,
\begin{equation}\label{eq:OA_q_curvature}
Q(v)\ge c_1\|v\|_\Sigma^2
\end{equation}
for fixed positive constants $c_0,c_1$.

The conditional quantile restriction and the coordinate
moment bounds imply that
$X_s\{\mathbf1(Y_{s+1}\le q_{0s})-\alpha\}$
is centered and has uniformly bounded coordinate
$L^q$ norms. Lemma~\ref{lem:OA_uniform_FN}, together
with sufficiently large fixed penalty constants, gives
\[
\left\|
\frac1{|I|}\sum_{s\in I}
X_s\{\mathbf1(Y_{s+1}\le q_{0s})-\alpha\}
\right\|_\infty
\le \frac{\lambda_{\gamma,t}}4
\]
simultaneously over all fits, with probability tending
to one. On this event, the basic inequality for partial
penalization gives
\[
\begin{aligned}
Q_I(\delta)+\frac{3\lambda}{4}\|\delta_{S^c}\|_1
&\le
\frac{\lambda}{4}\|\delta_{\mathcal J_Z}\|_1
+\frac{5\lambda}{4}
\|\delta_{S\setminus\mathcal J_Z}\|_1\\
&\le \frac{5\lambda}{4}\|\delta_S\|_1.
\end{aligned}
\]
Since $Q_I(\delta)\ge0$, the error belongs to
$\mathcal C(S)$. The same inequality holds for any
displacement whose penalized objective does not exceed
its value at $\gamma_0$.

For deterministic $B=o(1)$, the contraction lemma and
the score bound imply
\[
\max_{t,I}\sup_{\|v\|_1\le B}
|Q_I(v)-Q(v)|=O_p(B\lambda_T).
\]
On $\mathcal C(S)$, the restricted-eigenvalue condition gives
$\|v\|_1\le C\sqrt{s_\gamma}\|v\|_\Sigma$.
At $\|v\|_\Sigma=M a_{\gamma,T}$, the population lower
bound is of order $M^2s_\gamma\lambda_T^2$, whereas the
empirical fluctuation and penalty terms are
$O_p(Ms_\gamma\lambda_T^2)$.

For any prescribed error probability, choose $M$ large
enough that the quadratic term dominates. If a minimizer
were outside this prediction ball, convexity would place
a point on its boundary with objective no larger than
at $\gamma_0$, a contradiction. This proves the uniform
prediction rate. The cone inequality gives the uniform
$\ell_1$ rate in \eqref{eq:OA_qr_rates}.

The concentration lemma applied to the Gram coordinates
also gives
\[
\max_{I\in\mathcal W_T}
\|\widehat\Sigma_I-\Sigma\|_{\max}=O_p(g_T).
\]
Consequently, uniformly over the fitted coefficient errors
and the training windows,
\[
\|\delta\|_I^2
\le
\|\delta\|_\Sigma^2+
\|\delta\|_1^2
\|\widehat\Sigma_I-\Sigma\|_{\max}
=O_p(a_{\gamma,T}^2),
\]
because $b_{\gamma,T}^2g_T=o(a_{\gamma,T}^2)$.

For deterministic $v$, direct integration gives
\begin{equation}\label{eq:OA_U_drift_identity}
\E[
U_{s+1}(q_{0s}+X_s'v)-U_{s+1}(q_{0s})
\mid X_s]
=
\int_0^{X_s'v}
\{1-\alpha^{-1}F(q_{0s}+u\mid X_s)\}\,du.
\end{equation}
Its absolute value is at most $C(X_s'v)^2$.
Within the density neighborhood this follows from the
density upper bound; outside it, the Lipschitz bound
for $U$ and the fixed neighborhood radius give the
same inequality after enlarging $C$.

For $v\in\mathcal C(S_\gamma)$,
\eqref{eq:OA_local_moment_2} therefore implies
\[
\left\|
\E\!\left[
X_s\{U_{s+1}(q_{0s}+X_s'v)-U_{s+1}(q_{0s})\}
\right]
\right\|_\infty
\le C\|v\|_\Sigma^2.
\]
The contraction lemma bounds the centered empirical term
by $O_p(b_{\gamma,T}g_T)$ uniformly on the coefficient-rate
balls. Applying this bound to each fitted estimate relative
to $\gamma_0$ and subtracting proves
\eqref{eq:OA_generated_score}, since
\[
a_{\gamma,T}^2+b_{\gamma,T}g_T
=\zeta_{\gamma,T}.
\]
Finally, the pathwise Lipschitz bound for $U$ and the
empirical prediction rate prove \eqref{eq:OA_pseudo_L2}.
\end{proof}

\begin{OAlemma}[Partially penalized least squares with a generated response]
\label{lem:OA_generated_lasso}
For each $(t,I)$, suppose the training response has the
representation
\[
\widehat U_{t,I,s}
=X_s'b_0+u_s+r_{t,I,s},
\qquad s\in I.
\]
Let $S=S(b_0)$, $s=|S|$, and define
\[
\widehat b_{t,I}
\in\arg\min_{b\in\mathbb R^{p_T}}
\left\{
\frac1{2|I|}\sum_{s\in I}
(\widehat U_{t,I,s}-X_s'b)^2
+\lambda_t\|b_V\|_1
\right\},
\]
where $\lambda_t\asymp\lambda_T$ uniformly in $t$.
Here $\lambda_t$ denotes the penalty for the quadratic
stage under consideration, not a common penalty across stages.

Suppose the oracle-score bound
\[
\left\|
\frac1{|I|}\sum_{s\in I}X_su_s
\right\|_\infty\le\lambda_t/4
\]
and the sample restricted-eigenvalue condition on
$\mathcal C(S)$ hold simultaneously over $(t,I)$ with
probability tending to one. If, for a deterministic
rate $a_{r,T}$,
\[
\max_{t,I}R_{t,I}^2=O_p(a_{r,T}^2),
\qquad
R_{t,I}^2:=\frac1{|I|}\sum_{s\in I}r_{t,I,s}^2,
\]
then
\begin{align}
\max_{t,I}\|\widehat b_{t,I}-b_0\|_I^2
&=O_p(s\lambda_T^2+a_{r,T}^2),
\label{eq:OA_quad_pred}\\
\max_{t,I}\|\widehat b_{t,I}-b_0\|_1
&=O_p(s\lambda_T+a_{r,T}^2/\lambda_T).
\label{eq:OA_quad_l1}
\end{align}
If the sample sparse upper-eigenvalue bounds hold at
every fixed multiple of
$s+a_{r,T}^2/\lambda_T^2$, with a common upper bound, then
\begin{equation}\label{eq:OA_quad_support}
\max_{t,I}|S(\widehat b_{t,I})|
=O_p(s+a_{r,T}^2/\lambda_T^2).
\end{equation}
\end{OAlemma}

\begin{proof}
Fix $(t,I)$ and write
$\Delta=\widehat b_{t,I}-b_0$,
$d=\|\Delta\|_I$, $R_I=R_{t,I}$, and $\lambda=\lambda_t$.
The basic inequality, partial penalization, and
Cauchy--Schwarz give
\begin{equation}\label{eq:OA_quad_basic}
\frac12d^2+\frac{3\lambda}{4}\|\Delta_{S^c}\|_1
\le
\frac{5\lambda}{4}\|\Delta_S\|_1+R_Id.
\end{equation}
If $\Delta\in\mathcal C(S)$, restricted curvature gives
$\|\Delta_S\|_1\le\sqrt{s}\,d/\underline\kappa$ and
$\|\Delta\|_1\le4\|\Delta_S\|_1$. Hence
\[
d\le C\sqrt{s}\lambda+2R_I,\qquad
\|\Delta\|_1
\le C(s\lambda+R_I^2/\lambda).
\]
If $\Delta\notin\mathcal C(S)$, then
$\|\Delta_{S^c}\|_1>3\|\Delta_S\|_1$, so
\eqref{eq:OA_quad_basic} implies
\[
\frac12d^2+\frac{\lambda}{4}\|\Delta\|_1\le R_Id.
\]
Thus $d\le2R_I$ and
$\|\Delta\|_1\le2R_I^2/\lambda$.
The uniform response-error bound and penalty comparability
prove \eqref{eq:OA_quad_pred}--\eqref{eq:OA_quad_l1}.

For the support bound, let
\[
\widehat A_{t,I}
:=\{j>d_Z:\widehat b_{t,I,j}\ne0\}.
\]
Only these penalized coordinates satisfy the nonzero-penalty
KKT equality. For any $A\subseteq\widehat A_{t,I}$ of size $m$,
that equality and the oracle-score bound imply
\[
\frac{3\lambda}{4}\sqrt m
\le
\left\|
\frac1{|I|}\mathbf X_{I,A}'\mathbf X_I\Delta
\right\|_2
+
\left\|
\frac1{|I|}\mathbf X_{I,A}'\mathbf r_{t,I}
\right\|_2
\le
\sqrt{\phi_{\max,I}(m)}(d+R_I),
\]
where $\mathbf X_I$ is the training design matrix,
$\mathbf r_{t,I}=(r_{t,I,s})_{s\in I}$, and
$\phi_{\max,I}(m)$ is the largest eigenvalue over all
$m$-column subdesigns.

Fix $\epsilon>0$ and choose $C_\epsilon<\infty$ such that
\[
\Pp\!\left(
\max_{t,I}R_{t,I}^2\le C_\epsilon a_{r,T}^2
\right)\ge1-\epsilon
\]
for all sufficiently large $T$. The prediction bound above
then gives
$d+R_I\le C(\sqrt{s}\lambda+R_I)$ simultaneously over
$(t,I)$. Choose the deterministic integer
\[
m_\epsilon
=
\left\lceil
C_0\left(s+C_\epsilon a_{r,T}^2/\lambda_T^2\right)
\right\rceil,
\]
with $C_0$ sufficiently large. On the corresponding sparse
upper-eigenvalue event, the KKT bound is incompatible with
$|\widehat A_{t,I}|\ge m_\epsilon$.
If $m_\epsilon>p_T-d_Z$, the support bound is immediate.
Otherwise it follows simultaneously for all fits with
probability at least $1-\epsilon-o(1)$.
Since $\epsilon$ is arbitrary and
$|S(\widehat b_{t,I})|=d_Z+|\widehat A_{t,I}|$ with
$d_Z\le s$, this proves \eqref{eq:OA_quad_support}.
\end{proof}

\paragraph{ES and projection rates.}
For the ES regression, take $b_0=\beta_0$,
$u_s=\varepsilon_{s+1}$, and
\[
r_{t,I,s}
=
U_{s+1}(X_s'\widehat\gamma_{t,I})-U_{s+1}(q_{0s}).
\]
By \eqref{eq:OA_pseudo_L2}, the response-error rate is
$a_{r,T}=a_{\gamma,T}$.
Lemma~\ref{lem:OA_generated_lasso} therefore gives the
empirical prediction and coefficient rates. The uniform
Gram bound yields
\[
\|\widehat\beta_{t,I}-\beta_0\|_\Sigma^2
\le
\|\widehat\beta_{t,I}-\beta_0\|_I^2
+
\|\widehat\beta_{t,I}-\beta_0\|_1^2
\|\widehat\Sigma_I-\Sigma\|_{\max}.
\]
Since $(s_\beta+s_\gamma)g_T\to0$, we obtain
\begin{equation}\label{eq:OA_ES_rates}
\begin{aligned}
\max_{t,I}\|\widehat\beta_{t,I}-\beta_0\|_1
&=O_p(b_{\beta,T}),\\
\max_{t,I}\|\widehat\beta_{t,I}-\beta_0\|_\Sigma
&=O_p(a_{\beta,T}),
\end{aligned}
\end{equation}
and
\begin{equation}\label{eq:OA_ES_support}
\max_{t,I}|S(\widehat\beta_{t,I})|
=O_p(m_{\beta,T}).
\end{equation}

For projection coordinate $r$, the oracle response is
\[
w_{0s,r}=X_s'\rho_{0,r}+k_{0s,r},
\]
and its oracle score is controlled by the projection
normal equations and the uniform score bound.
For a fixed test function the response has no generated
error, so $a_{w,T}=0$.

For the scalar generated test function, put
$v_{t,I,s}:=X_s'(\widehat\beta_{t,I}-\beta_0)$. Then
\[
\widehat w_{t,I,s}-w_{0s}
=2e_{0s}v_{t,I,s}+v_{t,I,s}^2.
\]
By \eqref{eq:OA_ES_support}, the ES coefficient errors
belong, with arbitrarily high probability, to a
deterministic class of vectors supported on at most
$Ms_\circ$ coordinates, for a sufficiently large fixed $M$.
The bounds in
Lemma~\ref{lem:OA_primitive_design} therefore imply
\[
\begin{aligned}
\max_{t,I}\frac1{|I|}
\sum_{s\in I}|\widehat w_{t,I,s}-w_{0s}|^2
&\le
C\max_{t,I}\frac1{|I|}
\sum_{s\in I}
\{e_{0s}^2v_{t,I,s}^2+v_{t,I,s}^4\}\\
&=O_p(a_{\beta,T}^2+a_{\beta,T}^4)
=O_p(a_{\beta,T}^2).
\end{aligned}
\]
Thus the deterministic reference rate is
$a_{w,T}=a_{\beta,T}$ in this case.
Applying Lemma~\ref{lem:OA_generated_lasso} columnwise,
followed by the same Gram comparison, gives
\[
\begin{aligned}
\max_{t,I}\max_{1\le r\le d_w}
\|\widehat\rho_{t,I,r}-\rho_{0,r}\|_1
&=O_p(b_{\rho,T}),\\
\max_{t,I}\max_{1\le r\le d_w}
\|\widehat\rho_{t,I,r}-\rho_{0,r}\|_\Sigma
&=O_p(a_{\rho,T}),
\end{aligned}
\]
where
\[
a_{\rho,T}^2=s_\rho\lambda_T^2+a_{w,T}^2,
\qquad
b_{\rho,T}=s_\rho\lambda_T+a_{w,T}^2/\lambda_T.
\]
The effective projection support sizes are
$O_p(s_\rho+a_{w,T}^2/\lambda_T^2)$ uniformly over fits.
All support orders in these applications are $O(s_\circ)$,
so the sparse-eigenvalue conditions stated above apply.

\paragraph{Prediction bounds at evaluation observations.}

The preceding Lasso basic inequalities establish uniform
training and population prediction rates. To control
prediction errors at evaluation observations, we combine
these rates with delayed coupling and weighted deletion.

Define
\[
M_t:=\|X_t\|_\infty,
\qquad
Q_t:=(1+|e_{0t}|)M_t^2,
\qquad
K_{\mathrm{pred},T}:=\|Q_t\|_{L^q}.
\]
Assumption~\ref{ass:OA_primitive_arrays} implies
\[
\|M_t\|_{L^{2q}}\le Cp_T^{1/(2q)},
\qquad
\||e_{0t}|M_t\|_{L^{2q}}\le Cp_T^{1/(2q)}.
\]
Indeed, the coordinate $4q$ moments and the $4q$ moment
of $e_{0t}$ give
$\sup_j\E|e_{0t}X_{jt}|^{2q}\le C$.
Consequently,
\begin{equation}
\label{eq:OA_prediction_envelope}
K_{\mathrm{pred},T}
\le
\|M_t\|_{L^{2q}}
\left(
\|M_t\|_{L^{2q}}+
\||e_{0t}|M_t\|_{L^{2q}}
\right)
\le Cp_T^{1/q}.
\end{equation}
We retain $K_{\mathrm{pred},T}$ in the bounds because
its actual order may be smaller.

\begin{OAlemma}[Evaluation prediction products]
\label{lem:OA_evaluation_products}
Let $u_t$ and $v_t$ be coefficient errors from the
quantile, ES, or projection estimators, treating projection
coordinates separately. Let $a_{u,T}$ and $a_{v,T}$ be
their population prediction rates established above.

For rolling or recursive estimation, let
$\widetilde u_t$ and $\widetilde v_t$ denote the
corresponding shortened-sample errors. Suppose the
uniform population prediction rates hold for both the
operational and shortened fits, and their uniform
$\ell_1$ rates are bounded by $b_{\circ,T}$.

For $\chi_t=1$ or $\chi_t=|e_{0t}|$, and any
$0<\eta<1-1/q$,
\begin{equation}
\label{eq:OA_delayed_prediction_products}
\frac1P\sum_{t\in\mathcal T_{\mathrm{test}}}
\chi_t
|X_t'\widetilde u_t|
|X_t'\widetilde v_t|
=
O_p\!\left(
a_{u,T}a_{v,T}
+
K_{\mathrm{pred},T}b_{\circ,T}^2\beta(h)^\eta
\right).
\end{equation}

Define the weighted deletion discrepancy
\[
\mathcal R_T(u,v)
:=
\frac1P\sum_{t\in\mathcal T_{\mathrm{test}}}
Q_t
\left(
\|u_t-\widetilde u_t\|_1+
\|v_t-\widetilde v_t\|_1
\right).
\]
If a deterministic sequence $D_T^{\mathrm{del}}(h)$
satisfies
\[
\mathcal R_T(u,v)
=
O_p\!\left(
K_{\mathrm{pred},T}D_T^{\mathrm{del}}(h)
\right),
\]
then
\begin{equation}
\label{eq:OA_evaluation_prediction_products}
\begin{aligned}
&\frac1P\sum_{t\in\mathcal T_{\mathrm{test}}}
\chi_t|X_t'u_t|\,|X_t'v_t|\\
&\quad=
O_p\!\left(
a_{u,T}a_{v,T}
+
K_{\mathrm{pred},T}b_{\circ,T}^2\beta(h)^\eta
+
K_{\mathrm{pred},T}b_{\circ,T}
D_T^{\mathrm{del}}(h)
\right).
\end{aligned}
\end{equation}
The required bound for $\mathcal R_T(u,v)$ is supplied
by the weighted deletion argument below.

For fixed estimation,
\begin{equation}
\label{eq:OA_fixed_prediction_products}
\frac1P\sum_{t\in\mathcal T_{\mathrm{test}}}
\chi_t|X_t'u_t|\,|X_t'v_t|
=
O_p(a_{u,T}a_{v,T}).
\end{equation}
\end{OAlemma}

\begin{proof}
We first establish a population bound for deterministic
directions. On the classes $\mathcal V_M$ in
Assumption~\ref{ass:OA_primitive_arrays},
H\"older's inequality gives
\[
\begin{aligned}
\E\!\left[
(1+|e_{0t}|)|X_t'u|\,|X_t'v|
\right]
&\le
\|X_t'u\|_{L^2}\|X_t'v\|_{L^2}\\
&\quad+
\|e_{0t}\|_{L^2}
\|X_t'u\|_{L^4}\|X_t'v\|_{L^4}\\
&\le C_M\|u\|_\Sigma\|v\|_\Sigma.
\end{aligned}
\]
The final inequality follows from
\eqref{eq:OA_directional_primitive}.

For quantile errors, the relevant deterministic class
is the truth cone. For ES and projection errors, the
support bounds established above place the errors in
a class with support size at most $Ms_\circ$, with
arbitrarily high probability for sufficiently large
fixed $M$.

Consider rolling or recursive estimation.
For a fixed localization constant, truncate each
shortened-sample error to zero outside its own
$\mathcal F_{t-h}$-measurable coefficient, prediction,
and class-membership event. Denote the truncated
errors by $\widetilde u_t^C,\widetilde v_t^C$.
They satisfy
\[
\|\widetilde u_t^C\|_\Sigma\le Ca_{u,T},
\qquad
\|\widetilde v_t^C\|_\Sigma\le Ca_{v,T},
\qquad
\|\widetilde u_t^C\|_1+
\|\widetilde v_t^C\|_1
\le Cb_{\circ,T}.
\]
These inequalities now hold deterministically.

Couple $X_t$ to a copy $X_t^*$ with the same marginal
distribution, independent of $\mathcal F_{t-h}$,
such that
\[
\Pp(X_t\ne X_t^*)\le C\beta(h).
\]
Let $e_{0t}^*=X_t^{*\prime}\beta_0$, and define
$\chi_t^*$ and $Q_t^*$ from $X_t^*$ in the same way
as $\chi_t$ and $Q_t$.

Conditional on the truncated coefficients, independence
and the population bound give
\[
\E\!\left[
\chi_t^*
|X_t^{*\prime}\widetilde u_t^C|
|X_t^{*\prime}\widetilde v_t^C|
\mid\mathcal F_{t-h}
\right]
\le Ca_{u,T}a_{v,T}.
\]
On a coupling mismatch, the absolute difference between
the original and copied products is bounded by
\[
Cb_{\circ,T}^2(Q_t+Q_t^*).
\]
H\"older's inequality therefore gives
\[
\E\!\left[
b_{\circ,T}^2(Q_t+Q_t^*)
\mathbf1\{X_t\ne X_t^*\}
\right]
\le
CK_{\mathrm{pred},T}b_{\circ,T}^2
\beta(h)^{1-1/q}.
\]
Averaging these expectation bounds and applying
Markov's inequality proves
\eqref{eq:OA_delayed_prediction_products}
for the truncated coefficients.
The uniform estimation rates allow the probability
that any truncation changes a coefficient to be made
arbitrarily small. This removes the truncations.
At no point do we condition on the simultaneous
estimation-rate event.

To restore the operational coefficients, use
\[
\bigl||a||b|-|c||d|\bigr|
\le |a-c||b|+|c||b-d|.
\]
On a simultaneous $\ell_1$ rate event, this gives
\[
\begin{aligned}
&\chi_t
\left|
|X_t'u_t|\,|X_t'v_t|
-
|X_t'\widetilde u_t|\,|X_t'\widetilde v_t|
\right|\\
&\qquad\le
Cb_{\circ,T}Q_t
\left(
\|u_t-\widetilde u_t\|_1+
\|v_t-\widetilde v_t\|_1
\right).
\end{aligned}
\]
Its average is bounded by
$Cb_{\circ,T}\mathcal R_T(u,v)$.
Combining this inequality with
\eqref{eq:OA_delayed_prediction_products}
and the stated weighted deletion bound proves
\eqref{eq:OA_evaluation_prediction_products}.

For fixed estimation, the coefficient errors are
constant throughout the test interval.
Lemma~\ref{lem:OA_primitive_design} applies uniformly
over the relevant deterministic classes on that
interval. Thus
\[
\begin{aligned}
P_{\mathrm{test}}|X_t'u|\,|X_t'v|
&\le
\|u\|_{\mathrm{test}}\|v\|_{\mathrm{test}}\\
&\le C_M\|u\|_\Sigma\|v\|_\Sigma,
\end{aligned}
\]
and
\[
\begin{aligned}
&P_{\mathrm{test}}
|e_{0t}|\,|X_t'u|\,|X_t'v|\\
&\quad\le
\left\{
P_{\mathrm{test}}e_{0t}^2(X_t'u)^2
\right\}^{1/2}
\|v\|_{\mathrm{test}}\\
&\quad\le C_M\|u\|_\Sigma\|v\|_\Sigma.
\end{aligned}
\]
Substituting the uniform population prediction rates
proves \eqref{eq:OA_fixed_prediction_products}.
Independence between the training and test samples
is not required.
\end{proof}

The leading term in
\eqref{eq:OA_evaluation_prediction_products}
is the product of the population prediction rates.
The maximum-coordinate envelope enters only through
the coupling and deletion errors.

\begin{OAlemma}[Predictable nonnegative multipliers]
\label{lem:OA_predictable_multiplier}
Let $H_t\ge0$ be $\mathcal F_t$-measurable and let
$Z_{t+1}\ge0$ satisfy
\[
\E[Z_{t+1}\mid\mathcal F_t]\le C.
\]
For a deterministic positive sequence $c_T$, if
\[
\sum_{t\in\mathcal T_{\mathrm{test}}}H_t=O_p(c_T),
\]
then
\[
\sum_{t\in\mathcal T_{\mathrm{test}}}
H_tZ_{t+1}=O_p(c_T).
\]
\end{OAlemma}

\begin{proof}
For fixed $A>0$, define
\[
C_t:=\sum_{\substack{s\in\mathcal T_{\mathrm{test}}\\s\le t}}H_s,
\qquad
H_t^*:=H_t\mathbf1\{C_t\le Ac_T\}.
\]
The variable $H_t^*$ is $\mathcal F_t$-measurable,
and $\sum_tH_t^*\le Ac_T$ deterministically.
Therefore
\[
\E\sum_tH_t^*Z_{t+1}
\le C\E\sum_tH_t^*
\le CAc_T.
\]
On the event $\{\sum_tH_t\le Ac_T\}$, all the
original and truncated weights agree. Hence, for
$B>0$,
\[
\Pp\!\left(\sum_tH_tZ_{t+1}>Bc_T\right)
\le
\Pp\!\left(\sum_tH_t>Ac_T\right)+\frac{CA}{B}.
\]
First choose $A$ using the assumed stochastic order,
and then choose $B$ sufficiently large.
This proves the assertion.
\end{proof}

The conditional innovation-moment bound allows
Lemma~\ref{lem:OA_predictable_multiplier} to be applied
with $Z_{t+1}=|\varepsilon_{t+1}|$ or
$Z_{t+1}=\varepsilon_{t+1}^2$.
In particular, a bound for the average squared ES
prediction error also controls its product with
$|\varepsilon_{t+1}|$, without a conditional
prediction-domination assumption.
\subsection{Deletion stability and delayed separation}

For recursive or rolling estimation, let
\[
\widetilde{\mathcal I}_t^{(h)}
=\mathcal I_t\setminus\{t-h,\ldots,t-1\}.
\]
The associated proof-only fits use the same deterministic penalties as
the operational fits and are $\mathcal F_{t-h}$-measurable.
\paragraph{Quantile score expansion and deletion.}

For the rolling and recursive schemes, choose a
proof-only block length $\ell_T\to\infty$, distinct
from the block length $a_T$ in
\eqref{eq:Lambda_main}. Define
\[
L_T:=\log(eTp_T),
\qquad
\tau_T:=\sqrt{\frac{\ell_TL_T}{T}},
\qquad
u_T:=T^{-3},
\]
and
\[
B_T^X:=(Tp_T)^{1/(2q)}\log(eT),
\qquad
t_T^X:=\{p_T(B_T^X)^{-2q}\}^{1-1/(2q)},
\qquad
\nu_T:=(s_\gamma+1)L_T.
\]
Using the reference prediction rate
$a_{\gamma,T}=\sqrt{s_\gamma}\lambda_T$, set
\begin{equation}
\label{eq:OA_quantile_score_remainder}
r_T^{\mathrm{sc}}
:=
a_{\gamma,T}^2
+
\sqrt{
\frac{\ell_T(a_{\gamma,T}+u_T)\nu_T}{T}
}
+
\frac{\ell_TB_T^X\nu_T}{T}
+t_T^X+u_T.
\end{equation}
Require
\begin{equation}
\label{eq:OA_quantile_score_growth}
\frac{T}{\ell_T}\beta(\ell_T)\to0,
\qquad
\frac{s_\gamma\tau_T}{\lambda_T}\to0,
\qquad
\frac{r_T^{\mathrm{sc}}}{\lambda_T}\to0.
\end{equation}
These restrictions involve only dimension, sparsity,
moments, mixing, and the deterministic penalty rate.
They are used to prove the score expansion below.

Write $P_I f_s:=|I|^{-1}\sum_{s\in I}f_s$, and define
\[
H_\gamma
:=
\E[f(q_{0t}\mid X_t)X_tX_t'],
\qquad
\psi_I
:=
P_I X_s\{\mathbf1(Y_{s+1}\le q_{0s})-\alpha\}.
\]
At $\gamma_0+v$, a quantile-loss subgradient has the
form
\[
G_I(v)=P_I X_s(\xi_s-\alpha),
\]
where
\[
\xi_s\in
\left[
\mathbf1\{Y_{s+1}<q_{0s}+X_s'v\},
\mathbf1\{Y_{s+1}\le q_{0s}+X_s'v\}
\right].
\]
This definition includes arbitrary admissible
subgradient values at ties.

\begin{OAlemma}[Quantile score expansion and deletion]
\label{lem:OA_quantile_deletion}
Under Assumption~\ref{ass:OA_primitive_arrays} and
\eqref{eq:OA_quantile_score_growth}, the unsmoothed
partially penalized quantile estimators satisfy
\[
\max_{t,I}|S(\widehat\gamma_{t,I})|
=O_p(s_\gamma).
\]
For every KKT-compatible choice of subgradients,
\begin{equation}
\label{eq:OA_quantile_score_expansion}
\max_{t,I}
\left\|
G_I(\widehat\gamma_{t,I}-\gamma_0)
-\psi_I
-H_\gamma(\widehat\gamma_{t,I}-\gamma_0)
\right\|_\infty
=O_p(r_T^{\mathrm{sc}}).
\end{equation}

Suppose also that
$p_Td_h\beta(a_h)\to0$.
For any nonnegative weights satisfying
$\sup_t\|\omega_t\|_{L^q}\le K_{\omega,T}$,
\begin{equation}
\label{eq:OA_quantile_weighted_deletion}
\frac1P\sum_{t\in\mathcal T_{\mathrm{test}}}
\omega_t
\|\widehat\gamma_t-\widetilde\gamma_t^{(h)}\|_1
=
O_p\!\left(
K_{\omega,T}D_{\gamma,T}(h)
\right),
\end{equation}
where
\begin{equation}
\label{eq:OA_quantile_deletion_rate}
D_{\gamma,T}(h)
:=
s_\gamma
\left[
\frac hR
\{\overline\Lambda_h(p_T,q)+\lambda_T\}
+r_T^{\mathrm{sc}}
\right].
\end{equation}
\end{OAlemma}

\begin{proof}
We first record the probability bounds used in the
score argument, then establish a coarse score
approximation, selected-support control, the refined
score expansion, and deletion stability.

\emph{Block probability bounds.}
Partition the observed segment into alternating blocks
of length $\ell_T$ and couple each family of whole
vector blocks to independent copies. The total failure
probability is
\[
O\!\left(\frac{T}{\ell_T}\beta(\ell_T)\right)=o(1).
\]

Consider functions
$c_{j,s}\varphi_s(X_s'v)$, where
$\varphi_s(0)=0$, $\varphi_s$ is $L$-Lipschitz, and
$|c_{j,s}|\le |X_{js}|$.
Uniformly over long intervals and $\|v\|_1\le B$,
the contraction argument in
Lemma~\ref{lem:OA_contraction}, applied with block
length $\ell_T$, gives
\begin{equation}
\label{eq:OA_score_contraction}
\sup_{I,j,\|v\|_1\le B}
\left|
(P_I-\E)c_j\varphi(X'v)
\right|
=O_p(LB\tau_T).
\end{equation}
Here and below, expectations are evaluated at fixed
$v$.

To verify the required uniformity, for an independent
block family put
\[
F_b(I,j,v)
:=
\sum_{s\in B_b\cap I}c_{j,s}\varphi_s(X_s'v).
\]
Its increments satisfy
\[
|F_b(I,j,v)-F_b(I,j,v')|
\le
L\sqrt{\ell_T}
\left\{
\sum_{s\in B_b\cap I}
c_{j,s}^2[X_s'(v-v')]^2
\right\}^{1/2}.
\]
Vector contraction, followed by
$\ell_1$--$\ell_\infty$ duality, therefore yields
\[
\begin{aligned}
&\E_\epsilon
\sup_{I,j,\|v\|_1\le B}
\left|\sum_b\epsilon_bF_b(I,j,v)\right|\\
&\quad\le
CLB\sqrt{\ell_TL_T}
\max_{j,k}
\left(\sum_s c_{j,s}^2X_{ks}^2\right)^{1/2}.
\end{aligned}
\]
Different intervals, output coordinates, and signs
can be embedded in disjoint coordinate copies.
There are at most $T^2$ intervals, so their contribution
is absorbed by $L_T$.
Lemma~\ref{lem:OA_primitive_design} bounds the final
empirical factor by $O_p(\sqrt T)$.
The same bound holds for independent ghost copies.
Moreover,
\[
\sup_{I,j,\|v\|_1\le B}
\sum_b\E F_b(I,j,v)^2
\le CT\ell_TL^2B^2.
\]
Independent-block symmetrization in probability and
conditional Markov's inequality now prove
\eqref{eq:OA_score_contraction}.

We also need a finite-class bound.
If a finite deterministic class $\mathcal H$ satisfies
$|f|\le M$ and $\E f^2\le V$, then
\begin{equation}
\label{eq:OA_score_Bernstein}
\sup_{I,f\in\mathcal H}|(P_I-\E)f|
=
O_p\!\left(
\sqrt{\frac{\ell_TVx}{T}}
+
\frac{\ell_TMx}{T}
\right),
\qquad
x\ge C\log(eT|\mathcal H|).
\end{equation}
Indeed, each centered block sum is bounded by
$2\ell_TM$, and its variance is at most
$C\ell_T^2V$. Bernstein's inequality and a union
bound over the class, both block families, and all
intervals give \eqref{eq:OA_score_Bernstein}.
Both probability bounds are applied to independent
copies before adding the coupling error.

\emph{A coarse score approximation.}
By Lemma~\ref{lem:OA_quantile_lasso}, localize the
quantile errors in
\[
v\in\mathcal C(S_\gamma),
\qquad
\|v\|_1\le B:=Cb_{\gamma,T},
\qquad
\|v\|_\Sigma\le r:=Ca_{\gamma,T},
\]
where $C$ is fixed and can be chosen arbitrarily large.
Define the clipped displacement
\[
d_v(X)
:=
\max\{-r_f/2,\min(X'v,r_f/2)\}.
\]
Clipping changes a coordinate of an empirical
subgradient by at most
\[
\begin{aligned}
P_I|X_j|\mathbf1\{|X'v|>r_f/2\}
&\le
C(P_I X_j^2)^{1/2}
\{P_I(X'v)^4\}^{1/2}\\
&\le Cr^2,
\end{aligned}
\]
uniformly on this class, by
Lemma~\ref{lem:OA_primitive_design}.
The population counterpart has the same bound.
The local density Lipschitz condition consequently gives
\begin{equation}
\label{eq:OA_quantile_population_expansion}
\left\|
\E X\{
\mathbf1(Y\le q_0+d_v(X))
-\mathbf1(Y\le q_0)
\}
-H_\gamma v
\right\|_\infty
\le Cr^2.
\end{equation}

Let
\[
R_0(z):=\max\{0,\min(z,1)\}.
\]
For $\epsilon>0$, the functions
\[
R_0((z-Y)/\epsilon),
\qquad
R_0((z-Y+\epsilon)/\epsilon)
\]
are lower and upper bounds for every admissible
indicator subgradient at $z$.
After subtracting their values at $v=0$, with
$z=q_0+d_v(X)$, these functions vanish at zero and
are $1/\epsilon$-Lipschitz in $X'v$.
Equation~\eqref{eq:OA_score_contraction} therefore
bounds their centered increments by
$O_p(B\tau_T/\epsilon)$.
For signed coordinate weights, apply the bounds to
their positive and negative parts.

The baseline bands at $v=0$ also require control.
Uniform coordinate $2q$ moments imply
\[
\Pp\!\left(\max_{t,j}|X_{jt}|>B_T^X\right)
\le CTp_T(B_T^X)^{-2q}=o(1)
\]
and
\begin{equation}
\label{eq:OA_quantile_truncation_tail}
\max_j
\E\!\left[
|X_j|\mathbf1\{\|X\|_\infty>B_T^X\}
\right]
\le Ct_T^X.
\end{equation}
After truncation at $\|X\|_\infty\le B_T^X$,
the baseline band
$|X_j|\mathbf1\{|Y-q_0|\le\epsilon\}$
has mean $O(\epsilon)$, second moment
$O(\epsilon)$, and envelope $B_T^X$.
Equation~\eqref{eq:OA_score_Bernstein} bounds its
empirical average uniformly by
\[
O_p\!\left(
\epsilon+\tau_T\sqrt\epsilon
+\frac{\ell_TB_T^XL_T}{T}+t_T^X
\right).
\]
The population gaps between the ramps and the
indicator are also $O(\epsilon)$, by the local
density upper bound.

Combining these bounds, the clipping error, and
\eqref{eq:OA_quantile_population_expansion}, gives
\[
\begin{aligned}
&\max_{t,I}
\|G_I(\widehat\gamma_{t,I}-\gamma_0)
-\psi_I-H_\gamma(\widehat\gamma_{t,I}-\gamma_0)\|_\infty\\
&\quad=
O_p\!\left(
r^2+\epsilon+\frac{B\tau_T}{\epsilon}
+\tau_T\sqrt\epsilon
+\frac{\ell_TB_T^XL_T}{T}+t_T^X
\right).
\end{aligned}
\]
Choose $\epsilon=\sqrt{B\tau_T}$.
Since $s_\gamma\tau_T/\lambda_T\to0$,
\[
\frac{\epsilon}{\lambda_T}
+
\frac{B\tau_T}{\epsilon\lambda_T}
\to0.
\]
Also $r^2/\lambda_T=O(s_\gamma\lambda_T)\to0$,
and the remaining terms are $o(\lambda_T)$ by
\eqref{eq:OA_quantile_score_growth}.
Thus the score remainder is uniformly
$o_p(\lambda_T)$.

\emph{Selected-support control.}
Let $v=\widehat\gamma_{t,I}-\gamma_0$, and let $A$
be any subset of $m$ selected penalized coordinates.
The KKT equations, the oracle score bound, and the
coarse score approximation imply
\[
\sqrt m
\{\lambda_{\gamma,t}
-\|\psi_I\|_\infty-o_p(\lambda_T)\}
\le \|(H_\gamma v)_A\|_2.
\]
Since $H_\gamma$ is positive semidefinite,
\[
\begin{aligned}
\|(H_\gamma v)_A\|_2
&\le
\sqrt{\lambda_{\max}(H_{\gamma,AA})}
\sqrt{v'H_\gamma v}\\
&\le C\sqrt{s_\gamma}\lambda_T.
\end{aligned}
\]
The last inequality uses the density upper bound,
the population sparse upper eigenvalue bound, and
the uniform prediction rate.

After fixing the localization constant, choose
$m=\lceil Ms_\gamma\rceil$ with sufficiently large
fixed $M$. The preceding inequalities are then
incompatible with the existence of $m$ selected
penalized coordinates. If $m$ exceeds their ambient
dimension, the support bound is immediate.
Adding the fixed unpenalized block proves
\[
\max_{t,I}|S(\widehat\gamma_{t,I})|=O_p(s_\gamma).
\]

\emph{The refined score expansion.}
The preceding support bound allows localization in
the deterministic class
\[
\mathcal D_T
=
\left\{
v\in\mathcal C(S_\gamma):
\|v\|_0\le m,\ 
\|v\|_1\le B,\ 
\|v\|_\Sigma\le r
\right\},
\]
where $m=Cs_\gamma$, $B=Cb_{\gamma,T}$, and
$r=Ca_{\gamma,T}$.

Consider
\[
f_{v,j}
:=
X_j\mathbf1\{\|X\|_\infty\le B_T^X\}
\{
\mathbf1(Y\le q_0+d_v(X))
-\mathbf1(Y\le q_0)
\}.
\]
Its envelope is $B_T^X$, and the local density
upper bound gives
\[
\begin{aligned}
\E f_{v,j}^2
&\le C\E[X_j^2|X'v|]\\
&\le
C\|X_j\|_{L^2}\|X_jX'v\|_{L^2}
\le Cr.
\end{aligned}
\]

On each support, choose an $\ell_1$ net with centers
in $\mathcal D_T$ and mesh $u_T/B_T^X$.
Its total cardinality is at most
\[
\sum_{k=0}^{\min(m,p_T)}
\binom{p_T}{k}
\left(1+\frac{CBB_T^X}{u_T}\right)^k.
\]
The logarithm of this quantity is $O(\nu_T)$.
On the truncated set, replacing a vector by its
net center changes its clipped threshold by at
most $u_T$.
Indicators at the center shifted by $u_T$ in
either direction therefore bracket every admissible
subgradient, including values at ties.
The associated $|X_j|$-weighted bands have mean
$O(u_T)$, second moment $O(u_T)$, and envelope
$B_T^X$.

Apply \eqref{eq:OA_score_Bernstein} to the net
centers and their bracketing bands, all coordinates,
and all long intervals. This gives
\[
\sup_{I,j,v\in\mathcal D_T}
|(P_I-\E)f_{v,j}|
=
O_p\!\left(
\sqrt{\frac{\ell_T(r+u_T)\nu_T}{T}}
+
\frac{\ell_TB_T^X\nu_T}{T}
+u_T
\right).
\]
The empirical truncation error is zero with
probability tending to one.
The population truncation error is $O(t_T^X)$ by
\eqref{eq:OA_quantile_truncation_tail}.
Unclipping and the population Taylor expansion
each contribute $O(r^2)$.
Combining these bounds and removing localization
proves \eqref{eq:OA_quantile_score_expansion}.

\emph{Deletion stability.}
Fix an origin and write
\[
I=\mathcal I_t,
\qquad
J=\widetilde{\mathcal I}_t^{(h)},
\qquad
D=I\setminus J,
\qquad
n=|I|.
\]
Let
$\Delta=\widehat\gamma_t-\widetilde\gamma_t^{(h)}$.
The union of the two effective fitted supports has
size $O_p(s_\gamma)$.
The population sparse lower eigenvalue bound and
the density lower bound imply
\[
\Delta'H_\gamma\Delta\ge c\|\Delta\|_2^2
\]
on this union.

Subtract the two KKT equations, using the same
penalty in the original and shortened fits.
Monotonicity of the penalty subgradient and
\eqref{eq:OA_quantile_score_expansion} give
\[
c\|\Delta\|_2^2
\le
\|\Delta\|_1
\{\|\psi_I-\psi_J\|_\infty+2\mathcal E_T\},
\qquad
\mathcal E_T=O_p(r_T^{\mathrm{sc}})
\]
uniformly over origins.
Since $\|\Delta\|_1\le C\sqrt{s_\gamma}\|\Delta\|_2$,
\[
\|\Delta\|_1
\le
Cs_\gamma
\{\|\psi_I-\psi_J\|_\infty+2\mathcal E_T\}.
\]
The normalization identity is
\[
\psi_I-\psi_J
=
\frac hn(\psi_D-\psi_J).
\]
The long-window oracle score bound gives
$\max_{t}\|\psi_J\|_\infty=O_p(\lambda_T)$.
The weighted moving-block part of
Lemma~\ref{lem:OA_uniform_FN} gives
\[
\frac1P\sum_t\omega_t\|\psi_D\|_\infty
=
O_p\!\left(
K_{\omega,T}\overline\Lambda_h(p_T,q)
\right).
\]
Finally,
$P^{-1}\sum_t\omega_t=O_p(K_{\omega,T})$
by Markov's inequality.
Multiply the coefficient bound by $\omega_t$,
average, and use $n\ge R$.
This proves
\eqref{eq:OA_quantile_weighted_deletion}.
\end{proof}
\begin{OAlemma}[KKT perturbation]
\label{lem:OA_KKT}
For $a=1,2$, let
\(Q_a(b)=L_a(b)+\lambda\|b_V\|_1\), where $L_a$ is convex and
differentiable, and let $\widehat b_a$ minimize $Q_a$.  If the union
of their effective supports (including all unpenalized coordinates)
has size at most $m$, and $L_1$ has curvature at
least $\underline\kappa_L$ on that union, then, with
\(g=\|\nabla L_1(\widehat b_2)-\nabla L_2(\widehat b_2)\|_\infty\),
\[
\|\widehat b_1-\widehat b_2\|_2
\le\underline\kappa_L^{-1}\sqrt m\,g,
\qquad
\|\widehat b_1-\widehat b_2\|_1
\le\underline\kappa_L^{-1}m g.
\]
\end{OAlemma}

\begin{proof}
Choose $z_a$ with zero unpenalized coordinates and
penalized coordinates in $\partial\|\widehat b_{a,V}\|_1$, satisfying
\(\nabla L_a(\widehat b_a)+\lambda z_a=0\), and put
\(\Delta=\widehat b_1-\widehat b_2\).  Monotonicity of the subgradient
gives $\Delta'(z_1-z_2)\ge0$.  Hence
\[
\underline\kappa_L\|\Delta\|_2^2
\le\Delta'\{\nabla L_2(\widehat b_2)-\nabla L_1(\widehat b_2)\}
\le\sqrt m\|\Delta\|_2g.
\]
Division and $\|\Delta\|_1\le\sqrt m\|\Delta\|_2$ prove the claims.
\end{proof}

\begin{OAlemma}[Joint weighted deletion stability]
\label{lem:OA_deletion}
Under the conditions of the uniform estimation lemmas
and Lemma~\ref{lem:OA_quantile_deletion}, suppose
$p_T^2d_h\beta(a_h)\to0$.
For $j\in\{\gamma,\beta,\rho\}$ and any nonnegative
weights satisfying
$\sup_t\|\omega_t\|_{L^q}\le K_{\omega,T}$,
\begin{equation}\label{eq:OA_weighted_stability}
\frac1P\sum_t\omega_t
\|\widehat\eta_{j,t}-\widetilde\eta_{j,t}^{(h)}\|_1
=
O_p\!\left(K_{\omega,T}D_{\circ,T}(h)\right).
\end{equation}
Projection bounds hold columnwise, uniformly over the
fixed number of columns. In particular,
\begin{equation}\label{eq:OA_stability}
\frac1P\sum_t
\|\widehat\eta_{j,t}-\widetilde\eta_{j,t}^{(h)}\|_1
=O_p\!\left(D_{\circ,T}(h)\right).
\end{equation}
\end{OAlemma}

\begin{proof}
The quantile conclusion follows from
Lemma~\ref{lem:OA_quantile_deletion}, because
$D_{\gamma,T}(h)\le C D_{\circ,T}(h)$.
We prove the two quadratic-regression bounds.

Fix an origin, write
$I=\mathcal I_t$, $J=\widetilde{\mathcal I}_t^{(h)}$,
$D=I\setminus J$, and $n=|I|$.
Suppress the origin index on fitted coefficients.
All inequalities below hold on simultaneous estimation,
support, and design events whose probability can be
made arbitrarily close to one.

For
$L_A(b;V)=P_A(V-X'b)^2/2$, direct algebra gives
\begin{align}
&\nabla L_I(\widetilde b;\widehat V)
-\nabla L_J(\widetilde b;\widetilde V)
\nonumber\\
&\quad=
-\frac hn P_DX(\widehat V-X'\widetilde b)
-\frac{|J|}{n}P_JX(\widehat V-\widetilde V)
-\frac hn\nabla L_J(\widetilde b;\widetilde V).
\label{eq:OA_gradient_identity}
\end{align}
The last term has maximum norm at most
$(h/n)\lambda_{j,t}$ by the shortened-fit KKT equations.

\emph{ES regression.}
Let $Z_T=O_p(\zeta_{\gamma,T})$ denote the uniform
bound in \eqref{eq:OA_generated_score}.
Put $b=b_{\circ,T}$ and define
\[
B_t:=
\frac hR\left\{
\|P_DX\varepsilon\|_\infty
+b\|P_D|XX'|\|_{\max}
+\lambda_T
\right\}+Z_T.
\]
The identity
$U_{s+1}(q_{0s})=X_s'\beta_0+\varepsilon_{s+1}$,
the Lipschitz property of $U$, and
\eqref{eq:OA_gradient_identity} bound the ES gradient
discrepancy by $CB_t$.
The union of the fitted ES supports has size
$O_p(s_\circ)$.
Lemma~\ref{lem:OA_KKT} therefore gives
\begin{equation}\label{eq:OA_ES_deletion_pointwise}
\|\widehat\beta-\widetilde\beta\|_2
\le C\sqrt{s_\circ}B_t,
\qquad
\|\widehat\beta-\widetilde\beta\|_1
\le Cs_\circ B_t.
\end{equation}

The weighted moving-block bound in
Lemma~\ref{lem:OA_uniform_FN}, retaining the nonzero
mean of $|XX'|$, implies
\[
\begin{aligned}
\frac1P\sum_t\omega_t B_t
=O_p\Bigg(
K_{\omega,T}\Bigg[
&\frac hR\Big\{
\overline\Lambda_h(p_T^2,q)+\lambda_T\\
&\quad+
\big(1+\overline\Lambda_h(p_T^2,q)\big)b
\Big\}
+\zeta_{\gamma,T}
\Bigg]\Bigg).
\end{aligned}
\]
For the $Z_T$ term, use
$P^{-1}\sum_t\omega_t=O_p(K_{\omega,T})$.
This proves the ES assertion.

\emph{Projection regression.}
Treat each projection column separately.
Let $\Delta_\beta=\widehat\beta-\widetilde\beta$
and $\Delta_\rho=\widehat\rho-\widetilde\rho$.
The deleted-block and normalization contributions
to the projection gradient are bounded by $CC_t$, where
\[
\begin{aligned}
C_t:=\frac hR\Big\{
&\|P_DXk_0'\|_{\max}
+\lambda_T+b\|P_D|XX'|\|_{\max}\\
&+\mathbf1_{\mathrm{gen}}
\big[
b\|P_D|e_0XX'|\|_{\max}
+b^2P_DM_s^3
\big]
\Big\},
\qquad M_s=\|X_s\|_\infty.
\end{aligned}
\]
For the generated instrument, this follows from
\[
\widehat w-w_0
=
2e_0X'(\widehat\beta-\beta_0)
+\{X'(\widehat\beta-\beta_0)\}^2.
\]

On the common observations,
\[
\widehat w-\widetilde w
=
2e_0X'\Delta_\beta
+(X'\Delta_\beta)
X'(\widehat\beta+\widetilde\beta-2\beta_0).
\]
The weighted second-moment and fourth-moment bounds
in Lemma~\ref{lem:OA_primitive_design}, followed by
Cauchy--Schwarz, give
\[
\begin{aligned}
\|\widehat w-\widetilde w\|_J
&\le
C(1+a_{\beta,T})\|\Delta_\beta\|_\Sigma\\
&\le C\|\Delta_\beta\|_2.
\end{aligned}
\]
The last inequality uses population sparse upper
curvature on the union support.
For a fixed instrument, the common-observation
response difference is zero.

Subtract the projection KKT equations and take the
inner product with $\Delta_\rho$.
Sample sparse lower curvature on $I$ and sparse
upper curvature on $J$ imply
\[
c\|\Delta_\rho\|_2^2
\le
C\sqrt{s_\circ}\|\Delta_\rho\|_2 C_t
+
C\|\Delta_\rho\|_2
\|\widehat w-\widetilde w\|_J.
\]
Combining this inequality with
\eqref{eq:OA_ES_deletion_pointwise} yields
\[
\|\Delta_\rho\|_1
\le Cs_\circ(C_t+B_t).
\]

All coordinates of the score and design arrays in
$C_t$ have bounded $L^q$ norms under
Assumption~\ref{ass:OA_primitive_arrays}.
Their weighted block averages are therefore controlled
by Lemma~\ref{lem:OA_uniform_FN}, after centering and
retaining their means.

For the remaining cubic term, H\"older's inequality
and the coordinate $2q$ moments give, uniformly in
$s,t$,
\[
\E[\omega_tM_s^3]
\le
K_{\omega,T}
\|M_s^3\|_{L^{q/(q-1)}}
\le CK_{\omega,T}p_T^{3/(2q)}.
\]
Consequently, Markov's inequality gives
\[
\frac1P\sum_t
\omega_t\frac hR b^2P_DM_s^3
=
O_p\!\left(
K_{\omega,T}\frac hR b^2p_T^{3/(2q)}
\right).
\]
Averaging the coefficient bound therefore produces
exactly the additional generated-instrument term
in \eqref{eq:OA_Dcirc}.
This proves the projection assertion.
Removing localization completes the proof.
\end{proof}

In Lemma~\ref{lem:OA_evaluation_products}, take
$\omega_t=Q_t$.
Equation~\eqref{eq:OA_weighted_stability} then gives
\[
\mathcal R_T(u,v)
=
O_p\!\left(
K_{\mathrm{pred},T}D_{\circ,T}(h)
\right).
\]
Thus the evaluation prediction bound holds with
$D_T^{\mathrm{del}}(h)=D_{\circ,T}(h)$.
\begin{OAlemma}[Delayed separation]
\label{lem:OA_separation}
Let $\mathsf V_t$ be a centered stationary
$d\times p_T$ array, with fixed $d$, measurable with
respect to $\sigma(X_t,Y_{t+1})$, and satisfying
$\|\|\mathsf V_t\|_{\max}\|_{L^q}\le K_T$.
Let $\widetilde\delta_t$ be
$\mathcal F_{t-h}$-measurable, with
$\max_t\|\widetilde\delta_t\|_1=O_p(b_T)$.
If
\[
\frac1P\sum_t\|\mathsf V_t\|_{\max}
\|\widehat\delta_t-\widetilde\delta_t\|_1
=O_p(c_T),
\]
then, for $0<\eta<1-1/q$,
\begin{equation}\label{eq:OA_separation}
\left\|
P^{-1/2}\sum_t\mathsf V_t\widehat\delta_t
\right\|
=
O_p\!\left(
K_Tb_T\sqrt h+
\sqrt P\,K_Tb_T\beta(h)^\eta+
\sqrt P\,c_T
\right).
\end{equation}
\end{OAlemma}

\begin{proof}
Restoring the operational coefficients costs
$O_p(\sqrt P\,c_T)$.
For fixed $C$, truncate each delayed coefficient
separately:
\[
\delta_t^C
:=
\widetilde\delta_t
\mathbf1\{\|\widetilde\delta_t\|_1\le Cb_T\}.
\]
This truncation is $\mathcal F_{t-h}$-measurable.
Decompose
\[
\mathsf V_t\delta_t^C
=
\{\mathsf V_t-\E[\mathsf V_t\mid\mathcal F_{t-h}]\}
\delta_t^C
+
\E[\mathsf V_t\mid\mathcal F_{t-h}]\delta_t^C.
\]
Split the first sum into $h+1$ residue classes.
Successive terms within each class are martingale
differences: the preceding term is measurable with
respect to $\mathcal F_{t-h}$.
Each term has second moment bounded by $CK_T^2b_T^2$.
Orthogonality within classes and Cauchy--Schwarz across
classes therefore give $O_p(K_Tb_T\sqrt h)$ after
normalization by $\sqrt P$.

For the second sum, coupling to an independent-of-the-past
copy and applying H\"older's inequality give
\[
\E\left\|
\E[\mathsf V_t\mid\mathcal F_{t-h}]
\right\|_{\max}
\le CK_T\beta(h)^{1-1/q}.
\]
Since $\|\delta_t^C\|_1\le Cb_T$, Markov's inequality
gives the second term in \eqref{eq:OA_separation}.
The uniform coefficient rate allows the probability
that any truncation changes a coefficient to be made
arbitrarily small. Removing the truncations proves
the result.
\end{proof}

\begin{OAlemma}[Pseudo-outcome linearization]
\label{lem:OA_dynamic_pseudo}
Under the maintained conditions and the
scheme-specific restrictions,
\begin{equation}\label{eq:OA_dynamic_pseudo}
\begin{aligned}
P^{-1/2}\sum_t k_{0t}G_{t+1}
={}&
P^{-1/2}\sum_t
k_{0t}A_{t+1}X_t'(\widehat\gamma_t-\gamma_0)\\
&+O_p(\sqrt P\,a_{\gamma,T}^2)+o_p(1).
\end{aligned}
\end{equation}
\end{OAlemma}

\begin{proof}
For every scalar $v$,
\[
U_{t+1}(q_{0t}+v)-U_{t+1}(q_{0t})
=A_{t+1}v+R_{t+1}(v),
\]
where
\[
|R_{t+1}(v)|
\le
\alpha^{-1}|v|
\mathbf1\{|Y_{t+1}-q_{0t}|\le|v|\}.
\]
The remainder is globally Lipschitz in $v$.
For a deterministic direction $u$ in the quantile cone,
the local density bound, followed by Cauchy--Schwarz,
gives
\[
\begin{aligned}
\E[\|k_{0t}\||R_{t+1}(X_t'u)|]
&\le C\E[\|k_{0t}\|(X_t'u)^2]\\
&\le
C\|k_{0t}\|_{L^2}\|X_t'u\|_{L^4}^2\\
&\le C\|u\|_\Sigma^2.
\end{aligned}
\]
For displacements outside the density neighborhood,
the first inequality follows from
$|v|\le C v^2$ when $|v|>r_f$.

For rolling and recursive estimation, weighted deletion
replaces the operational coefficient by its delayed
version at cost
\[
O_p\!\left(\sqrt P\,K_TD_{\circ,T}(h)\right)=o_p(1).
\]
Truncate each delayed coefficient separately on its
own coefficient and prediction-rate event.
On a copy of $(X_t,Y_{t+1})$ independent of the
delayed information, the preceding population bound
gives expectation $O(a_{\gamma,T}^2)$.
On a coupling mismatch, the Lipschitz envelope has
$L^q$ norm at most $CK_Tb_{\gamma,T}$.
H\"older and Markov therefore give the normalized bound
\[
O_p\!\left(
\sqrt P\,a_{\gamma,T}^2+
\sqrt P\,K_Tb_{\gamma,T}\beta(h)^\eta
\right).
\]
The second term vanishes by
\eqref{eq:OA_delay_rates}. Remove the truncations.

For fixed estimation, apply the contraction argument
of Lemma~\ref{lem:OA_contraction} on the test interval
to
$\|k_{0t}\||R_{t+1}(X_t'u)|$.
This class vanishes at $u=0$ and is uniformly
Lipschitz in $X_t'u$. Its empirical design factor obeys
\[
\max_jP_{\mathrm{test}}\|k_{0t}\|^2X_{jt}^2
\le
P_{\mathrm{test}}\max_j\|k_{0t}\|^2X_{jt}^2
=O_p(K_T^2),
\]
by Markov's inequality.
The population factor has the same bound.
Thus its empirical fluctuation is
$O_p(K_Tb_{\gamma,T}\Lambda_P(d_wp_T,q))$.
Together with the population bound and
\eqref{eq:OA_fixed_rate}, this proves the fixed case.
\end{proof}

\subsection{Proof of Theorem~\ref{thm:primitive_lasso}}

\begin{proof}
The uniform estimation assertions follow from
Lemmas~\ref{lem:OA_quantile_lasso},
\ref{lem:OA_generated_lasso}, and
\ref{lem:OA_primitive_design}, and the ES and
projection applications given above.

Write
\[
\begin{gathered}
V_{\gamma t}:=X_t'(\widehat\gamma_t-\gamma_0),
\qquad
V_{\beta t}:=X_t'(\widehat\beta_t-\beta_0),\\
V_{\rho t}:=(\widehat\rho_t-\rho_0)'X_t,
\qquad
a_T^\circ:=\sqrt{s_\circ}\lambda_T,
\qquad b_T^\circ:=b_{\circ,T}.
\end{gathered}
\]
For rolling and recursive estimation, set
\[
e_T:=
K_T\left\{
(b_T^\circ)^2\beta(h)^\eta+
b_T^\circ D_{\circ,T}(h)
\right\};
\]
for fixed estimation set $e_T=0$.
The rate restrictions imply $\sqrt P\,e_T\to0$.

Lemma~\ref{lem:OA_evaluation_products} and the joint
weighted deletion bound imply that every pair of
prediction errors satisfies
\begin{equation}\label{eq:OA_eval_product}
\frac1P\sum_t
(1+|e_{0t}|)|V_{jt}|\,|V_{\ell t}|
=
O_p\!\left((a_T^\circ)^2+e_T\right),
\end{equation}
with projection coordinates treated separately.
In particular,
\begin{equation}\label{eq:OA_eval_second}
\frac1P\sum_t
\left(
V_{\gamma t}^2+V_{\beta t}^2+\|V_{\rho t}\|^2
\right)
=
O_p\!\left((a_T^\circ)^2+e_T\right).
\end{equation}

Recall the exact score decomposition
\[
\widehat S_{t+1}^{\perp}-S_{t+1}^{\perp}
=
k_{0t}G_{t+1}-k_{0t}V_{\beta t}
+D_t\varepsilon_{t+1}
+D_t(G_{t+1}-V_{\beta t}).
\]

\emph{Linear terms.}
Lemma~\ref{lem:OA_dynamic_pseudo} reduces the first
term to the centered array $k_{0t}A_{t+1}X_t'$
times the quantile coefficient error, plus
$O_p(\sqrt P\,a_{\gamma,T}^2)+o_p(1)$.
The other linear arrays are
\[
k_{0t}X_t',
\qquad
X_t\varepsilon_{t+1},
\qquad
e_{0t}X_t\varepsilon_{t+1}.
\]
They are centered by the projection normal equations
and the conditional null given $X_t$.
For rolling and recursive estimation, their normalized
sums vanish by
Lemmas~\ref{lem:OA_deletion} and
\ref{lem:OA_separation}, together with
\eqref{eq:OA_delay_rates}.
For fixed estimation, factor out the common coefficient
errors and apply Lemma~\ref{lem:OA_uniform_FN}.
The resulting order is at most
\[
O_p\!\left(
K_Tb_T^\circ\sqrt P\,\Lambda_P(d_wp_T,q)
\right)=o_p(1).
\]
The nonlinear quantile remainder vanishes because $\sqrt P\,s_\circ\lambda_T^2\to0$, together with the scheme-specific remainder bounds established above.

\emph{Generated-instrument innovation term.}
For a fixed instrument, $D_t=-V_{\rho t}$.
For the generated instrument,
\[
D_t
=
2e_{0t}V_{\beta t}+V_{\beta t}^2-V_{\rho t}.
\]
After the linear terms, the remaining contribution
to $D_t\varepsilon_{t+1}$ is
$\varepsilon_{t+1}V_{\beta t}^2$.
By \eqref{eq:OA_eval_second} and
Lemma~\ref{lem:OA_predictable_multiplier},
\[
\frac1{\sqrt P}\sum_t
|\varepsilon_{t+1}|V_{\beta t}^2
=
O_p\!\left(
\sqrt P\{(a_T^\circ)^2+e_T\}
\right)=o_p(1).
\]
Here the multiplier lemma applies because
$V_{\beta t}^2$ is $\mathcal F_t$-measurable and
$\E[|\varepsilon_{t+1}|\mid\mathcal F_t]\le C$.

\emph{Products of prediction errors.}
The global Lipschitz bound gives
$|G_{t+1}|\le C|V_{\gamma t}|$.
The quadratic part of
$\|D_t(G_{t+1}-V_{\beta t})\|$ is bounded by
\[
C(1+|e_{0t}|)
(|V_{\beta t}|+\|V_{\rho t}\|)
(|V_{\gamma t}|+|V_{\beta t}|).
\]
Equation~\eqref{eq:OA_eval_product} makes its
normalized sum $o_p(1)$.
For the generated instrument, the remaining cubic
part is bounded on the uniform coefficient-rate event by
\[
C|V_{\beta t}|^2
(|V_{\gamma t}|+|V_{\beta t}|)
\le C(b_T^\circ)^3M_t^3.
\]
Stationarity and Markov's inequality therefore bound
its normalized sum by
\[
O_p\!\left(
\sqrt P\,(b_T^\circ)^3C_{3,T}
\right)=o_p(1)
\]
under \eqref{eq:OA_cubic_rate}.
This verifies
\eqref{eq:OA_HL1}--\eqref{eq:OA_HL4}.

\emph{Empirical score distance.}
Put $N_t:=|e_{0t}|M_t$ and recall
$Q_t=(1+|e_{0t}|)M_t^2$.
On the uniform coefficient-rate event,
\[
\|D_t\|
\le
C\{b_T^\circ(M_t+N_t)+(b_T^\circ)^2M_t^2\},
\]
and
\[
\|D_t(G_{t+1}-V_{\beta t})\|
\le
C\{(b_T^\circ)^2Q_t+(b_T^\circ)^3M_t^3\}.
\]
The terms involving $k_{0t}$ have empirical
$L^2$ norm $O_p(b_T^\circ K_T)$ by the envelope
definition and Markov's inequality.
For $D_t\varepsilon_{t+1}$, apply
Lemma~\ref{lem:OA_predictable_multiplier} with
$H_t=\|D_t\|^2$ and
$Z_{t+1}=\varepsilon_{t+1}^2$.
Its conditional mean is bounded.

Since
\[
\|M_t\|_{L^{2q}}^2\le\|Q_t\|_{L^q}\le K_T,
\qquad
\|N_t\|_{L^{2q}}\le K_T,
\]
squaring the exact decomposition and using $2q>6$
gives
\begin{equation}\label{eq:OA_score_distance_bound}
\begin{aligned}
r_{S,T}
&=
O_p\!\left(
b_T^\circ K_T+
(b_T^\circ)^2K_T+
(b_T^\circ)^3K_T^{3/2}
\right)\\
&=O_p(b_{\circ,T}K_T),
\end{aligned}
\end{equation}
because $b_{\circ,T}K_T\to0$.
This is a sufficient score-distance bound for HAC.

Finally, conditioning on $\mathcal F_t$ yields
\[
\E\|S_{t+1}^{\perp}\|^q
\le C\E\|k_{0t}\|^q\le C.
\]
Polynomial beta mixing with $\mu>2$ and $q>4$
implies
$\sum_j\beta(j)^{1-2/q}<\infty$.
The oracle central limit theorem and the remainder
bounds therefore give the asserted Gaussian limit.
Studentization follows from the HAC argument in the
proof of Theorem~\ref{thm:main_asymptotics}, using
\eqref{eq:OA_score_distance_bound}.
\end{proof}
\subsection{Proof of Corollary~\ref{cor:primitive_nonvacuous}}

\begin{proof}
Throughout, $R,P\asymp T$ and $s_\circ=O(1)$.
Under the coordinate-moment alternative, population
curvature gives
$\|v\|_1\le C_M\|v\|_\Sigma$ on the cones and sparse
supports in $\mathcal V_M$. Minkowski's inequality yields
\[
\|X_t'v\|_{L^{4q}}
\le \sum_j|v_j|\|X_{jt}\|_{L^{4q}}
\le C_M\|v\|_\Sigma.
\]
Thus \eqref{eq:OA_directional_primitive} holds.
The maintained population moment conditions imply
\[
K_T=O(p_T^{1/q})=O(T^{\kappa/q}),
\qquad
C_{3,T}=O(p_T^{3/(2q)}),
\]
for the envelopes in \eqref{eq:OA_inference_envelope}.

Set $\mu'=2$ and $h=\lfloor T^{1/2}\rfloor$.
Then $a_n\asymp n^{1/3}$ and $d_n\asymp n^{2/3}$.
Define
\[
r_\lambda
:=\min\left\{\frac13,
\frac{2(q-2)-6\kappa}{3q}\right\},
\qquad
c:=\min\left\{\frac13,
\frac{2(q-2)-12\kappa}{3q}\right\}.
\]
Direct substitution into the reference rates gives
\[
\lambda_T=O(T^{-r_\lambda}\sqrt{\log T}),
\qquad
\lambda_T+g_T=O(T^{-c}\sqrt{\log T}),
\qquad
\lambda_T\gtrsim T^{-1/3}\sqrt{\log T}.
\]
The first two bounds in \eqref{eq:OA_dimension_regime}
imply
\[
r_\lambda\ge c>\frac14+\frac{\kappa}{q},
\qquad
\frac{\kappa}{q}<\frac1{15}.
\]
Moreover,
\[
g_{4,T}
=O\left(
T^{-1/3}\sqrt{\log T}
+T^{\{24\kappa-2(q-2)\}/(3q)}
\right)=o(1),
\]
because $\kappa<(q-2)/15<(q-2)/12$.
The third dimension bound implies
$\kappa<(\mu-2)/12$, and hence
\[
p_T^2d_h\beta(a_h)
=O(T^{2\kappa-(\mu-2)/6})=o(1).
\]
The corresponding condition at sample size $T$ follows
as well. These bounds verify
\eqref{eq:OA_primitive_growth}; the sample design
conclusions follow from Lemma~\ref{lem:OA_primitive_design}.

We next verify the quantile-score restrictions.
Choose the proof block length
$\ell_T=\lfloor T^a\rfloor$, where
\[
\frac1{\mu+1}<a<
\min\left\{
\frac13,\,
r_\lambda-\frac{2\kappa}{q},\,
\frac12-\frac{1+3\kappa}{2q}
\right\}.
\]
This interval is nonempty. First, $\mu>2$ implies
$1/(\mu+1)<1/3$, and the first dimension bound gives
$\kappa<(q-3)/9$, so the last upper bound exceeds $1/3$.
For the two branches of
$r_\lambda-2\kappa/q$, use
\[
\mu-2>\frac{6(2q-3)\kappa}{q-2}.
\]
The comparison for the branch $1/3-2\kappa/q$ follows
from
\[
(2q-3)(q-6\kappa)>3(q-2).
\]
For the other branch, the required comparison follows
from
\[
\begin{aligned}
&24\kappa(q-3)
-\frac{72\kappa^2(2q-3)}{q-2}
+3(q-4)\\
&\qquad>
(q-4)\left(3+\frac{72\kappa}{5}\right)>0.
\end{aligned}
\]
Both inequalities use $\kappa<(q-2)/15$ and $q>4$.

For this choice of $a$,
\[
\frac{T}{\ell_T}\beta(\ell_T)
=O(T^{1-a(\mu+1)})=o(1),
\qquad
\frac{\tau_T}{\lambda_T}
=O(T^{(a-1/3)/2})=o(1).
\]
Write $r_T^{\mathrm{sc}}$ for the score-remainder rate
in \eqref{eq:OA_quantile_score_remainder}.
We show that $\sqrt P K_T r_T^{\mathrm{sc}}\to0$.
Its population term satisfies
\[
\sqrt P K_Ta_{\gamma,T}^2
=O(T^{1/2+\kappa/q-2c}\log T)=o(1).
\]
After multiplication by $\sqrt P K_T$, the powers of
$T$ in its two stochastic terms are, apart from
logarithmic factors,
\[
\frac{\kappa}{q}+\frac{a-r_\lambda}{2}<0,
\qquad
a-\frac12+\frac{1+3\kappa}{2q}<0.
\]
The truncation term satisfies
\[
\sqrt P K_Tt_T^X
=O\left(
T^{-1/2+(1+2\kappa)/(2q)}
\{\log(eT)\}^{-(2q-1)}
\right)=o(1),
\]
and the terms involving $u_T=T^{-3}$ are smaller.
Therefore $\sqrt P K_T r_T^{\mathrm{sc}}\to0$.
Since $K_T\ge1$ and $\sqrt P\lambda_T\to\infty$,
this also gives $r_T^{\mathrm{sc}}/\lambda_T\to0$.
Bounded sparsity now verifies
\eqref{eq:OA_quantile_score_growth}.

For the deletion restrictions, substitution into the
short-block rate gives
\[
\begin{aligned}
K_T\overline\Lambda_h(p_T^2,q)
=O\Bigl(&
T^{\kappa/q-1/6}\sqrt{\log T}
+T^{\{15\kappa-(q-2)\}/(3q)}\\
&+
T^{\{(2q-3)\kappa-(q-2)(\mu-2)/6\}/q}
\Bigr)=o(1).
\end{aligned}
\]
All three exponents are negative by
\eqref{eq:OA_dimension_regime}.
Since $b_{\circ,T}=O(\lambda_T)$,
\[
K_Tb_{\circ,T}\sqrt h
=O(T^{1/4+\kappa/q-c}\sqrt{\log T})=o(1).
\]
Taking $\eta=1/2<1-1/q$ also gives
\[
\sqrt P K_Tb_{\circ,T}\beta(h)^\eta
=O(T^{1/2+\kappa/q-c-\mu/4}\sqrt{\log T})
=o(1).
\]
Furthermore,
\[
\sqrt P K_T\zeta_{\gamma,T}
=O(T^{1/2+\kappa/q-2c}\log T)=o(1).
\]
Because $\sqrt P h/R=O(1)$, these bounds control all
terms in \eqref{eq:OA_Dcirc} except its quantile-score
and generated-instrument additions. The former is
controlled by $\sqrt P K_T r_T^{\mathrm{sc}}\to0$.
For the latter,
\[
b_{\circ,T}p_T^{3/(2q)}
=O(T^{-c+3\kappa/(2q)}\sqrt{\log T})=o(1).
\]
Consequently,
\[
\sqrt P K_T\frac hR
b_{\circ,T}^2p_T^{3/(2q)}=o(1).
\]
Thus $\sqrt P K_TD_{\circ,T}(h)\to0$, completing
the verification of \eqref{eq:OA_delay_rates}.

For fixed estimation, $d_w$ is fixed and
$\Lambda_P(d_wp_T,q)=O(\lambda_T)$. Hence
\[
K_Tb_{\circ,T}\sqrt P\Lambda_P(d_wp_T,q)
=O(T^{1/2+\kappa/q-2c}\log T)=o(1),
\]
which verifies \eqref{eq:OA_fixed_rate}.
The same bounds imply
$\sqrt P s_\circ\lambda_T^2\to0$.
For the generated instrument,
\[
\sqrt P b_{\circ,T}^3C_{3,T}
=O\left(
T^{1/2-3c+3\kappa/(2q)}(\log T)^{3/2}
\right)=o(1),
\]
verifying \eqref{eq:OA_cubic_rate}.

Finally, $b_{\circ,T}K_T=o(T^{-1/4})$.
Writing $d=\mu(1-4/q)>0$, polynomial mixing gives
\[
A_P=
\begin{cases}
O(P^{1-d}),&0<d<1,\\
O(\log P),&d=1,\\
O(1),&d>1.
\end{cases}
\]
Thus $L_P=\lfloor\log P\rfloor$ satisfies
\[
\frac{L_P^3}{P}\to0,
\qquad
\frac{L_P^2A_P}{P}\to0,
\qquad
L_Pb_{\circ,T}K_T\to0.
\]
The choice $L_P=\lfloor P^{1/8}\rfloor$ satisfies
these restrictions whenever $d>1/4$.
Theorem~\ref{thm:primitive_lasso} therefore gives
all the stated conclusions.
\end{proof}
\section{Additional Details for the Monte Carlo Experiments}
\label{sec:OA_mc}

\subsection{Population targets}
\label{sec:OA_mc_targets}

Under the homoskedastic design in \eqref{eq:dgp}, the conditional
quantile and ES are
\begin{align}
q_{0t}&=X_t'\gamma_0,
&\gamma_0&=\delta+\sigma q_\alpha^u\iota,
\label{eq:OA_mc_var}\\
e_{0t}&=X_t'\beta_0,
&\beta_0&=\delta+\sigma e_\alpha^u\iota,
\label{eq:OA_mc_es}
\end{align}
where $\iota=(1,0,\ldots,0)'$. For the unit-variance $t_5$
innovation, let $a_\alpha=F_5^{-1}(\alpha)$, with $F_5$ and
$f_5$ the distribution and density functions of a standard
$t_5$ variable. Then
\[
q_\alpha^u=\sqrt{\frac35}\,a_\alpha,
\qquad
e_\alpha^u=-\sqrt{\frac35}\,
\frac{5+a_\alpha^2}{4\alpha}\,f_5(a_\alpha).
\]
Thus the conditional mean, quantile, and ES share their slope
coefficients but have different intercepts. In the omission
experiments, replace $\delta$ by $\delta^{(\eta)}$.
The full-information targets remain linear; $H_2$ excludes
predictors from both fitted models, whereas $H_3$ excludes
them from ES only.

\subsection{Estimation and tuning}
\label{sec:OA_mc_tuning}

The omission experiments use a fixed initial training sample.
VaR is estimated by $\ell_1$-penalized smoothed quantile regression.
The second-stage ES estimators are ordinary Lasso, adaptive Lasso,
and OLS; the first-stage quantile procedure is held fixed across
these comparisons. All fitted components remain fixed during
the evaluation period, and intercepts are unpenalized.

Penalty candidates are ranked by ten-fold block cross-validation
within the training sample. Each contiguous block is held out
in turn, and estimation uses the remaining training observations.
Validation uses quantile loss for VaR and squared loss for the
ES and projection regressions. The generated pseudo-outcomes
are held fixed during second-stage cross-validation.

For penalized ES estimation, an additional coefficient-count
screen is applied to the first 60 candidates in validation-loss
order. The implementation selects the first fit whose number
of coefficients exceeding $10^{-6}$ in absolute value, including
the intercept, differs from one. If none passes, it selects the
best-ranked candidate. This rule is intended to avoid
intercept-only fits and is an additional tuning choice.

Adaptive Lasso uses a ridge pilot with slope-penalty weights
proportional to
\[
\{\max(|\widetilde b_j|,10^{-6})\}^{-1}.
\]
The pilot coefficients and weights are held fixed during
adaptive-Lasso cross-validation. Instrument projections use
adaptive Lasso with a ridge pilot, including when ES is estimated
by ordinary Lasso or OLS. Under $H_2$ and $H_3$, these projections
use the retained ES regressors $X_{-k,t}$.

\subsection{Test functions and nondegeneracy}
\label{sec:OA_mc_instruments}

In addition to the omitted-predictor function $w_1$ used in the
main figure, we consider squared fitted restricted ES,
\[
\widehat w_{\mathrm{sq},t}
=\bigl(X_{-k,t}'\widehat\beta_t\bigr)^2,
\]
and the unsquared full-model ES function $\widehat w_{2,t}$
defined in Section~\ref{sec:mc}. The latter is obtained from an
auxiliary fit using the full predictor set in both forecasting
stages. Its ES estimator matches the estimator used for the
restricted model. All generated functions and projections
are estimated using training observations only.

The unsquared full-model function requires separate treatment.
At $\eta=0$, its population target is linear in the retained
predictors. Writing $\Pi_{-k}$ for population linear projection
on their span gives
\[
w_{0t}^{\mathrm{full}}=X_{-k,t}'\beta_{0,-k},
\qquad
k_{0t}=w_{0t}^{\mathrm{full}}
       -\Pi_{-k}w_{0t}^{\mathrm{full}}=0.
\]
The oracle score therefore has zero long-run variance, violating
the nondegeneracy condition in
Assumption~\ref{ass:general_pipeline}. Results for this function
are reported as a stress test outside the maintained assumptions.
They are not evidence of valid size or a size-adjusted comparison
of power. Squaring the fitted ES does not entail this automatic
degeneracy.

\subsection{HAC studentization}
\label{sec:OA_mc_hac}

The $H_2$ and $H_3$ experiments use Bartlett HAC studentization
with automatic bandwidth selection through
\texttt{sandwich::NeweyWest}, setting
\texttt{prewhite = FALSE}. The two-sided rejection rule uses
the critical value 1.96. All figures plot rejection frequencies
against the actual omission-strength parameter $\eta$.

\subsection{Monte Carlo uncertainty}
\label{sec:OA_mc_measures}

The omission experiments attempt 200 replications per design.
Nonfinite score observations are removed before studentization;
replications without a computable test statistic are omitted
from the reported rejection frequency. The aggregate results
do not record the number of valid replications by cell.

For $B$ valid independent replications and rejection frequency
$\widehat\pi$, the estimated Monte Carlo standard error is
\[
\left\{\frac{\widehat\pi(1-\widehat\pi)}{B}\right\}^{1/2}.
\]
If all 200 replications are valid, this is approximately
1.5 percentage points at a rejection probability of 5\%
and at most 3.5 percentage points. Accordingly, small
differences between curves should be interpreted cautiously.
The reported curves use the nominal critical value and are
not adjusted for differences in null rejection frequencies.

\subsection{Additional results}
\label{sec:OA_mc_results}

Table~\ref{tab:OA_mc_size_all} extends the main size comparison
to adaptive Lasso and OLS. The Lasso panel is repeated to
facilitate comparison. Orthogonalization reduces overrejection
in most configurations, although OLS-based LR tests also
overreject substantially
in some configurations with many predictors.

\begin{table}[!htbp]
\centering
\caption{Empirical size across ES estimators}
\label{tab:OA_mc_size_all}
\small
\begin{tabular}{@{}r*{8}{r}@{}}
\toprule
& \multicolumn{2}{c}{$p_T-1=25$} & \multicolumn{2}{c}{$p_T-1=50$}
& \multicolumn{2}{c}{$p_T-1=100$} & \multicolumn{2}{c}{$p_T-1=150$} \\
\cmidrule(lr){2-3}\cmidrule(lr){4-5}\cmidrule(lr){6-7}\cmidrule(l){8-9}
$T$ & LR & NLR & LR & NLR & LR & NLR & LR & NLR \\
\midrule
\multicolumn{9}{l}{\textit{Lasso}} \\
250 & 0.055 & 0.115 & 0.065 & 0.120 & 0.030 & 0.195 & 0.075 & 0.395 \\
500 & 0.050 & 0.145 & 0.050 & 0.140 & 0.075 & 0.215 & 0.040 & 0.280 \\
1000 & 0.060 & 0.130 & 0.035 & 0.190 & 0.085 & 0.220 & 0.065 & 0.305 \\
5000 & 0.070 & 0.155 & 0.065 & 0.200 & 0.060 & 0.285 & 0.045 & 0.335 \\
\midrule
\multicolumn{9}{l}{\textit{Adaptive Lasso}} \\
250 & 0.045 & 0.090 & 0.070 & 0.120 & 0.035 & 0.230 & 0.080 & 0.430 \\
500 & 0.055 & 0.110 & 0.050 & 0.110 & 0.030 & 0.175 & 0.055 & 0.285 \\
1000 & 0.070 & 0.105 & 0.045 & 0.150 & 0.075 & 0.175 & 0.050 & 0.230 \\
5000 & 0.070 & 0.145 & 0.050 & 0.100 & 0.060 & 0.135 & 0.050 & 0.095 \\
\midrule
\multicolumn{9}{l}{\textit{OLS}} \\
250 & 0.040 & 0.085 & 0.090 & 0.170 & 0.150 & 0.200 & 0.195 & 0.330 \\
500 & 0.075 & 0.105 & 0.070 & 0.105 & 0.100 & 0.200 & 0.150 & 0.305 \\
1000 & 0.070 & 0.110 & 0.090 & 0.115 & 0.110 & 0.170 & 0.120 & 0.215 \\
5000 & 0.070 & 0.115 & 0.055 & 0.085 & 0.060 & 0.105 & 0.075 & 0.060 \\
\bottomrule
\end{tabular}
\par\medskip
\begin{minipage}{0.98\textwidth}\footnotesize
\textit{Notes.} Rejection frequencies at the nominal 5\% level,
with test function $(\delta'X_t)^2$ and 200 replications.
Only the second-stage ES estimator varies across panels.
The first-stage estimator and the design are those in the main size table.
LR and NLR denote orthogonal and baseline tests.
\end{minipage}
\end{table}

\FloatBarrier

Table~\ref{tab:OA_mc_omission_nulls} reports the $\eta=0$
points of the omission experiments for all three test functions
with ordinary Lasso ES estimation. These are separate designs
from the main size table, which uses $(\delta'X_t)^2$.
The table documents both remaining null overrejection with
the omitted-predictor function and the behavior of the
degenerate full-model ES function.

\begin{table}[!htbp]
\centering
\caption{Null rejection frequencies within the omission experiments}
\label{tab:OA_mc_omission_nulls}
\small
\begin{tabular}{@{}rr*{6}{r}@{}}
\toprule
& & \multicolumn{3}{c}{$H_2$} & \multicolumn{3}{c}{$H_3$} \\
\cmidrule(lr){3-5}\cmidrule(l){6-8}
$p_T-1$ & $T$ & Squared ES & Omitted & Full ES & Squared ES & Omitted & Full ES \\
\midrule
\multicolumn{8}{l}{\textit{One omitted predictor}} \\
25 & 250 & 0.065 & 0.090 & 0.130 & 0.065 & 0.075 & 0.130 \\
 & 500 & 0.095 & 0.060 & 0.140 & 0.095 & 0.060 & 0.145 \\
 & 1000 & 0.070 & 0.130 & 0.250 & 0.065 & 0.100 & 0.275 \\
 & 5000 & 0.080 & 0.105 & 0.325 & 0.095 & 0.095 & 0.315 \\
\addlinespace[2pt]
50 & 250 & 0.100 & 0.070 & 0.285 & 0.110 & 0.070 & 0.260 \\
 & 500 & 0.055 & 0.100 & 0.205 & 0.065 & 0.105 & 0.195 \\
 & 1000 & 0.055 & 0.145 & 0.345 & 0.055 & 0.140 & 0.350 \\
 & 5000 & 0.060 & 0.150 & 0.330 & 0.065 & 0.155 & 0.330 \\
\addlinespace[2pt]
100 & 250 & 0.090 & 0.140 & 0.210 & 0.085 & 0.125 & 0.185 \\
 & 500 & 0.085 & 0.140 & 0.315 & 0.090 & 0.145 & 0.315 \\
 & 1000 & 0.025 & 0.120 & 0.380 & 0.030 & 0.100 & 0.385 \\
 & 5000 & 0.050 & 0.095 & 0.360 & 0.050 & 0.095 & 0.355 \\
\addlinespace[2pt]
150 & 250 & 0.035 & 0.150 & 0.205 & 0.035 & 0.145 & 0.225 \\
 & 500 & 0.065 & 0.140 & 0.385 & 0.065 & 0.140 & 0.355 \\
 & 1000 & 0.065 & 0.125 & 0.465 & 0.065 & 0.110 & 0.450 \\
 & 5000 & 0.035 & 0.080 & 0.435 & 0.045 & 0.080 & 0.430 \\
\midrule
\multicolumn{8}{l}{\textit{Three omitted predictors}} \\
25 & 250 & 0.060 & 0.060 & 0.140 & 0.070 & 0.075 & 0.125 \\
 & 500 & 0.105 & 0.070 & 0.150 & 0.090 & 0.060 & 0.160 \\
 & 1000 & 0.065 & 0.135 & 0.220 & 0.055 & 0.105 & 0.240 \\
 & 5000 & 0.075 & 0.105 & 0.285 & 0.085 & 0.095 & 0.290 \\
\addlinespace[2pt]
50 & 250 & 0.090 & 0.085 & 0.235 & 0.090 & 0.095 & 0.235 \\
 & 500 & 0.050 & 0.095 & 0.195 & 0.065 & 0.105 & 0.205 \\
 & 1000 & 0.055 & 0.150 & 0.330 & 0.055 & 0.140 & 0.350 \\
 & 5000 & 0.070 & 0.140 & 0.285 & 0.060 & 0.150 & 0.290 \\
\addlinespace[2pt]
100 & 250 & 0.090 & 0.135 & 0.165 & 0.080 & 0.130 & 0.160 \\
 & 500 & 0.095 & 0.145 & 0.305 & 0.080 & 0.155 & 0.305 \\
 & 1000 & 0.040 & 0.130 & 0.335 & 0.035 & 0.100 & 0.365 \\
 & 5000 & 0.050 & 0.085 & 0.355 & 0.050 & 0.095 & 0.355 \\
\addlinespace[2pt]
150 & 250 & 0.035 & 0.170 & 0.200 & 0.030 & 0.145 & 0.220 \\
 & 500 & 0.075 & 0.140 & 0.330 & 0.075 & 0.145 & 0.350 \\
 & 1000 & 0.065 & 0.120 & 0.450 & 0.065 & 0.115 & 0.440 \\
 & 5000 & 0.035 & 0.085 & 0.435 & 0.045 & 0.080 & 0.435 \\
\bottomrule
\end{tabular}
\par\medskip
\begin{minipage}{0.98\textwidth}\footnotesize
\textit{Notes.} LR rejection frequencies at $\eta=0$ in the
omission experiments, with ordinary Lasso ES estimation.
Squared ES is squared fitted restricted ES;
Omitted is the first omitted active predictor;
Full ES is the unsquared fitted full-model ES.
All three functions are projected on the retained ES regressors.
For Full ES, the population residualized function vanishes under the null,
violating the nondegeneracy condition of the stated asymptotic theory.
These designs and test functions differ from the main size experiment.
\end{minipage}
\end{table}

\FloatBarrier

\begin{figure}[!htbp]
\centering
\includegraphics[width=\textwidth,height=0.68\textheight,keepaspectratio]{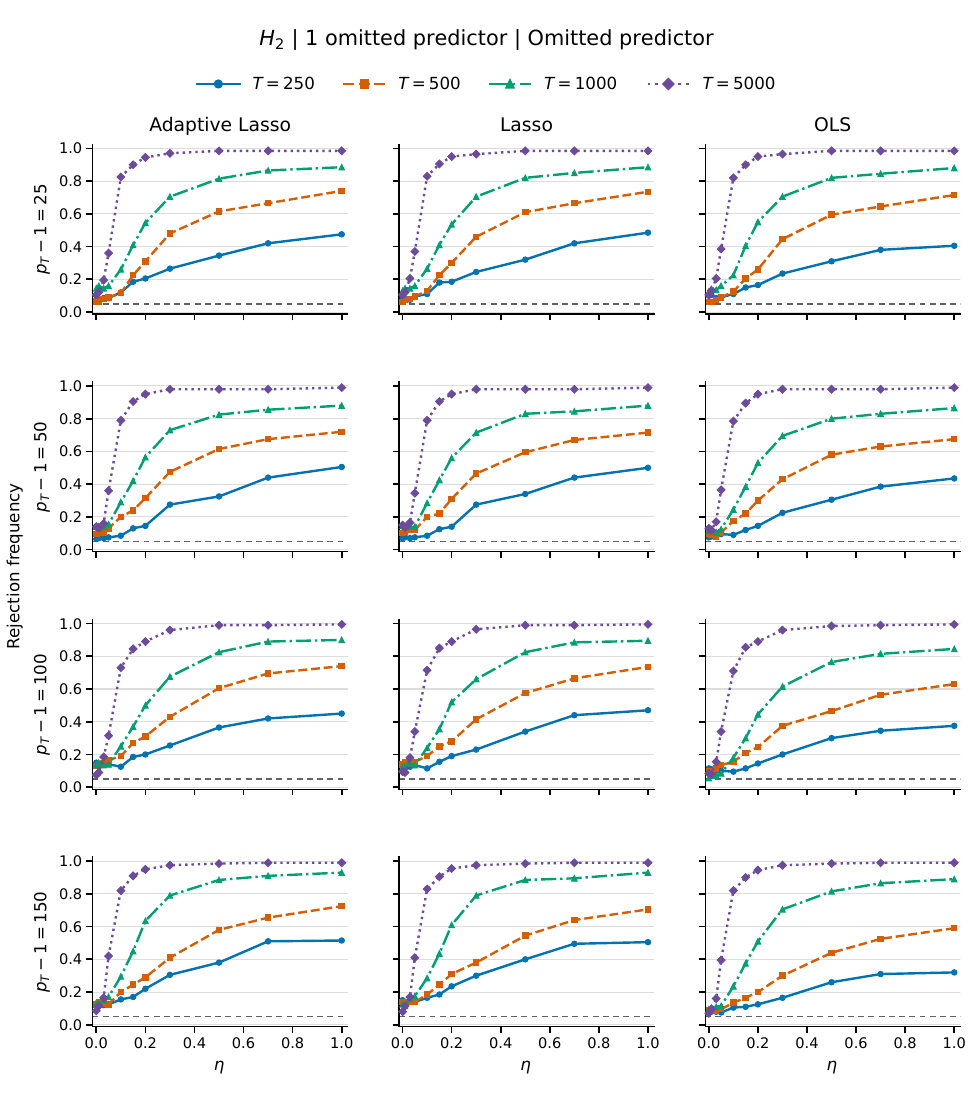}
\caption{$H_2$ with 1 omitted predictor and test function the first omitted active predictor. Rows vary the predictor dimension; columns vary the ES estimator.
All test functions are residualized against the retained ES regressors.
Curves show unadjusted LR rejection frequencies for four sample sizes.
The dashed line marks the nominal 5\% level; $\eta=0$ is retained.
}
\label{fig:OA_mc_A1_k1_w2}
\end{figure}

\begin{figure}[p]
\centering
\includegraphics[width=\textwidth]{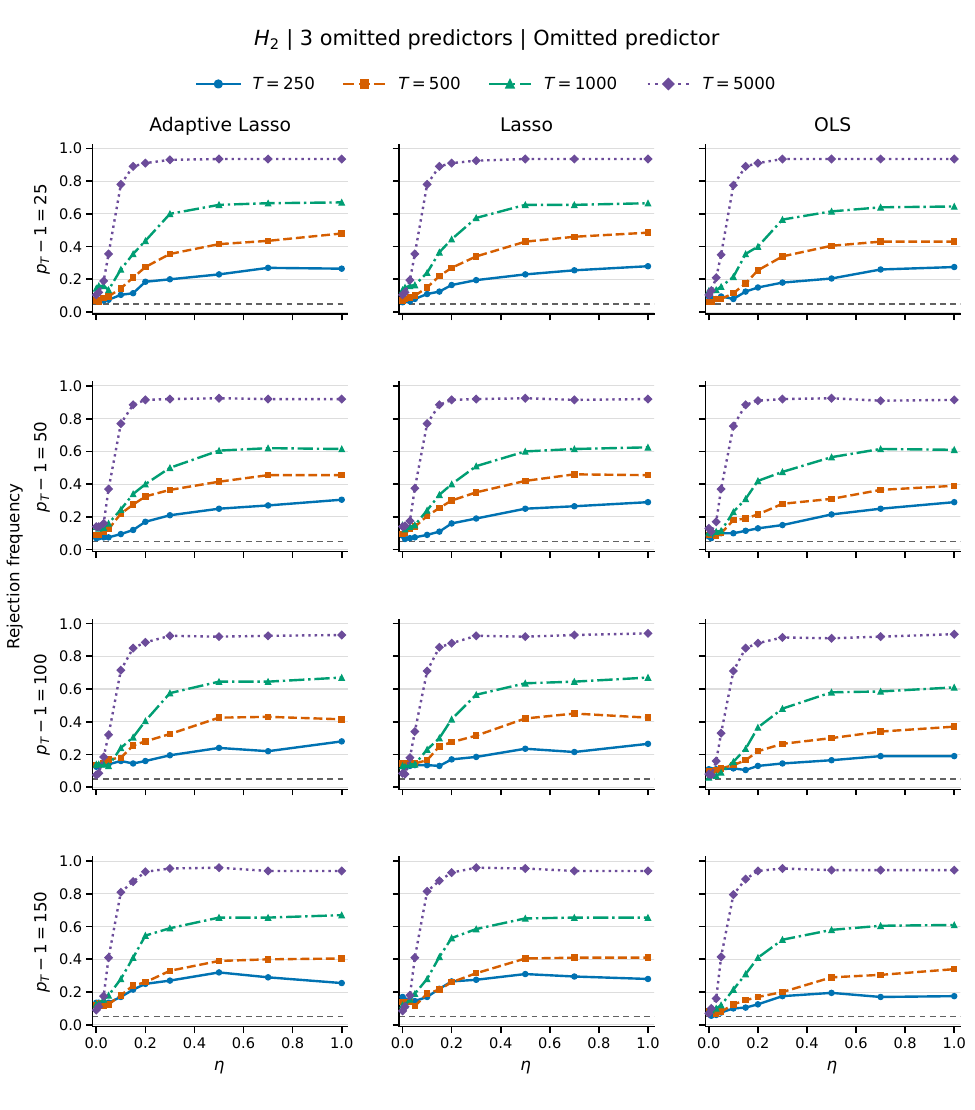}
\caption{$H_2$ with 3 omitted predictors and test function the first omitted active predictor. Rows vary the predictor dimension; columns vary the ES estimator.
All test functions are residualized against the retained ES regressors.
Curves show unadjusted LR rejection frequencies for four sample sizes.
The dashed line marks the nominal 5\% level; $\eta=0$ is retained.
}
\label{fig:OA_mc_A1_k3_w2}
\end{figure}

\begin{figure}[p]
\centering
\includegraphics[width=\textwidth]{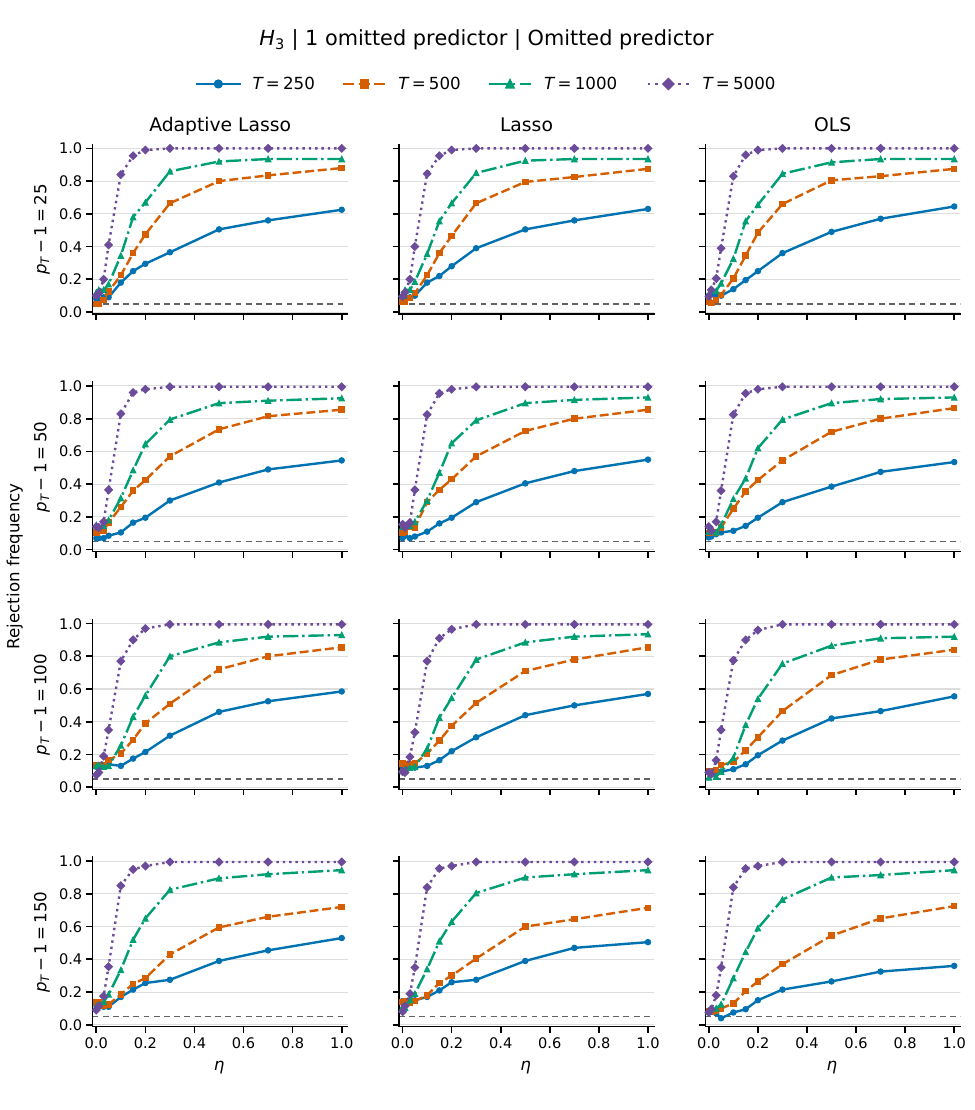}
\caption{$H_3$ with 1 omitted predictor and test function the first omitted active predictor. Rows vary the predictor dimension; columns vary the ES estimator.
All test functions are residualized against the retained ES regressors.
Curves show unadjusted LR rejection frequencies for four sample sizes.
The dashed line marks the nominal 5\% level; $\eta=0$ is retained.
}
\label{fig:OA_mc_A2_k1_w2}
\end{figure}

\begin{figure}[p]
\centering
\includegraphics[width=\textwidth]{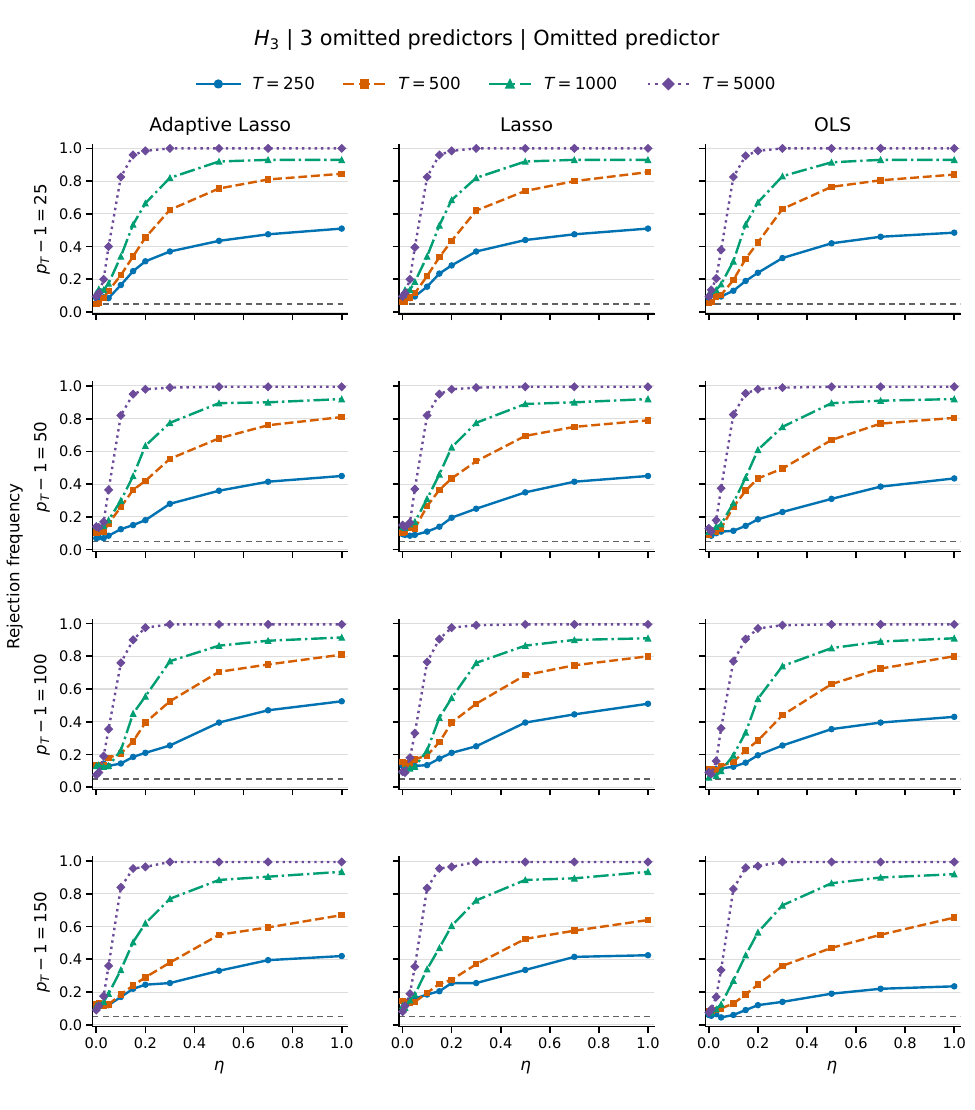}
\caption{$H_3$ with 3 omitted predictors and test function the first omitted active predictor. Rows vary the predictor dimension; columns vary the ES estimator.
All test functions are residualized against the retained ES regressors.
Curves show unadjusted LR rejection frequencies for four sample sizes.
The dashed line marks the nominal 5\% level; $\eta=0$ is retained.
}
\label{fig:OA_mc_A2_k3_w2}
\end{figure}

\begin{figure}[p]
\centering
\includegraphics[width=\textwidth]{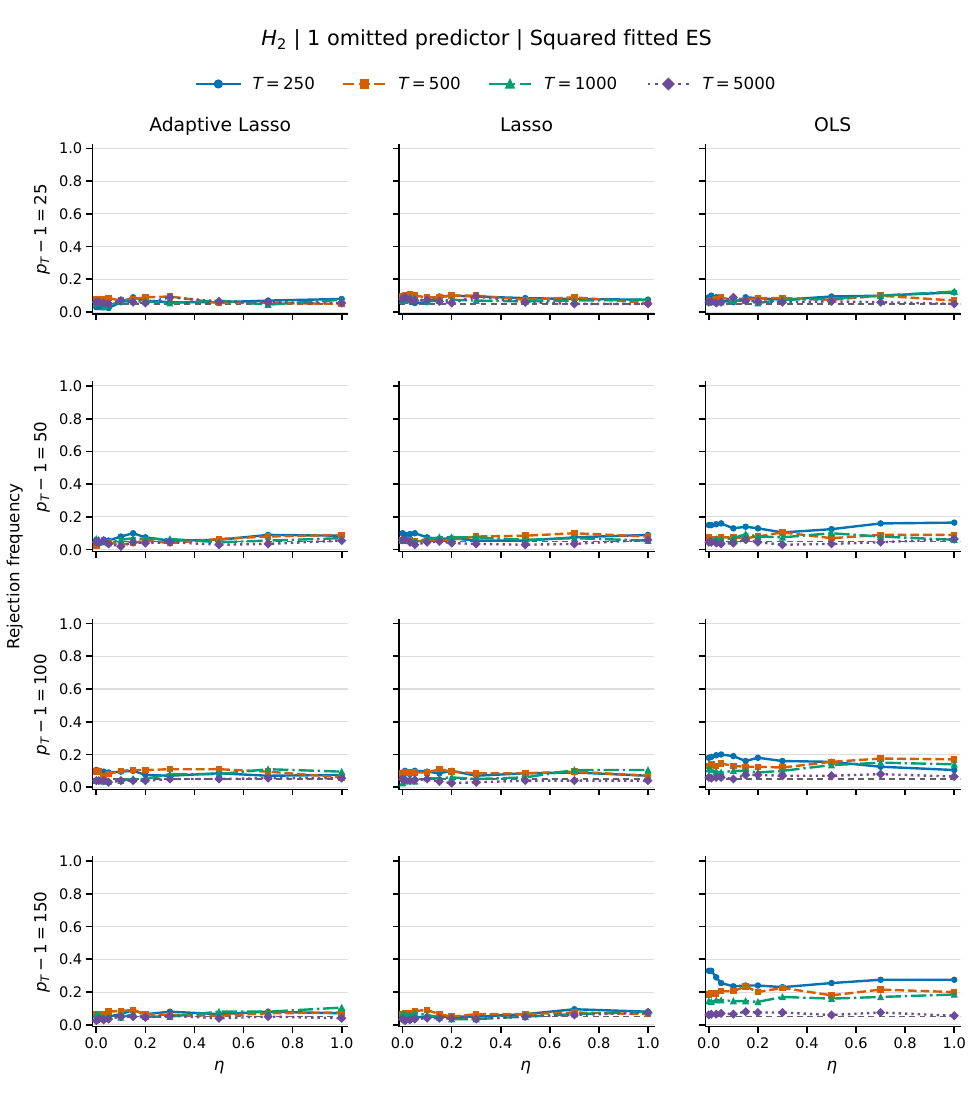}
\caption{$H_2$ with 1 omitted predictor and test function squared fitted restricted ES. Rows vary the predictor dimension; columns vary the ES estimator.
All test functions are residualized against the retained ES regressors.
Curves show unadjusted LR rejection frequencies for four sample sizes.
The dashed line marks the nominal 5\% level; $\eta=0$ is retained.
}
\label{fig:OA_mc_A1_k1_w1}
\end{figure}

\begin{figure}[p]
\centering
\includegraphics[width=\textwidth]{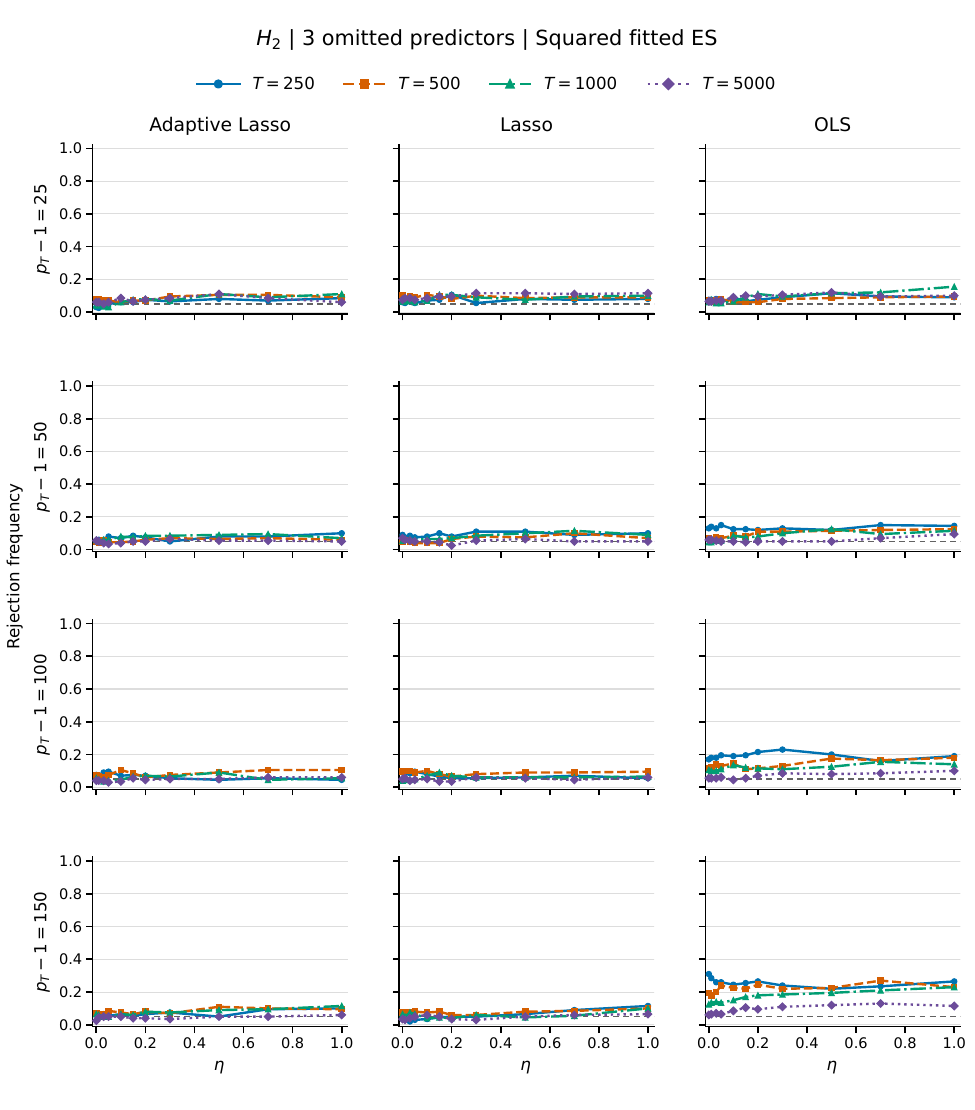}
\caption{$H_2$ with 3 omitted predictors and test function squared fitted restricted ES. Rows vary the predictor dimension; columns vary the ES estimator.
All test functions are residualized against the retained ES regressors.
Curves show unadjusted LR rejection frequencies for four sample sizes.
The dashed line marks the nominal 5\% level; $\eta=0$ is retained.
}
\label{fig:OA_mc_A1_k3_w1}
\end{figure}

\begin{figure}[p]
\centering
\includegraphics[width=\textwidth]{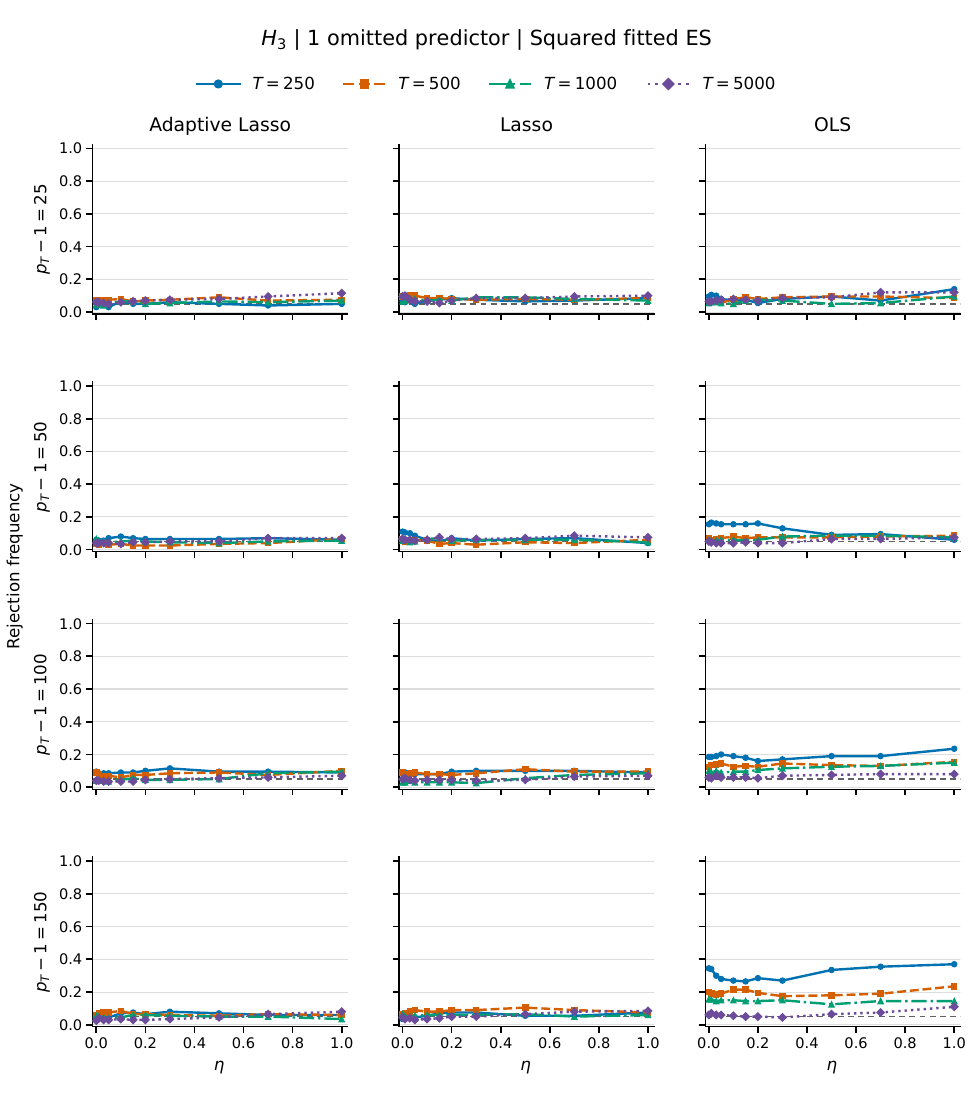}
\caption{$H_3$ with 1 omitted predictor and test function squared fitted restricted ES. Rows vary the predictor dimension; columns vary the ES estimator.
All test functions are residualized against the retained ES regressors.
Curves show unadjusted LR rejection frequencies for four sample sizes.
The dashed line marks the nominal 5\% level; $\eta=0$ is retained.
}
\label{fig:OA_mc_A2_k1_w1}
\end{figure}

\begin{figure}[p]
\centering
\includegraphics[width=\textwidth]{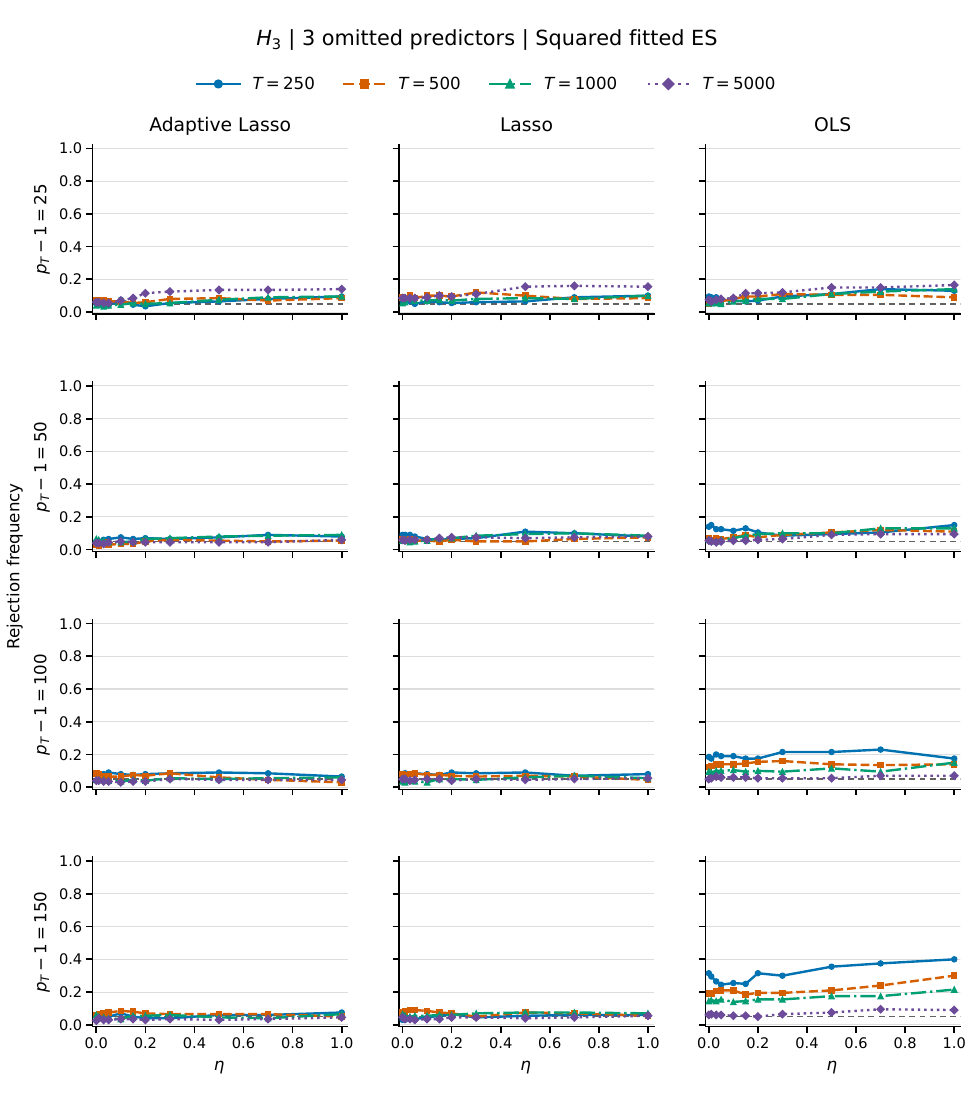}
\caption{$H_3$ with 3 omitted predictors and test function squared fitted restricted ES. Rows vary the predictor dimension; columns vary the ES estimator.
All test functions are residualized against the retained ES regressors.
Curves show unadjusted LR rejection frequencies for four sample sizes.
The dashed line marks the nominal 5\% level; $\eta=0$ is retained.
}
\label{fig:OA_mc_A2_k3_w1}
\end{figure}

\begin{figure}[p]
\centering
\includegraphics[width=\textwidth]{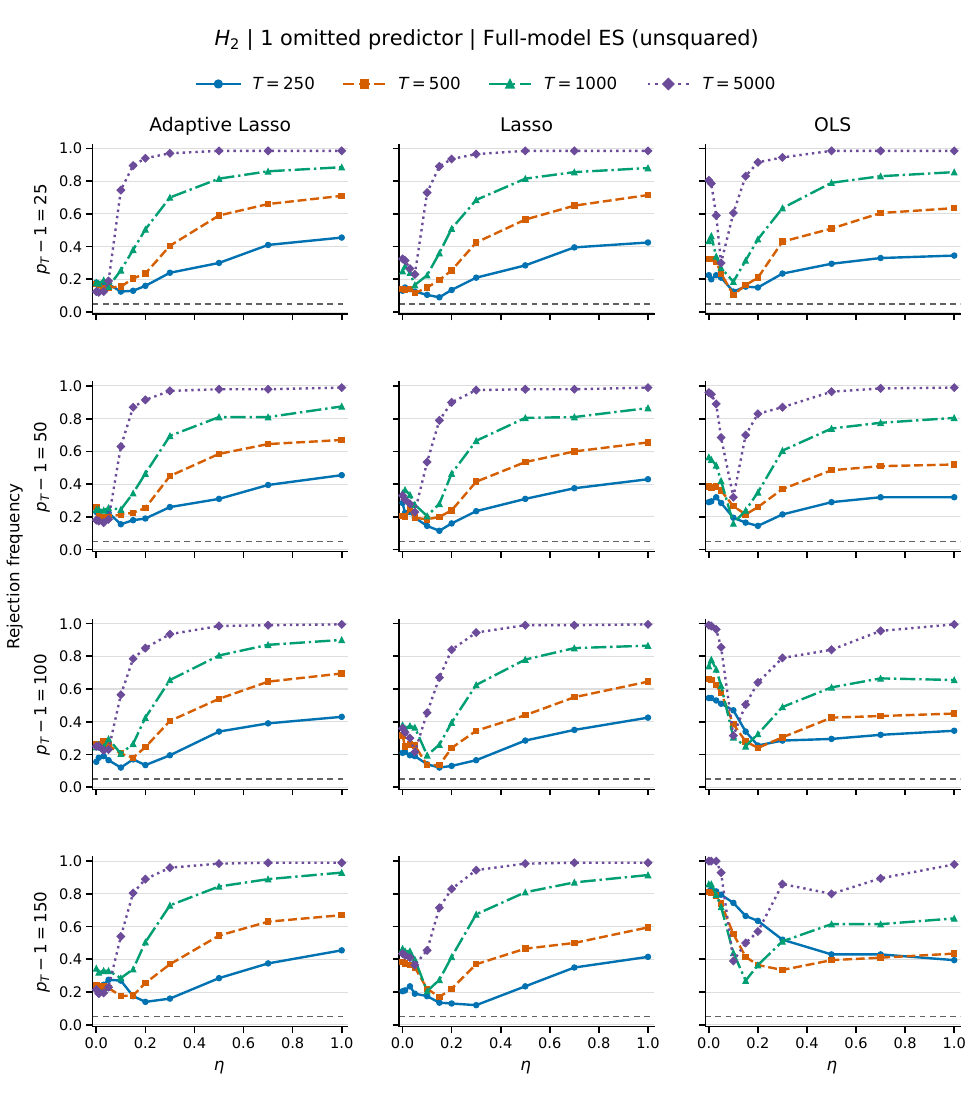}
\caption{$H_2$ with 1 omitted predictor and test function the unsquared fitted full-model ES. Rows vary the predictor dimension; columns vary the ES estimator.
All test functions are residualized against the retained ES regressors.
Curves show unadjusted LR rejection frequencies for four sample sizes.
The dashed line marks the nominal 5\% level; $\eta=0$ is retained.
 Under the null, this test function has a zero population residual after projection. This is a nondegeneracy stress test outside the maintained assumptions; the curves do not establish size control or size-adjusted power.}
\label{fig:OA_mc_A1_k1_w3}
\end{figure}

\begin{figure}[p]
\centering
\includegraphics[width=\textwidth]{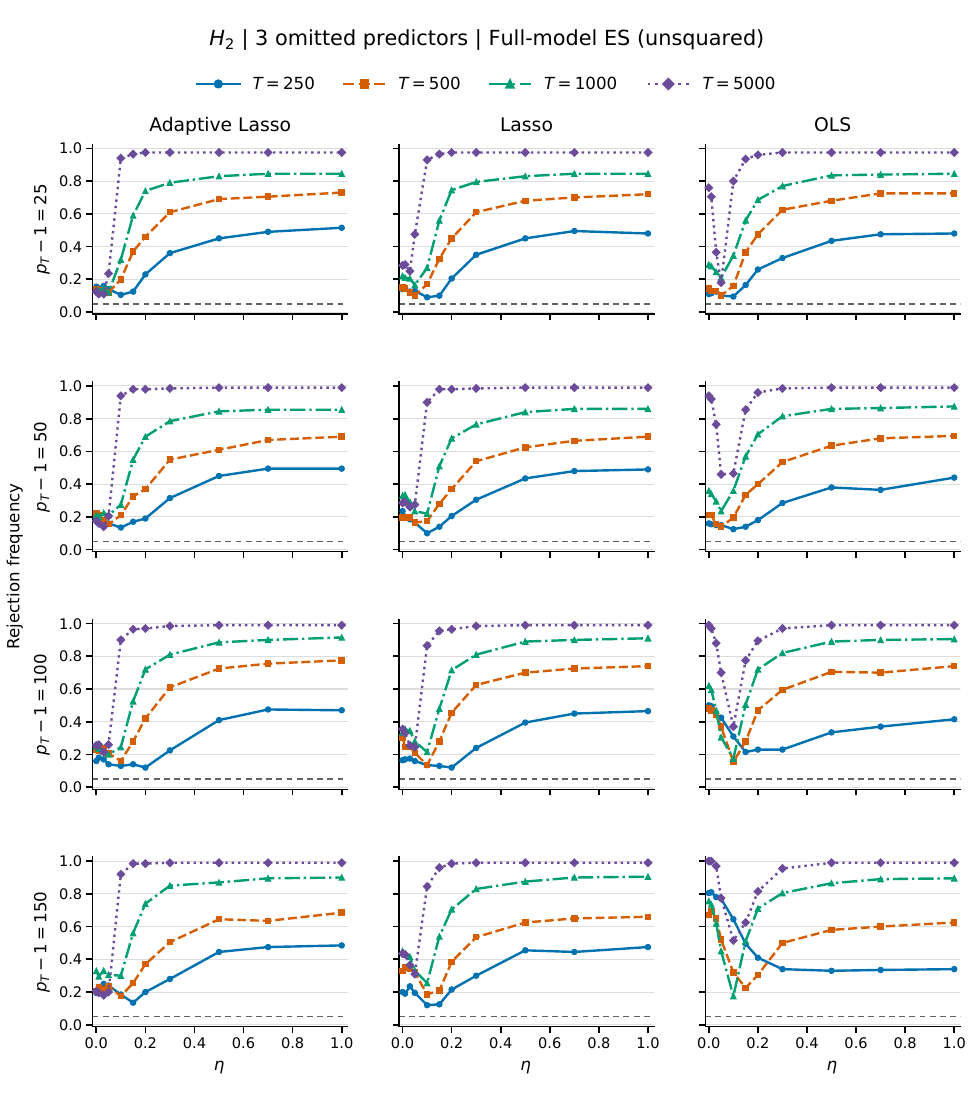}
\caption{$H_2$ with 3 omitted predictors and test function the unsquared fitted full-model ES. Rows vary the predictor dimension; columns vary the ES estimator.
All test functions are residualized against the retained ES regressors.
Curves show unadjusted LR rejection frequencies for four sample sizes.
The dashed line marks the nominal 5\% level; $\eta=0$ is retained.
 Under the null, this test function has a zero population residual after projection. This is a nondegeneracy stress test outside the maintained assumptions; the curves do not establish size control or size-adjusted power.}
\label{fig:OA_mc_A1_k3_w3}
\end{figure}

\begin{figure}[p]
\centering
\includegraphics[width=\textwidth]{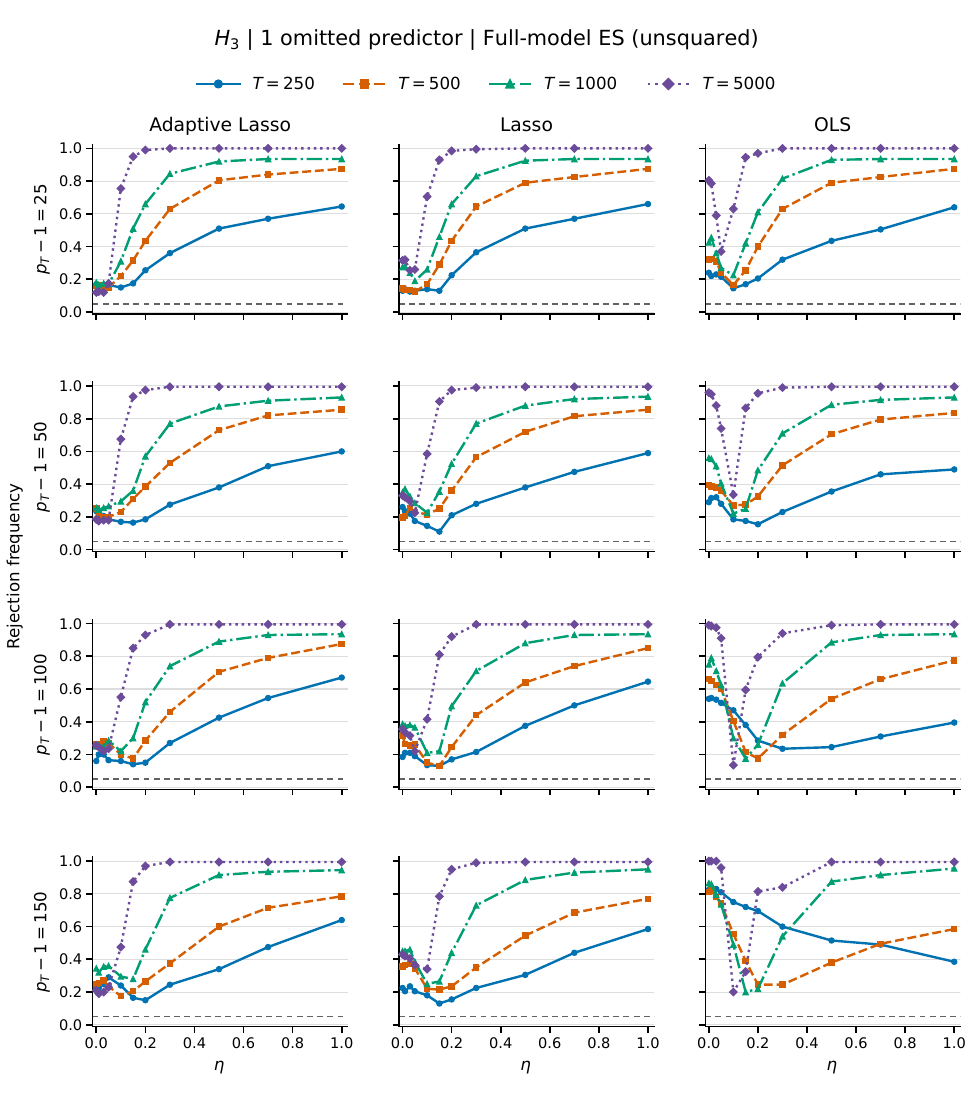}
\caption{$H_3$ with 1 omitted predictor and test function the unsquared fitted full-model ES. Rows vary the predictor dimension; columns vary the ES estimator.
All test functions are residualized against the retained ES regressors.
Curves show unadjusted LR rejection frequencies for four sample sizes.
The dashed line marks the nominal 5\% level; $\eta=0$ is retained.
 Under the null, this test function has a zero population residual after projection. This is a nondegeneracy stress test outside the maintained assumptions; the curves do not establish size control or size-adjusted power.}
\label{fig:OA_mc_A2_k1_w3}
\end{figure}

\begin{figure}[p]
\centering
\includegraphics[width=\textwidth]{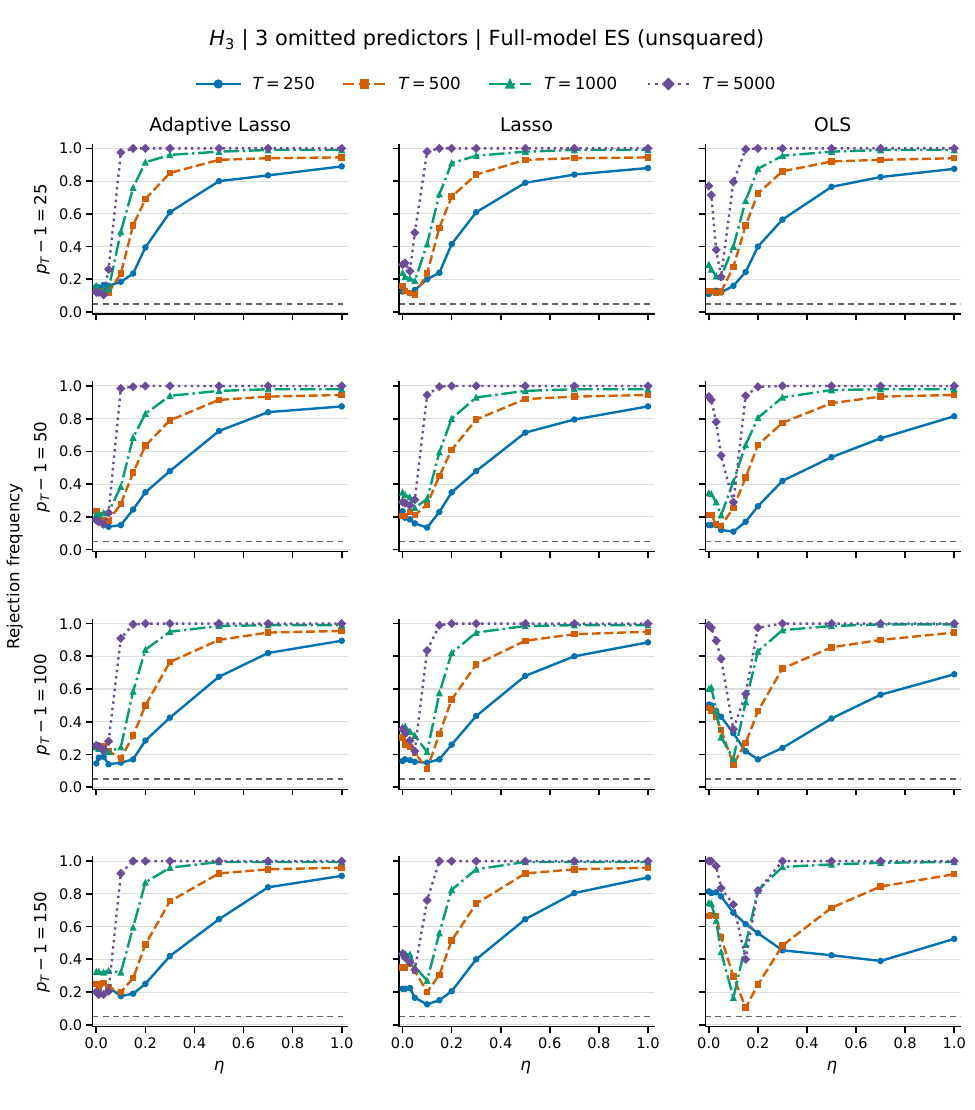}
\caption{$H_3$ with 3 omitted predictors and test function the unsquared fitted full-model ES. Rows vary the predictor dimension; columns vary the ES estimator.
All test functions are residualized against the retained ES regressors.
Curves show unadjusted LR rejection frequencies for four sample sizes.
The dashed line marks the nominal 5\% level; $\eta=0$ is retained.
 Under the null, this test function has a zero population residual after projection. This is a nondegeneracy stress test outside the maintained assumptions; the curves do not establish size control or size-adjusted power.}
\label{fig:OA_mc_A2_k3_w3}
\end{figure}

Figures~\ref{fig:OA_mc_A1_k1_w2}--\ref{fig:OA_mc_A2_k3_w2}
report results for the omitted-predictor function across all
dimensions, ES estimators, and omission counts.
Figures~\ref{fig:OA_mc_A1_k1_w1}--\ref{fig:OA_mc_A2_k3_w1}
report the corresponding squared-ES results, which show
limited sensitivity to these omission alternatives.
Figures~\ref{fig:OA_mc_A1_k1_w3}--\ref{fig:OA_mc_A2_k3_w3}
report the unsquared full-model ES stress test.
Taken together, the results distinguish robustness to
estimation effects from sensitivity to the particular
departure represented by a test function.

\clearpage

\clearpage

\clearpage

\clearpage

\clearpage

\clearpage

\clearpage

\clearpage

\clearpage

\clearpage

\clearpage

\clearpage

\section{Additional Details for the Empirical Application}
\label{sec:OA_empirical}

\subsection{Predictors and sample accounting}
\label{sec:OA_empirical_predictors}

The 15 series in $V_t$ are the supplied measures of geopolitical risk,
world uncertainty, stock returns, stock-return volatility, exchange-rate
returns, CPI inflation, credit-to-GDP growth, investment growth, the
credit-to-GDP gap, leverage, loan growth, credit growth, the term spread,
house-price growth, and financial conditions. Four lags of each are
combined with four growth lags and an intercept. The source families are
FRED, OECD, and the macro-financial data of \citet{jorda2017macrofinancial},
as documented in the original application. The supplied country-specific
volatility proxies are retained. Original transformation code and
release-vintage documentation are incomplete; the exercise uses the
supplied series and does not claim real-time availability of every regressor.

The supplied aligned predictor--outcome matrices contain 232 observations
from 1962Q1--2019Q4, except Canada, with 228 from 1963Q1--2019Q4. The first
eight rows are excluded to retain the original effective estimation and
evaluation samples. Thus $T=224=154+70$ in six countries and
$T=220=152+68$ in Canada. The effective training periods begin in 1964Q1
and 1965Q1, respectively. Dates are supported by lag identities and
comparisons with public GDP-growth series; revised public values are not
substituted into estimation. The forecast horizon is one quarter, while
the year-on-year outcome entails overlap in the underlying quarterly levels.

\subsection{Estimation and temporal tuning}
\label{sec:OA_empirical_design}

Let $n_t=|\mathcal I_t|$. At each training window, standardize the 60
components of $V_s$ by their training means and sample standard deviations.
Keep $Z_s$ in raw units. Constant columns are omitted, and the same
transformation is applied to the prediction observation. Conditional on
the training transformation, this is a linear reparameterization of the
model with an intercept. Scaling is recomputed separately within every
validation fold.

The VaR fit minimizes mean check loss plus
$\lambda_{\gamma,t}\|\gamma_V\|_1$, leaving the five coefficients on $Z_s$
unpenalized. The M1 fit minimizes half the mean squared pseudo-residual
plus $\lambda_{\beta,t}\|\beta_V\|_1$. This loss normalization
rescales the penalty by a factor of two relative to
\eqref{eq:linear_ES_lasso}, without changing the fitted program
after the corresponding rescaling. Its unpenalized coefficients are
handled by partialling out $Z_s$ from both the response and standardized
macro-financial predictors. M0 is the corresponding unpenalized least-squares
regression on $Z_s$ alone. Each component of the M1 projection uses the
same partially penalized least-squares procedure. M0 projections are
ordinary least squares on $Z_s$.

Three validation blocks follow training prefixes of lengths
$\lfloor n_t/2\rfloor$, $\lfloor2n_t/3\rfloor$, and
$\lfloor5n_t/6\rfloor$. The candidate penalty ratios are
$\{\infty,1,0.5,0.2,0.1,0.05,0.02\}$; the infinite-penalty candidate
sets all additional coefficients to zero. Each finite penalty is a
ratio times the corresponding training-window zero-selection threshold.
For quantile regression this threshold is computed from a valid dual
subgradient at the growth-only fit, including fractional values at tied
zero residuals. For least squares it is the largest absolute sample
covariance of the residualized predictors with the residualized response.
Ties in validation loss favor the larger penalty.

In ES validation, each fold's training and validation pseudo-outcomes use
a quantile model fitted and temporally tuned only on that fold's training
prefix. For the squared-GES projection, each fold likewise constructs
its generated signal from its own earlier M1 fit. Final estimates use
all observations in $\mathcal I_t$. No final evaluation outcome is used
to estimate its own forecast or projection. Selected coefficients are
counted using a $10^{-6}$ threshold in standardized macro-financial units.

The score construction follows
Corollary~\ref{cor:residualized_orthogonality}; the inferential reference
requires the high-level conditions of Theorem~\ref{thm:main_asymptotics}.
Partial penalization and temporal cross-validation are empirical
implementation choices, not an independent verification of the primitive
penalty and nuisance-rate conditions. All model interpretations maintain
conditional quantile correctness. The redesign uses an evaluation sample
already examined in earlier specifications; it is an exploratory
application rather than a new untouched validation sample.

\subsection{Covariance estimation and sensitivity}
\label{sec:OA_empirical_inference}

For either LR or baseline score vectors, write
$\widetilde S_{t+1}=\widehat S_{t+1}-\bar S_P$ and define
\[
\widehat\Gamma_\ell
=\frac1P\sum_{t=R+\ell}^{T-1}
 \widetilde S_{t+1}\widetilde S_{t+1-\ell}',
\qquad
\widehat\Omega_L
=\widehat\Gamma_0+
\sum_{\ell=1}^{L}\left(1-\frac{\ell}{L+1}\right)
(\widehat\Gamma_\ell+\widehat\Gamma_\ell').
\]
The main convention is $L=4$, without prewhitening or a degrees-of-freedom
adjustment. This specifies the finite-sample centered Bartlett version
of the HAC estimator. Table~\ref{tab:OA_empirical_sensitivity} reports $L=2,4,8$;
Table~\ref{tab:OA_empirical_fixed} reports fixed estimation on the
same target dates. Every joint test
uses the full covariance matrix. Its variance-normalized minimum eigenvalue
must exceed $10^{-8}$ and its condition number must be below $10^8$;
no pseudoinverse is used. All reported joint matrices pass these numerical
criteria, which do not themselves establish a reliable asymptotic approximation.

The primary expanding family contains the seven M0 financial-exclusion
joint tests and seven M1 nonlinear joint tests. Holm adjustment is reported
both separately over each seven-country family and jointly over all 14.
The baseline counterparts and fixed-estimation results have analogous,
separate families. The scalar squared-GES diagnostic is a distinct
seven-country reference family. These adjustments have their usual
interpretation only when the marginal $p$-values are valid.

The US M0 rejection survives the combined adjustment at all three
bandwidths: adjusted $p$-values are 0.0329, 0.0120, and 0.0010 for
$L=2,4,8$. Italy's primary unadjusted $p=0.0152$ becomes 0.0913 under
the seven-country adjustment and 0.1979 under the combined adjustment.
The UK M0 test rejects under expanding estimation only at $L=8$,
not at the primary bandwidth.
Under fixed estimation, M0 rejects for the USA, UK, Germany, and Italy
within that scheme's combined family; only the USA and UK survive
adjustment across the 28 primary LR tests from both schemes.

For Canada, the M1 squared-GES LR $p$-value is 0.0322 at $L=2$,
0.0061 at $L=4$, and 0.0007 at $L=8$. At $L=4$, its separate
seven-country adjusted value is 0.0428. The primary nonlinear LR
nonrejection should not be interpreted as contradicting this scalar
finding: the instruments differ. Some scores are concentrated in crisis
observations; a single observation accounts for approximately 54\% of
Canada's centered squared-GES LR score variation and 98\% of Germany's
centered NFCI-squared LR score variation. All observations are retained.

\begin{table}[!htbp]
\centering
\begin{threeparttable}
\caption{LR joint-test sensitivity to the HAC lag length}
\label{tab:OA_empirical_sensitivity}
\small
\setlength{\tabcolsep}{5pt}
\begin{tabular}{lrrrrrr}
\toprule
Country & M0: $L=2$ & M0: $L=4$ & M0: $L=8$ & M1: $L=2$ & M1: $L=4$ & M1: $L=8$ \\
\midrule
USA & 0.0024 & 0.0009 & 0.0001 & 0.8006 & 0.7775 & 0.6991 \\
CAN & 0.3588 & 0.4393 & 0.5051 & 0.7825 & 0.8125 & 0.7997 \\
DEU & 0.2226 & 0.2432 & 0.2537 & 0.4071 & 0.4047 & 0.3208 \\
UK & 0.0945 & 0.0874 & 0.0023 & 0.5498 & 0.5078 & 0.5727 \\
ITA & 0.0294 & 0.0152 & 0.0075 & 0.0968 & 0.0954 & 0.0740 \\
JPN & 0.1933 & 0.1650 & 0.1351 & 0.4212 & 0.3098 & 0.2048 \\
FRA & 0.2014 & 0.2360 & 0.2164 & 0.3802 & 0.4326 & 0.4834 \\
\bottomrule
\end{tabular}
\begin{tablenotes}[flushleft]\footnotesize
\item Expanding estimation. Entries are unadjusted $p$-values for the three-component M0 exclusion test and the two-component M1 nonlinear test. Forecasts and scores are unchanged across lag lengths.
\end{tablenotes}
\end{threeparttable}
\end{table}

\begin{table}[!htbp]
\centering
\begin{threeparttable}
\caption{Fixed-estimation LR joint tests}
\label{tab:OA_empirical_fixed}
\small
\setlength{\tabcolsep}{5pt}
\begin{tabular}{lrrrr}
\toprule
Country & M0: LR & M0: Holm 14 & M1: LR & M1: Holm 14 \\
\midrule
USA & 0.0001 & 0.0007 & 0.8256 & 1.0000 \\
CAN & 0.1893 & 1.0000 & 0.1892 & 1.0000 \\
DEU & 0.0040 & 0.0476 & 0.5645 & 1.0000 \\
UK & $<0.0001$ & 0.0001 & 0.1126 & 0.9011 \\
ITA & 0.0043 & 0.0476 & 0.0623 & 0.6225 \\
JPN & 0.0712 & 0.6407 & 0.4769 & 1.0000 \\
FRA & 0.7933 & 1.0000 & 0.3386 & 1.0000 \\
\bottomrule
\end{tabular}
\begin{tablenotes}[flushleft]\footnotesize
\item Parameters and projection coefficients are estimated on the initial training sample and held fixed. Predictor observations update and evaluation dates match the expanding design. Bartlett HAC uses $L=4$. Holm adjustment combines both joint vectors across seven countries within this fixed-estimation family.
\end{tablenotes}
\end{threeparttable}
\end{table}
\FloatBarrier

\subsection{Forecast losses, benchmarks, and the financial increment}
\label{sec:OA_empirical_performance}

For outcomes and forecasts measured in percentage points, let
$I=\mathbf1\{y\le q\}$, $g(e)=(1+\exp(-e))^{-1}$, and
$G(e)=\log(1+\exp(e))$. The reported joint loss is
\begin{equation}
\label{eq:OA_empirical_loss}
\mathcal L_\alpha(y,q,e)
=(I-\alpha)(q-y)
+g(e)\left[e-q+\frac{I(q-y)}{\alpha}\right]-G(e).
\end{equation}
This is a member of the jointly consistent VaR--ES score class in
\citet{FisslerZiegel2016}. Unlike negative-ES-only specifications, it
is defined for all real forecasts, including positive or unordered
candidate actions. Its logistic scale is fixed at one percentage point
for all countries and procedures. Lower average loss is better; negative
values are admissible. Under integrability and a unique quantile, the
expected score is minimized first in $q$ at the true quantile and then
in $e$ at the true ES. Rankings are specific to this score and need not
hold for every consistent loss. The separate quantile loss is
$(I-\alpha)(q-y)$.

The empirical-AR benchmark estimates a constant and four growth lags by
least squares and uses the empirical distribution of training residuals.
For $m$ sorted residuals $r_{(1)},\ldots,r_{(m)}$, its innovation quantile
is $r_{(\lceil\alpha m\rceil)}$ and its innovation ES is
\[
\frac{\sum_{j=1}^{\lfloor\alpha m\rfloor}r_{(j)}
 +(\alpha m-\lfloor\alpha m\rfloor)
  r_{(\lfloor\alpha m\rfloor+1)}}{\alpha m}.
\]
The fitted conditional mean is added to both quantities. This benchmark
is distinct from M0, which uses the high-dimensional quantile-generated
pseudo-outcome rather than an empirical-AR predictive distribution.

The AR--GARCH benchmark combines the same growth-lag mean specification
with GARCH(1,1) conditional variance and standardized Student-$t$
innovations. Let $\nu>2$, $a_\alpha$ be the ordinary Student-$t_\nu$
quantile, and $f_\nu$ its density. With conditional mean $\mu$ and
standard deviation $\sigma$, its forecasts are
\[
q=\mu+\sigma\sqrt{\frac{\nu-2}{\nu}}\,a_\alpha,
\qquad
e=\mu-\frac{\sigma}{\alpha}\sqrt{\frac{\nu-2}{\nu}}
       \frac{\nu+a_\alpha^2}{\nu-1}f_\nu(a_\alpha).
\]
The two benchmarks use the same effective training and evaluation dates
as the nested models. Under fixed estimation, estimated parameters are
frozen while observed growth lags and the GARCH variance state update.

Table~\ref{tab:OA_empirical_benchmarks} uses common dates for the expanding
comparison. Five Japanese GARCH forecasts failed, leaving 65 common dates
for that panel; the main M0/M1 comparison retains all 70. Negative-curvature
Hessian flags affect 19 German and 29 Japanese expanding forecasts,
and all 70 forecasts from the fixed Japanese GARCH fit. These comparisons
are provisional; no replacement forecasts are substituted. Empirical-AR
forecasts are available at every target date. No fitted GES is clipped
or reordered to enforce $e\le q$.

The empirical-AR benchmark has lower average joint loss than both nested
models in six countries; Japan is the exception. This comparison evaluates
different complete predictive procedures. By contrast, the paired
M1-minus-M0 comparison holds the estimated VaR fixed. Its Japanese gain
is $-0.0151$ over all 70 expanding targets, but becomes a loss increase
of 0.0366 under fixed estimation. In the other six countries the fixed
M0/M1 forecasts coincide to numerical precision.

\begin{figure}[!htbp]
\centering
\includegraphics[width=0.95\textwidth,height=0.66\textheight,keepaspectratio]{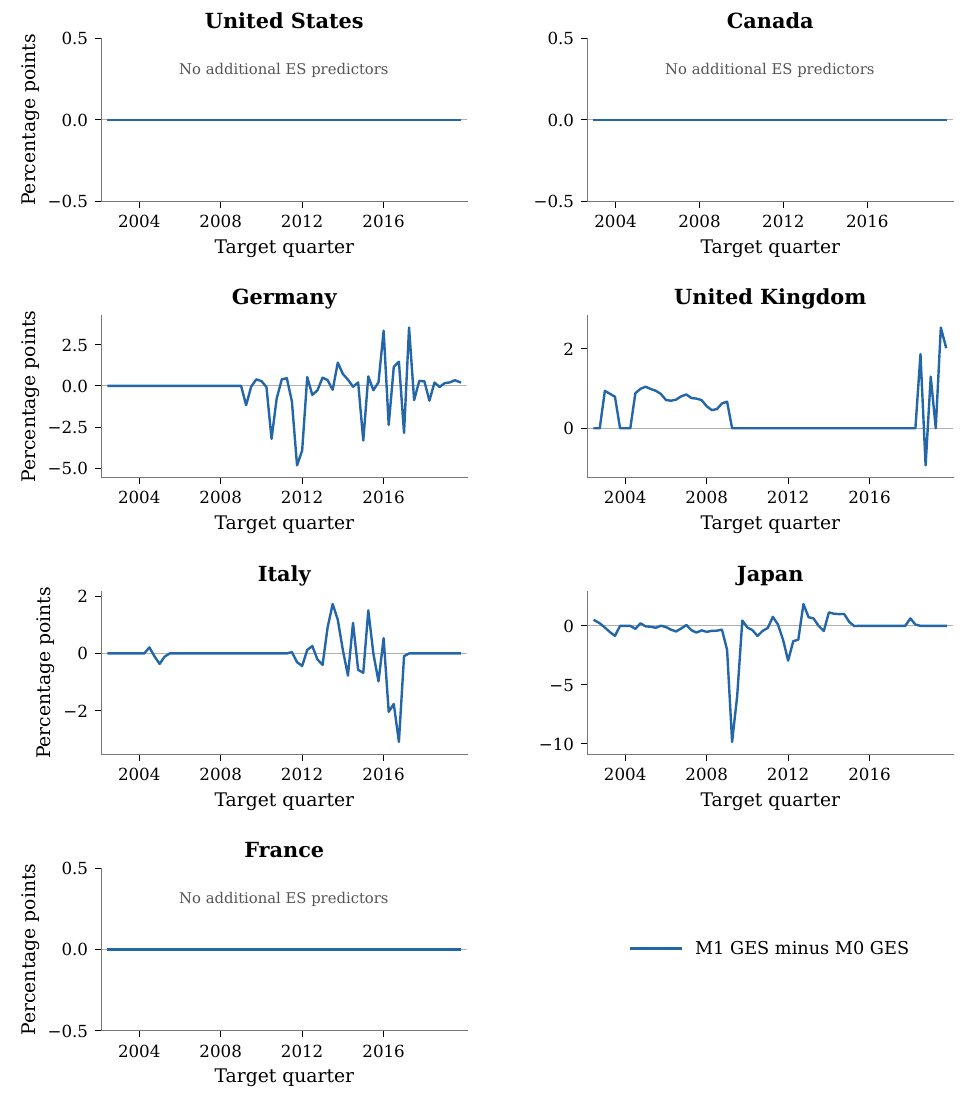}
\caption{Incremental fitted macro-financial contribution to GES. Curves show $\widehat e_{t,\alpha}^{(1)}-\widehat e_{t,\alpha}^{(0)}$ in percentage points under expanding estimation, including the adjustment of unpenalized growth coefficients. Country scales differ and all extremes are retained. The increment is zero in the USA, Canada, and France. It is not a causal effect or a separately studentized signal.}
\label{fig:empirical_increment}
\end{figure}

Because the two ES models share their training pseudo-outcome and leave
$Z_s$ unpenalized, projecting the fitted M1 training signal on $Z_s$
recovers the M0 signal. Thus
$\widehat e_{t,\alpha}^{(1)}-\widehat e_{t,\alpha}^{(0)}$ is the fitted
macro-financial contribution after accounting for the adjustment of the
growth coefficients. Figure~\ref{fig:empirical_increment} plots it without
clipping. It is not a causal effect. We do not assign it a separate
studentized test: under the restricted null its oracle projection
residual can vanish, violating the nondegenerate covariance condition.

\begin{table}[!htbp]
\centering
\begin{threeparttable}
\caption{Joint forecast losses on common evaluation dates}
\label{tab:OA_empirical_benchmarks}
\small
\setlength{\tabcolsep}{5pt}
\begin{tabular}{lrrrrr}
\toprule
Country & $N$ & M0 & M1 & Empirical AR & AR--GARCH \\
\midrule
USA & 70 & -1.2724 & -1.2724 & -1.3269 & -1.2909 \\
CAN & 68 & -1.2722 & -1.2722 & -1.2921 & -1.2755 \\
DEU & 70 & -0.6635 & -0.5786 & -0.7390 & -0.6591 \\
UK & 70 & -0.7127 & -0.6987 & -0.9240 & -1.3111 \\
ITA & 70 & -0.0644 & -0.0259 & -0.3256 & -0.4241 \\
JPN & 65 & 0.2856 & 0.2614 & 0.3706 & 0.2679 \\
FRA & 70 & -0.3971 & -0.3971 & -0.7210 & -0.9482 \\
\bottomrule
\end{tabular}
\begin{tablenotes}[flushleft]\footnotesize
\item Expanding estimation. Lower values of \eqref{eq:OA_empirical_loss} are better. Japan has 65 dates shared by all four procedures because five GARCH forecasts failed. The principal M0/M1 comparison in Table~\ref{tab:empirical_forecasts} retains all 70 Japanese targets. GARCH likelihood-curvature flags affect 19 German and 29 Japanese forecasts; those benchmark values are provisional. No claim of significance or score-independent dominance is made.
\end{tablenotes}
\end{threeparttable}
\end{table}

\begin{table}[!htbp]
\centering
\begin{threeparttable}
\caption{Quantile losses and GES coherence diagnostics}
\label{tab:OA_empirical_calibration}
\small
\setlength{\tabcolsep}{5pt}
\begin{tabular}{lrrrrr}
\toprule
Country & M0/M1: loss & AR: loss & GARCH: loss & M0: $e>q$ & M1: $e>q$ \\
\midrule
USA & 0.1357 & 0.1299 & 0.1345 & 1 & 1 \\
CAN & 0.1324 & 0.1291 & 0.1267 & 0 & 0 \\
DEU & 0.2270 & 0.2078 & 0.2197 & 3 & 9 \\
UK & 0.1933 & 0.1762 & 0.1214 & 3 & 11 \\
ITA & 0.1892 & 0.1744 & 0.1420 & 28 & 30 \\
JPN & 0.3150 & 0.3270 & 0.3337 & 8 & 7 \\
FRA & 0.1473 & 0.1477 & 0.1114 & 21 & 21 \\
\bottomrule
\end{tabular}
\begin{tablenotes}[flushleft]\footnotesize
\item Expanding estimation; all available dates for each procedure. Quantile loss is $(\mathbf1\{Y_{t+1}\le q\}-\alpha)(q-Y_{t+1})$. The GARCH Japanese average uses 65 dates, while its AR and nested-model averages use 70. Counts refer to same-level 10\% GES above GaR; no forecasts are repaired. M0 and M1 share their quantile forecasts.
\end{tablenotes}
\end{threeparttable}
\end{table}

\FloatBarrier

\subsection{Interpretation of baseline and LR differences}
\label{sec:OA_empirical_interp}

The Canadian nonlinear-vector comparison is consistent with the pattern
of baseline overrejection in the null simulations. However, in observed
data, the null is unknown and changing instruments also changes sensitivity
to alternatives. It is therefore not possible to identify a baseline
rejection as false from the LR nonrejection alone. Conversely, the
Canadian squared-GES result shows that orthogonalization can strengthen
evidence. High-dimensional prediction, restriction testing, and comparative
forecast loss answer different questions; none of the reported
nonrejections establishes complete calibration or model adequacy.

\FloatBarrier

\subsection{Forecasting with a small set of tested predictors}
\label{sec:OA_empirical_targeted}

We retain the common high-dimensional GaR first step and consider
three candidate additions to the growth-only ES regression:
the latest included financial-conditions index, investment growth
and term spread, as used in the M0 backtest.
For each subset of these predictors, we estimate the ES coefficients
by unpenalized least squares on the common training pseudo-outcome.
M2 always includes all three predictors. M3 selects among their
eight subsets, including the empty subset corresponding to M0.

Selection uses only observations preceding each forecast target.
For a training sample of size $n$, the three training endpoints
are $\lfloor n/2\rfloor$, $\lfloor 2n/3\rfloor$ and
$\lfloor 5n/6\rfloor$; the corresponding validation blocks end
at $\lfloor 2n/3\rfloor$, $\lfloor 5n/6\rfloor$ and $n$.
Each fold uses its own training-prefix GaR fit to construct
both training and validation pseudo-outcomes.
All eight subsets share these fold-specific pseudo-outcomes.
M3 minimizes the sum of validation squared errors divided by
the total number of validation observations. Ties favor fewer
additions and then a fixed ordering. The selected subset is
re-estimated using the complete available training sample.

Table~\ref{tab:empirical_targeted} compares the procedures on
the original expanding-window evaluation dates. Joint loss uses
the same VaR--ES score as the main application. Squared
pseudo-outcome loss is
\[
\left\{
Y^*_{t+1}(\widehat\gamma_t)
-\widehat e_{t,\alpha}^{(j)}
\right\}^{2},
\]
with the same estimated GaR used for every ES procedure.
The comparisons are descriptive: the candidate set was motivated
by previously inspected backtesting results, and chronological
validation does not make the evaluation sample an untouched holdout.

M2 lowers both losses relative to M0 in Canada, Germany, the UK,
Italy and France; M3 does so in the USA, Canada, Germany, the UK
and France. M3 worsens both losses in Italy. In Japan, both
extensions lower squared loss but increase joint loss.
Panel B reports violations of the lower-tail restriction
$\widehat e_{t,\alpha}\leq\widehat q_{t,\alpha}$; forecasts
are not clipped or otherwise altered.
The empirical-AR benchmark, which uses its own GaR forecast,
has lower average joint loss than both extensions in six
countries, with Japan the exception.

\begin{table}[!htbp]
\centering
\caption{Forecast performance of small ES extensions}
\label{tab:empirical_targeted}
\small
\begin{tabular}{@{}lrrrrr@{}}
\toprule
\multicolumn{6}{l}{
Panel A: Average loss differences relative to M0} \\
\midrule
& & \multicolumn{2}{c}{Joint VaR--ES loss}
  & \multicolumn{2}{c}{Squared pseudo-outcome loss} \\
\cmidrule(lr){3-4}\cmidrule(lr){5-6}
Country & $P$ & M2 & M3 & M2 & M3 \\
\midrule
USA & 70 & -0.0092 & -0.0151 &  0.0115 & -0.1463 \\
CAN & 68 & -0.0008 & -0.0032 & -0.0241 & -0.0496 \\
DEU & 70 & -0.0423 & -0.0358 & -1.6653 & -1.0608 \\
UK  & 70 & -0.0653 & -0.0254 & -0.2422 & -0.3847 \\
ITA & 70 & -0.0561 &  0.0138 & -1.5539 &  0.1728 \\
JPN & 70 &  0.0579 &  0.0498 & -2.5152 & -2.8990 \\
FRA & 70 & -0.2185 & -0.1484 & -1.8766 & -1.2189 \\
\midrule
\multicolumn{6}{l}{
Panel B: Number of forecasts with ES above VaR} \\
\midrule
Country & $P$ & M0 & M1 & M2 & M3 \\
\midrule
USA & 70 &  1 &  1 & 25 &  8 \\
CAN & 68 &  0 &  0 &  1 &  0 \\
DEU & 70 &  3 &  9 & 18 & 12 \\
UK  & 70 &  3 & 11 & 37 & 13 \\
ITA & 70 & 28 & 30 & 38 & 38 \\
JPN & 70 &  8 &  7 & 22 & 23 \\
FRA & 70 & 21 & 21 & 40 & 34 \\
\bottomrule
\end{tabular}

\par\smallskip
\begin{minipage}{0.98\linewidth}
\footnotesize
\textit{Notes:}
All procedures use expanding estimation and the same GaR forecasts
at $\alpha=0.10$.
M2 adds the three M0 backtesting predictors without penalization;
M3 selects among their eight subsets by chronological validation.
Panel A reports the average loss of each extension minus that of M0;
negative values favor the extension.
Panel B counts violations on the same evaluation dates.
Comparisons are exploratory and carry no significance claims.
\end{minipage}
\end{table}

\FloatBarrier

\end{document}